\documentclass[twocolumn,english,amsmath,amssymb,nofootinbib,superscriptaddress,showpacs,showkeys]{revtex4}
\usepackage{ae,aecompl}
\usepackage[T1]{fontenc}
\usepackage[latin1]{inputenc}
\usepackage{color}
\usepackage{babel}
\usepackage{verbatim}
\usepackage{float}
\usepackage{textcomp}
\usepackage{amsmath}
\usepackage{graphicx}
\usepackage{wasysym}
\usepackage{esint}
\usepackage[unicode=true,
 bookmarks=false,
 breaklinks=false,pdfborder={0 0 1},backref=false,colorlinks=false]
 {hyperref}

\makeatletter

\providecommand{\tabularnewline}{\\}

\@ifundefined{textcolor}{}
{%
 \definecolor{BLACK}{gray}{0}
 \definecolor{WHITE}{gray}{1}
 \definecolor{RED}{rgb}{1,0,0}
 \definecolor{GREEN}{rgb}{0,1,0}
 \definecolor{BLUE}{rgb}{0,0,1}
 \definecolor{CYAN}{cmyk}{1,0,0,0}
 \definecolor{MAGENTA}{cmyk}{0,1,0,0}
 \definecolor{YELLOW}{cmyk}{0,0,1,0}
}

\usepackage{babel}
\usepackage{color,soul}

\usepackage{aecompl}
\def\O#1#2{O(#1)}
\def\Os#1{O(#1)}

\def\Ra{\Rightarrow}
\def\be{\begin{equation}}
\def\ee{\end{equation}}
\def\best{\begin{equation*}}
\def\eest{\end{equation*}}
\def\beqn{\begin{eqnarray}}
\def\eeqn{\end{eqnarray}}
\def\a{\alpha}
\def\b{\beta}
\def\g{\gamma}
\def\d{\delta}
\def\e{\epsilon}
\def\l{\lambda}
\def\s{\sigma}

\def\eb{{\bf e}}

\def\inn#1{\mathbf{#1}\!\cdot\!\mathbf{#1}}

\def\innst#1{\star\!\mathbf{#1}\!\cdot\!\mathbf{#1}}
\def\OO{{\cal O}}
\def\Op{{\cal O}'}
\def\Ou{{\cal O}({\bf u})}
\def\Oup{{\cal O}'({\bf u}')}
\def\M{\mathbb{M}}
\def\J{\mathbb{J}}

\def\Qop{\QQ^\a_{~\b}}

\def\EE{\mathbb{E}}
\def\HH{\mathbb{H}}

\def\QQ{\mathbb{Q}}
\def\Maxwell{Faraday }

\def\v{w}
\def\spacev{y}
\def\eb{\mathfrak{e}}
\def\bb{\mathfrak{b}}
\def\AD{\zeta_{\textrm{g}}}
\def\tI{t_{{\rm I}}}
\usepackage{babel}

\def\AC{\mathbb{A}_{{\rm C}}}
\def\BC{\mathbb{B}_{{\rm C}}}

\def\QC{{\cal Q}}

\def\IS{I_S}
\def\JS{J_S}
\def\AA{c}
\def\x{x}
\def\K{K}
\def\L{L}
\def\Rkk{R_{kk}}
\def\Rll{R_{ll}}
\def\TT{{\cal T}}

\def\hati{{\hat{\imath}}}

\def\hatl{{\hat{l}}}
\def\hatm{{\hat{m}}}

\def\kt{\tilde{k}}
\def\lt{\tilde{l}}

\usepackage{babel}

\makeatother

\begin{document}



\title{Gravitomagnetism and the significance of the curvature scalar invariants}

\author{L.~Filipe~O.~Costa}

\email{lfilipecosta@tecnico.ulisboa.pt}

\affiliation{CAMGSD, Departamento de Matemática, Instituto Superior Técnico, 1049-001
Lisboa, Portugal}

\affiliation{Centro de Física do Porto --- CFP, Departamento de Física e Astronomia,
Universidade do Porto, 4169-007 Porto, Portugal}

\author{Lode~Wylleman}

\email{lode.wylleman@uis.no}

\affiliation{Faculty of Science and Technology, University of Stavanger, N-4036
Stavanger, Norway}

\affiliation{Faculty of Applied Sciences TW16, Ghent University, Galglaan 2, 9000
Ghent, Belgium}

\author{José~Natário}

\email{jnatar@math.ist.utl.pt}

\affiliation{CAMGSD, Departamento de Matemática, Instituto Superior Técnico, 1049-001
Lisboa, Portugal}

\date{\today}
\begin{abstract}
The curvature invariants have been subject of interest due to the
debate concerning the notions of intrinsic/extrinsic frame-dragging,
the use of the electromagnetic analogy in such classification, and
the question of whether there is a fundamental difference between
the gravitomagnetic fields arising from the translational and rotational
motions of the sources (which have been subject of observational and
experimental tests, including the dedicated Gravity Probe-B and LARES
space missions). In this work we clarify both the algebraic and physical
meaning of the curvature invariants and their electromagnetic counterparts.
They are seen to yield conditions for the existence of observers measuring
vanishing electric/magnetic fields and gravitoelectric/gravitomagnetic
tidal tensors, respectively. We determine these observers %
(in the gravitational sector and in the presence of sources, for the
more relevant gravitomagnetic case) obtaining their velocities explicitly
in terms of the fields/tidal tensors as measured by an arbitrary observer.
The structure of the invariants of the astrophysical setups of interest
is studied in detail, and its relationship with the gravitomagnetic
effects is dissected. Finally, a new classification for intrinsic/extrinsic
gravitomagnetism is proposed. 
\end{abstract}
\maketitle
\tableofcontents{}

\section{Introduction}

In the last three decades different experiments succeeded in measuring
the so-called ``gravitomagnetic (GM) field'' --- which can be described
as the inertial force generated by mass/energy currents that is manifest
in the precession of gyroscopes, or in the Coriolis-like (apparent)
acceleration of a particle in geodesic motion relative to a frame
fixed to the distant stars. One can cast (e.g. \cite{KopeikinInv,Ciufolini LLR})
the effects detected in two main types: rotational gravitomagnetism,
arising from the rotation of a celestial body, and translational gravitomagnetism,
originated by bodies in translational motion with respect to the reference
frame.

Translational gravitomagnetism has been detected, to high precision,
in a number of ways. The observations of the binary pulsar PSR 1913
+16 (namely the effect of the gravitomagnetic field caused by the
motion of each star in the orbit of its companion) \cite{Nordtvedt1988}
form one example; other effects that can be cast as translational
gravitomagnetism (cf. \cite{AshbySaless,Review of GPB,OConnel_CQG2005,CiufoliniNature2007,PaperIAU,JantzenThomas})
are the different measurements of the ``geodetic'' (or de Sitter)
precession, namely the precession of the Earth-Moon system along its
orbit around the Sun due to the gravitomagnetic field generated by
the relative motion of the Sun, detected in the analysis of Lunar
Laser Ranging (LLR) data \cite{BertottiCiufoliniBender1987,Shapiro_et_alPRL1988,WilliamsNewhallDickey1996};
the precession of the gyroscopes in the Gravity Probe-B \cite{GPB}
due to the translational motion of the Earth relative to the probe;
and the precession of the pulsar's spin vector in the binary systems
PSR J0737\textminus 3039A/B \cite{BretonStairsScience2008} and (with
lower precision) PSR B1534+12 \cite{StairsThorsettArzoumanianPRL2004}.
It has also been claimed \cite{NordtvedtPRL1988,Nordtvedt2003,MurphyNordtvedtPRL1,SoffelKlioner}
that the influence on the lunar orbit of the gravitomagnetic field
generated by the translational motion of the Earth relative to the
Sun has been detected via LLR.\footnote{There is however a debate concerning such claim, see \cite{MurphyNordtvedtPRL1,KopeikinPRL,MurphyNordtvedtPRL2,SoffelKlioner,Ciufolini LLR}.
On the one hand, as argued in \cite{NordtvedtPRL1988,Nordtvedt2003,MurphyNordtvedtPRL1,MurphyNordtvedtPRL2,SoffelKlioner},
the constraints on the PPN parameters imposed by the measurements
performed so far imply the gravitomagnetic interaction; and once both
local Lorentz invariance and the Moon's orbit, as seen from the Earth,
have been tested with sufficient accuracy, then in the Sun's rest
frame one unavoidably needs to take into account the Earth's GM field
to obtain the Moon's correct trajectory. On the other hand, in agreement
with \cite{KopeikinPRL,Ciufolini LLR}, it is true that LLR cannot
(since the stations are based on Earth) \emph{directly} measure the
Earth's translational gravitomagnetic field (which is zero in the
Earth's rest frame). The only GM field that is being \emph{directly}
tested by LLR is in fact the one generated by the relative translation
of the Sun, which is essentially \cite{Ciufolini LLR} the one involved
in the geodetic precession of the Earth-Moon system detected in \cite{BertottiCiufoliniBender1987,Shapiro_et_alPRL1988,WilliamsNewhallDickey1996}.}

The gravitomagnetic field generated by the rotational motion of celestial
bodies has been more elusive, due to its typically smaller magnitude.
The only measurements performed to date concern the gravitomagnetic
field arising from the Earth's rotation, detected in the analysis
of the LAGEOS Satellites data \cite{Ciufolini Lageos,CiufoliniPavlisPeron},
by the Gravity Probe-B mission \cite{GPB,Review of GPB}, and by the
LARES mission \cite{LARES}, with announced accuracies of 10\%, 19\%,
and 2\%, respectively. Its detection to a 0.2\% accuracy is also the
primary goal of the planned LARES 2 mission \cite{LARES2}.

The curvature invariants have been subject of interest in this context
due to the ongoing debate concerning the notions of ``intrinsic''
vs.\ ``frame-dependent'' gravitomagnetism, and the question of
whether there is a fundamental difference between the gravitomagnetic
fields generated by the rotation and the translational motion of celestial
bodies. The use of the curvature invariants in such discussion is
motivated by an analogy with the quadratic invariants of the \Maxwell
tensor: $\inn{F}$ and $\innst{F}$. In electromagnetism these invariants
give conditions for the vanishing of the electric/magnetic fields
for some observers; in particular, when $\innst{F}=0$ and $-\inn{F}>0$,
there are observers $u^{\alpha}$ measuring zero magnetic field $\mathbf{B}$.
The latter is said to be ``frame-dependent'', and an example is
the field produced by a uniformly moving (non-spinning) point charge.
On the other hand, if $\innst{F}\ne0$ or $-\inn{F}<0$, then $\mathbf{B}\ne0$
for \emph{all} observers; $\mathbf{B}$ is said to be ``intrinsic'',
and an example is the field produced by a spinning charge. Based on
the formal analogy with the quadratic curvature invariants $\inn{R}$
and $\innst{R}$ (the Kretschmann and Chern-Pontryagin invariants,
respectively), a similar classification was (somewhat naively) proposed
in \cite{Gravitation and Inertia,Ciufolini LLR} (and then supported
in \cite{KopeikinInv,OConnel Inv,KopeikinFomlalont,PfisterKingBook,Pfister,Overduin,PascualSanchez})
for the \emph{gravitomagnetic field}: the non-vanishing or vanishing
of $\innst{R}$ would signal the presence of intrinsic or frame-dependent
gravitomagnetism, respectively. This scheme has then been used to
imply a fundamental distinction between the ``translational'' gravitomagnetic
fields mentioned above, and the gravitomagnetic field generated by
Earth's rotational motion implied in \cite{Ciufolini Lageos,GPB}.

On the other hand, the scalar invariants of the Weyl tensor (which
in vacuum becomes the Riemann tensor), have been studied in the context
of a hitherto separate research field \cite{matte,belRadiation,Zakharov,McIntosh et al 1994,Bonnor:1995zf,Lozanovski:99,cherubini:02,WyllePRD}.
In particular, the invariant criteria for the classification of a
vacuum Riemann tensor as purely electric/magnetic (such classification
implying the existence of some observer for which its electric/magnetic
part vanishes) are well established since the work by McIntosh \emph{et
al} \cite{McIntosh et al 1994} (see also \cite{FerrandoSaez2002,WyllePRD}).

In the present contribution, we start by discussing the rigorous mathematical
implications of the scalar invariants, closing the gap between these
two research fields. We show them to yield, in the electromagnetic
sector, (sufficient) conditions for the existence of observers for
which the electric or magnetic fields vanish; and in the gravitational
sector (insufficient) conditions for the vanishing of the gravitoelectric
or gravitomagnetic tidal tensors. We fully characterize such observers
in the electromagnetic and in the vacuum gravitational sectors; in
the presence of gravitational sources, we fully characterize the more
relevant case of observers measuring a vanishing gravitomagnetic tidal
tensor. We then study in detail and physically interpret the invariant
structure of the astrophysical setups of interest, and, finally, dissect
their actual implications on the motion of test particles, in particular
the relation with the gravitomagnetic effects.

\subsection{Notation and conventions\label{sub:Notation-and-conventions}}
\begin{enumerate}
\item We use the convention $G=c=1$, where $G$ is the gravitational constant
and $c$ the speed of light, and metric signature $(-+++)$; $\epsilon_{\alpha\beta\sigma\gamma}\equiv\sqrt{-g}[\alpha\beta\gamma\delta]$
is the Levi-Civita tensor, with the orientation $[1230]=1$ (i.e.,
in flat spacetime, $\epsilon_{1230}=1$); $\epsilon_{ijk}\equiv\epsilon_{ijk0}$.
\textcolor{black}{Greek letters $\alpha$, $\beta$, $\gamma$, ...
denote 4-D spacetime indices, running 0-3; Roman letters $i,j,k,...$
denote spatial indices, running 1-3}. The convention for the Riemann
tensor is $R_{\ \beta\mu\nu}^{\alpha}=\Gamma_{\beta\nu,\mu}^{\alpha}-\Gamma_{\beta\mu,\nu}^{\alpha}+...$. 
\item \emph{Tensors and vectors}. To refer to tensors (including 4-vectors)
we use either a bold font symbol $\mathbf{T}$ or abstract index notation
$T^{\alpha\beta\gamma\cdots}$. $\mathbf{S}\cdot\mathbf{T}$ stands
for the full contraction $S^{\alpha\beta\gamma\cdots}T_{\alpha\beta\gamma\cdots}$.
Round (square) brackets around indices indicate (anti)symmetrization.
$\star$ denotes the Hodge dual: $\star F_{\alpha\beta}\equiv\epsilon_{\alpha\beta}^{\ \ \mu\nu}F_{\mu\nu}/2$
for an antisymmetric tensor $F_{\alpha\beta}=F_{[\alpha\beta]}$,
while $\star R_{\alpha\beta\gamma\delta}\equiv\epsilon_{\alpha\beta}{}^{\mu\nu}R_{\mu\nu\gamma\delta}/2$
and $R\star_{\alpha\beta\gamma\delta}\equiv R_{\alpha\beta\mu\nu}\epsilon^{\mu\nu}{}_{\gamma\delta}/2$
are, respectively, the dual in the first and the second pair of indices
for Riemann-like tensors $R_{\alpha\beta\gamma\delta}=R_{[\alpha\beta]\gamma\delta}=R_{\alpha\beta[\gamma\delta]}$.
Arrow notation $\vec{V}$ denotes the collection of space components
of a vector $V^{\alpha}$ in a given frame. 
\item \emph{Observers and reference frames}. Following \cite{deFeliceClarkeBook,The many faces,GEM User Manual,PaperAnalogies},
an observer (of 4-velocity ${\bf u}\equiv u^{\alpha}$), denoted by
$\OO$ or $\Ou$, is an entity endowed with a worldline in spacetime
(tangent to $u^{\alpha}$), equipped with (besides other possible
measurement devices) a clock and a system of axes to perform measurements.
By reference frame ($\mathcal{S}$), over an extended spacetime region,
we understand a 4-D basis (which could be orthonormal, or any coordinate
basis) composed of a time-like plus 3 space-like vectors, continuously
defined therein; it embodies a \emph{congruence} of observers (whose
worldlines are the integral lines of the time-axis). The projector
onto the instantaneous rest space of $\Ou$ is 
\begin{equation}
h_{\beta}^{\alpha}\equiv\delta_{\beta}^{\alpha}+u^{\alpha}u_{\beta}\ .\label{eq:SpaceProjector}
\end{equation}

\item \emph{``Dyadic'' notation.}\label{enu:Diadic-notation} Let $\mathbf{e}_{\hat{\imath}}$
be an orthonormal basis in the rest space of $\Ou$, $\mathbf{e}_{\hat{\imath}}\cdot\mathbf{u}=0$.
Following \cite{CampbelMacekMorgan,CampbellMorganAjp}, sometimes
we shall denote the collection of space components $A_{\hat{\imath}\hat{\jmath}}$
of a symmetric tensor $\mathbf{A}$ by $\overleftrightarrow{A}$;
the following notation applies: $(\overleftrightarrow{A}\times\vec{v})_{\hat{k}\hat{l}}\equiv\epsilon_{\ \ \hat{l}}^{\hat{\imath}\hat{\jmath}}A_{\hat{k}\hat{\imath}}v_{\hat{\jmath}}$;
$(\vec{v}\times\overleftrightarrow{A})_{\hat{k}\hat{l}}\equiv\epsilon_{\ \ \hat{k}}^{\hat{\imath}\hat{\jmath}}v_{\hat{\imath}}A_{\hat{l}\hat{\jmath}}$;
$(\overleftrightarrow{A}\times\overleftrightarrow{B})_{\hat{\imath}}=\e_{\hat{\imath}\hat{\jmath}\hat{k}}A^{\hat{\jmath}\hat{l}}B_{\hat{l}\hat{k}}$.
Hats in the indices denote orthonormal tetrad components (dropped
in approximations where the distinction between coordinate and tetrad
indices is immaterial.) 
\end{enumerate}

\section{Electromagnetic Scalar Invariants\label{sec:Electromagnetic-Scalar-Invariant}}

As a preparation for the gravitational case, we start by discussing
the electromagnetic invariants and their physical meaning.

In terms of the \Maxwell tensor ${\bf F}\equiv F^{\a\b}$, the electric
and magnetic \emph{4-vector} fields as measured by an observer $\Ou$
of 4-velocity $u^{\alpha}$ are given by 
\begin{equation}
E^{\alpha}\equiv F_{\ \beta}^{\alpha}u^{\beta}\ ,\qquad B^{\alpha}\equiv\star F_{\ \beta}^{\alpha}u^{\beta}\ .\label{eq:ElectricFieldCov}
\end{equation}

Both $E^{\a}$ and $B^{\a}$ are spatial with respect to $u^{\alpha}$
($E^{\alpha}u_{\alpha}=B^{\alpha}u_{\alpha}=0$) and thus have 3 independent
components each, encoding the 6 independent components of $F^{\a\b}$
and assembled into associated 3-vectors $\vec{E}$ and $\vec{B}$.
The \Maxwell tensor and its Hodge dual decompose in terms of $E^{\a}$
and $B^{\a}$ as 
\begin{eqnarray}
F^{\alpha\beta} & = & 2u^{[\a}E^{\b]}+\epsilon^{\alpha\beta\gamma\delta}B_{\gamma}u_{\delta}\ ,\label{FaradayDecomp}\\
\star F^{\alpha\beta} & = & 2u^{[\a}B^{\b]}-\epsilon^{\alpha\beta\gamma\delta}E_{\gamma}u_{\delta}\ .\label{Fstar}
\end{eqnarray}

The {\em electromagnetic scalar invariants} are the two real, independent
relativistic Lorentz scalars that can be constructed from the \Maxwell
tensor: 
\begin{align}
 & -\tfrac{1}{2}\inn{F}\ \equiv\ -\tfrac{1}{2}F^{\alpha\beta}F_{\alpha\beta}=E^{\alpha}E_{\alpha}-B^{\alpha}B_{\alpha}\ ,
\label{eq:EBSquare}\\
 & -\tfrac{1}{4}\innst{F}\ \equiv\ -\tfrac{1}{4}\star\!F^{\alpha\beta}F_{\alpha\beta}=E^{\alpha}B_{\alpha}\ .
\label{eq:EB}
\end{align}
The final expressions in (\ref{eq:EBSquare})-(\ref{eq:EB}), which
read $\vec{E}^{2}-\vec{B}^{2}$ and $\vec{E}\cdot\vec{B}$ in 3-vector
notation, are thus independent of the observer $\Ou$. In particular,
if $\vec{E}\cdot\vec{B}=0$ or $\vec{E}^{2}$ is larger, smaller,
or equal to $\vec{B}^{2}$ for some observer, then this is true for
every observer.

At each point the \Maxwell tensor can be completely classified in
terms of its invariants, and one distinguishes the following cases~\cite{LandauLifshitz,Stephani}: 
\begin{enumerate}
\item[\textbf{(i)}] $\vec{E}\cdot\vec{B}\neq0\ [\Leftrightarrow\innst{F}\neq0]\,\Ra\,$
$\vec{E}$ and $\vec{B}$ are both non-vanishing for all observers. 
\item[\textbf{(ii)}] $\vec{E}\cdot\vec{B}=0$ and $\vec{E}^{2}-\vec{B}^{2}>0\,(<0)\,[\Leftrightarrow\innst{F}=0,\ -\inn{F}>0\,(<0)]$
$\Rightarrow$ one can \emph{always} find observers for which the
magnetic field $\vec{B}$ (electric field $\vec{E}$) vanishes. The
electromagnetic field is thus classified as \emph{purely electric}
(\emph{purely magnetic}). 
\item[\textbf{(iii)}] {\em null} case\footnote{\label{foot:pure radiation}If the \Maxwell tensor is null in an
open 4-D region, then the EM field there is a {\em pure radiation}
field, see \cite{Stephani} p. 68.}: $\vec{E}\cdot\vec{B}=\vec{E}^{2}-\vec{B}^{2}=0\ [\Leftrightarrow\innst{F}=\inn{F}=0]$
$\Ra$ either $\vec{E}=\vec{B}=0$, or $\vec{E}$ and $\vec{B}$ are
both non-vanishing for all observers. 
\end{enumerate}
The implications in \textbf{(i)} and \textbf{(iii)} are obvious. The
proof of statement \textbf{(ii)}, as well as the explicit construction
of the observers measuring no magnetic or electric field is given
in the next subsection; one conclusion is however immediate: the condition
$\vec{E}^{2}-\vec{B}^{2}>0\,(<0)$ implies the electric (magnetic)
field to be non-zero for all observers, such that for a non-zero \Maxwell
tensor there cannot, simultaneously, exist observers for which $\vec{B}=0$
and observers for which $\vec{E}=0$.

\subsection{Observers measuring no magnetic/electric fields\label{sub:Observers-with-no}}

Consider two observers $\Ou$ and $\Oup$; their 4-velocities are
related by 
\begin{equation}
u'^{\alpha}=\gamma(u^{\alpha}+v^{\alpha})\ ;\qquad\gamma\equiv-u^{\alpha}u'_{\alpha}=\frac{1}{\sqrt{1-v^{\alpha}v_{\alpha}}}\ ,\label{eq:u_u'}
\end{equation}
where $v^{\alpha}$ is a vector orthogonal to $u^{\alpha}$, $u^{\alpha}v_{\alpha}=0$,
interpreted as the spatial velocity of $\Oup$ relative to $\Ou$.
In a locally \emph{inertial} frame momentarily comoving with $\Ou$
(where $u^{i}=0$), $v^{i}=dx^{i}/dt$, yielding the ordinary 3-velocity
of $\Oup$.

By (\ref{eq:ElectricFieldCov},\,\ref{FaradayDecomp},\,\ref{Fstar})
the electric and magnetic fields measured by $\Oup$ are related to
the ones measured by $\Ou$ according to 
\begin{eqnarray}
E'^{\alpha} & = & \left[2E^{[\beta}u^{\alpha]}+\epsilon^{\alpha\beta\gamma\delta}B_{\gamma}u_{\delta}\right]u'_{\beta}\ ,\label{e2}\\
B'^{\alpha} & = & \left[2B^{[\beta}u^{\alpha]}-\epsilon^{\alpha\beta\gamma\delta}E_{\gamma}u_{\delta}\right]u'_{\beta}\ .\label{b2}
\end{eqnarray}
To make contact with the textbooks on classical electromagnetism,
consider, in flat spacetime, the inertial frames $\mathcal{S}$ and
$\mathcal{S}'$ momentarily comoving with the observers $\Ou$ and
$\Oup$, respectively; in this special case one obtains the well known
non-covariant expressions (e.g. Eqs. (11.149) of \cite{Jackson})
\begin{eqnarray}
\vec{E}' & = & \gamma\left(\vec{E}+\vec{v}\times\vec{B}\right)-\frac{\gamma^{2}}{\gamma+1}\vec{v}\left(\vec{v}\cdot\vec{E}\right)\ ,\label{Evec'}\\
\vec{B}' & = & \gamma\left(\vec{B}-\vec{v}\times\vec{E}\right)-\frac{\gamma^{2}}{\gamma+1}\vec{v}\left(\vec{v}\cdot\vec{B}\right)\ ,\label{Bvec'}
\end{eqnarray}
where $\vec{E}'$ and $\vec{B}'$ are space components of the electric
and magnetic fields measured by $\Oup$ \emph{and expressed} in the
coordinate system $\mathcal{S}'$ (the time components $E'^{0}$ and
$B'^{0}$ are zero in $\mathcal{S}'$).

Let us now prove that observers exist for which $B'^{\a}=0$ if, and
only if, $\vec{E}\cdot\vec{B}=0$ and $\vec{E}^{2}>\vec{B}^{2}$ (i.e.,~$-\mathbf{F}\cdot\mathbf{F}>0$
and $\star\mathbf{F}\cdot\mathbf{F}=0$). From \eqref{b2}, the equation
$B'^{\a}=0$ splits into two components, one parallel to $u^{\alpha}$:
$u^{\alpha}B^{\beta}u'_{\beta}=0$, i.e., 
\begin{equation}
B^{\beta}u'_{\beta}=B^{\beta}v_{\beta}=0\,,\label{eq1}
\end{equation}
plus one orthogonal to $u^{\alpha}$,

\begin{equation}
B^{\alpha}=-\frac{\epsilon^{\alpha\beta\gamma\delta}u_{\delta}E_{\gamma}u'_{\beta}}{u^{\mu}u'_{\mu}}=\epsilon^{\alpha\beta\gamma\delta}u_{\delta}E_{\gamma}v_{\beta}\ ,\label{eq22}
\end{equation}
which implies \eqref{eq1} on its turn. In the rest frame of $\Ou$,
and in 3-vector form, Eq. \eqref{eq22} reads 
\begin{equation}
\vec{B}=\vec{v}\times\vec{E}\ .\label{eq2}
\end{equation}
Hence, it is possible to find an observer for which $\vec{B}'$ vanishes
if and only if \eqref{eq2} admits a solution $\vec{v}$. Since $|\vec{v}|<c=1$,
this is the case if and only if $\vec{B}$ lies in the plane orthogonal
to $\vec{E}$ and is contained within a circle of radius $|\vec{E}|$,
which precisely means $\vec{E}\cdot\vec{B}=0$ and $\vec{E}^{2}>\vec{B}^{2}$.
This concludes the proof.

To obtain the velocities of the observers for which $B'^{\a}=0$ (i.e,
$\vec{B}'=0$), it is useful to decompose $v^{\alpha}$ into its projections
parallel and orthogonal to $E^{\alpha}$, 
\[
v_{\parallel E}^{\alpha}=\frac{E_{\beta}v^{\beta}}{E_{\nu}E^{\nu}}E^{\alpha}\ ;\qquad v_{\perp E}^{\alpha}=v^{\alpha}-v_{\parallel E}^{\alpha}\ ,
\]
and to recall the definition of the Poynting vector measured by $\Ou$,
\begin{equation}
p^{\alpha}=\frac{1}{4\pi}\epsilon_{\ \sigma\tau\beta}^{\alpha}E^{\sigma}B^{\tau}u^{\beta},\quad{\rm i.e.,}\quad\vec{p}=\frac{1}{4\pi}\vec{E}\times\vec{B}.\label{eq:poynting}
\end{equation}
Since $v^{\alpha}B_{\alpha}=0,$ cf. Eq. \eqref{eq1}, $v_{\perp E}^{\alpha}$
is also the component of $v^{\alpha}$ parallel to $p^{\alpha}$:
$v_{\perp E}^{\alpha}=v_{\parallel p}^{\alpha}$. As is clear from
\eqref{eq2}, $v_{\parallel E}^{\alpha}$ is arbitrary. Contracting
\eqref{eq22} with $\epsilon^{\lambda}{}_{\sigma\alpha\tau}u^{\tau}E^{\sigma}$
(or, equivalently, taking the cross product of \eqref{eq2} with $\vec{E}$)
we obtain\footnote{This agrees with results known (in different contexts) in the literature:
in Problem $\mathsection25$ of \cite{LandauLifshitz} and Exercise
20.6 of \cite{Gravitation}, \emph{implicit} expressions are obtained
for the velocity of the observers measuring aligned fields $\vec{E}'$
and $\vec{B}'$, or, equivalently, for which the Poynting vector vanishes.
For a purely electric field, one can transform them into explicit
expressions which match \eqref{eq:explicitv}.} 
\begin{equation}
v_{\parallel p}^{\alpha}=\frac{\epsilon_{\ \sigma\tau\beta}^{\alpha}E^{\sigma}B^{\tau}u^{\beta}}{E_{\nu}E^{\nu}},\quad\textrm{i.e.,}\quad\vec{v}_{\parallel p}=\frac{\vec{E}\times\vec{B}}{\vec{E}^{2}}\ ,\label{eq:explicitv}
\end{equation}

\begin{figure}
\includegraphics[width=1\columnwidth]{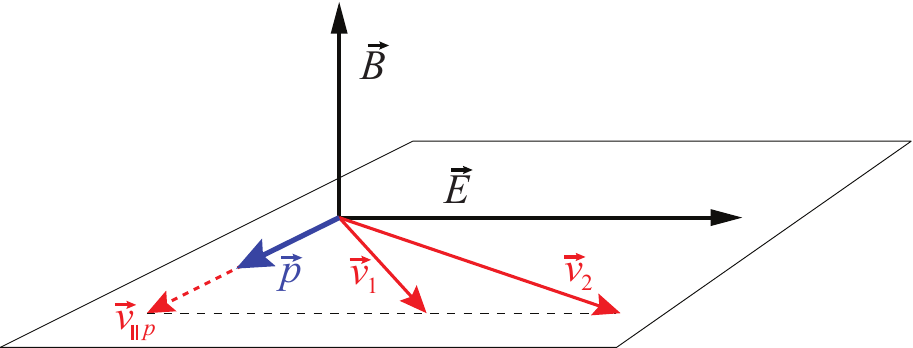}

\protect\protect\protect\protect\caption{\label{EBfig} Boosted observers that measure $\vec{B}'=0$. Their
velocities $\vec{v}_{i}$ lie in the planes orthogonal to $\vec{B}$,
and are such that their component orthogonal to $\vec{E}$ (that is,
along the Poynting vector $\vec{p}$) is $\vec{v}_{\parallel p}=\vec{E}\times\vec{B}/\vec{E}^{2}$.
The component $\vec{v}_{\parallel E}$ parallel to $\vec{E}$ is arbitrary.}
\end{figure}

Therefore, the observers $\Oup$ for which $\vec{B}'=0$ must move
with a velocity that is orthogonal to the magnetic field $\vec{B}$
as measured by $\Ou$, cf.~Eq.~\eqref{eq1}, must have a component
along the Poynting vector given by \eqref{eq:explicitv}, and may
have an arbitrary component $\vec{v}_{\parallel E}$ parallel to $\vec{E}$,
see Fig. \ref{EBfig}. In other words, the 4-velocities of these observers
are those contained in the timelike plane spanned by $E^{\alpha}$
and the timelike vector 

\begin{equation}
T^{\alpha}=u^{\alpha}+\frac{\epsilon_{\ \sigma\tau\beta}^{\alpha}E^{\sigma}B^{\tau}u^{\beta}}{E_{\nu}E^{\nu}}\label{eq:t}
\end{equation}
(with $T^{\a}T_{\a}=\vec{B}^{2}/\vec{E}^{2}-1<0$), i.e., those of
the form%
{} 
\begin{equation}
u'^{\alpha}=C\frac{T^{\alpha}}{\sqrt{-T^{\a}T_{\a}}}+D\frac{E^{\alpha}}{\sqrt{E^{\a}E_{\a}}},\quad D^{2}-C^{2}=-1\ .\label{eq:u'}
\end{equation}
Replacing $\{E^{\alpha},B^{\alpha}\}\rightarrow\{B^{\alpha},-E^{\alpha}\}$
in the above {[}compare (\ref{e2}) to (\ref{b2}){]} one shows that
the purely magnetic case ($\innst{F}=0$, $-\inn{F}<0$) yields a
class of observers measuring no electric field with 4-velocities given
by analogues of (\ref{eq:t}), (\ref{eq:u'}).

\begin{figure}
\includegraphics[width=0.6\columnwidth]{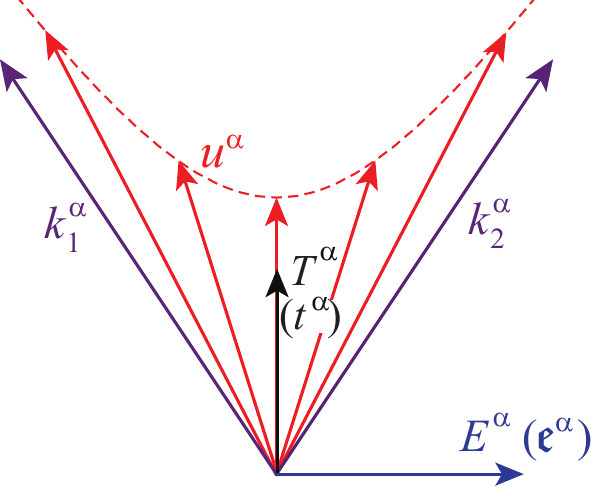}

\protect\protect\protect\protect\caption{\label{fig:PNDblade}The time-like principal plane, spanned by the
principal null directions ($k_{1}^{\alpha}$ and $k_{2}^{\alpha}$)
of a purely electric \Maxwell tensor. The unit time-like vectors
in this plane (and only those) are the 4-velocities of the observers
measuring no magnetic field. This plane is also spanned by the timelike
vector $T^{\alpha}$ and the electric field $E^{\alpha}$. The same
diagram holds for the observers measuring no gravitomagnetic tidal
tensor in vacuum Petrov type D spacetimes, by replacing $T^{\alpha}$
and $E^{\alpha}$ by the vectors $t^{\alpha}$ and $\eb^{\alpha}$
of Eqs. (\ref{eq:u'grav})-(\ref{eq:t-grav}).}
\end{figure}

To end this section we still mention some additional properties to
draw parallels and differences with the gravitational case below.
If the \Maxwell tensor is non-null {[}$(\innst{F},\inn{F})\neq(0,0)${]}
it has exactly two {\em principal null directions (PNDs)}, spanned
by null vectors $k^{\alpha}$ that satisfy 
\[
k^{[\alpha}F_{\ \ \gamma}^{\beta]}k^{\gamma}=0.
\]
The PNDs generate the {\em timelike principal plane}. Observers
$\Ou$ with 4-velocity $u^{\alpha}$ lying in this plane are precisely
those measuring a vanishing Poynting vector $p^{\alpha}$ (see e.g.
\cite{FerrandoSaezSE,WylCosNat15}). For a purely electric (purely
magnetic) \Maxwell tensor one has $E^{\alpha}B_{\alpha}=0$, and
by \eqref{eq:poynting} the vanishing of $p^{\alpha}$ implies the
vanishing of $B^{\alpha}$ ($E^{\alpha}$); hence the observers measuring
no magnetic (electric) field are \emph{those and only those} whose
4-velocity lies in the timelike principal plane. This is illustrated
in Fig. \ref{fig:PNDblade}. 

\section{Gravitational scalar invariants\label{sec:Gravitational-scalar-invariants}}

Analogously to electromagnetism, the scalar invariants of the curvature
tensor are related with the existence of observers for which its electric
or magnetic parts vanish; but by contrast with electromagnetism, the
invariants do not always yield sufficient conditions for that. We
shall first focus on the vacuum case; relevant comments on the non-vacuum
case are given at the end.

\subsection{Vacuum Riemann tensor\label{sub:Rie-vac}}

In the vacuum case, characterized by a vanishing Ricci tensor ($R_{\a\b}\equiv R_{~\a\g\b}^{\g}=0$),
the Riemann tensor ${\bf R}\equiv R_{\a\b\g\d}$ generically has 10
independent components in any frame, and exhibits the special property
that the dual in the first pair of indices equals the dual in the
second pair: 
\begin{equation}
\star\!{\bf R}={\bf R}\!\star\qquad(\e_{\a\b}{}^{\e\zeta}R_{\e\zeta\g\d}=R_{\a\b\e\zeta}\e^{\e\zeta}{}_{\g\d}).\label{eq:leftisright}
\end{equation}
Relative to an observer $\Ou$, the {\em gravitoelectric and gravitomagnetic
tidal tensors} \cite{CHPRD,PaperAnalogies} (or ``electric'' and
``magnetic'' parts of the Riemann tensor, e.g. \cite{MagneticCurvatures,nareshdadhich:00})
are pointwise defined by 
\begin{equation}
\EE_{\alpha\beta}\equiv R_{\alpha\gamma\beta\delta}u^{\gamma}u^{\delta}\ ,\qquad\HH_{\alpha\beta}\equiv\star\!R_{\alpha\gamma\beta\delta}u^{\gamma}u^{\delta}\ .\label{eq:EEHH}
\end{equation}
$\EE_{\alpha\beta}$ and $\HH_{\alpha\beta}$ are symmetric, tracefree,
and spatial with respect to ${\bf u}$ ($\EE_{\alpha\beta}=\EE_{(\alpha\beta)},\,\EE_{\ \alpha}^{\alpha}=0,\,\EE_{\alpha\beta}u^{\beta}=0$,
and similarly for $\HH_{\alpha\beta}$), thus having 5 independent
components each. In terms of $\EE_{\alpha\beta}$ and $\HH_{\alpha\beta}$
one has decompositions 
\begin{eqnarray}
R_{\a\b}{}^{\g\d} & = & 4\left(2u_{[\a}u^{[\g}+\d_{[\a}^{[\g}\right)\mathbb{E}_{\b]}{}^{\d]}\nonumber \\
 & + & 2\e_{\a\b\e\zeta}\mathbb{H}^{\e[\d}u^{\g]}u^{\zeta}+2\e^{\g\d\e\zeta}\mathbb{H}_{\e[\b}u_{\a]}u_{\zeta}\ ,\label{eq:Rie-decomp}\\
\star\!R_{\a\b}{}^{\g\d} & = & 4\left(2u_{[\a}u^{[\g}+\d_{[\a}^{[\g}\right)\mathbb{H}_{\b]}{}^{\d]}\nonumber \\
 & - & 2\e_{\a\b\e\zeta}\mathbb{E}^{\e[\d}u^{\g]}u^{\zeta}-2\e^{\g\d\e\zeta}\mathbb{E}_{\e[\b}u_{\a]}u_{\zeta}\ ,\label{eq:Riest-decomp}
\end{eqnarray}
which exhibit a certain analogy with (\ref{FaradayDecomp})-(\ref{Fstar}).

In contrast to the \Maxwell tensor, a vacuum Riemann tensor has {\em
four} real, independent invariants~\cite{Geheniau-Norden}: two
quadratic invariants, namely the Kretschmann scalar $\inn{R}$ and
Chern-Pontryagin scalar $\innst{R}$, which in terms of the tidal
tensors measured by an observer read 
\begin{align}
 & \tfrac{1}{8}\inn{R}\equiv\tfrac{1}{8}R^{\a\b\g\d}R_{\a\b\g\d}=\EE^{\a\b}\EE_{\a\b}-\HH^{\a\b}\HH_{\a\b}\ ;\label{eq:inv-A}\\
 & \tfrac{1}{16}\innst{R}\equiv\tfrac{1}{16}\star\!R^{\a\b\g\d}R_{\a\b\g\d}=\EE^{\a\b}\HH_{\a\b}\ ,\label{eq:inv-B}
\end{align}
and are formally analogous to the electromagnetic invariants (\ref{eq:EBSquare})-(\ref{eq:EB}),\footnote{\label{foot:signdiff} The formal analogy is up to a factor 8 and
a minus sign. The sign difference is due to the contraction of one
pair of antisymmetric indices within $\inn{F}$ and $\innst{F}$ whereas
two such pairs are contracted within $\inn{R}$ and $\innst{R}$.} but also two {\em cubic} invariants (e.g. \cite{belRadiation})
\begin{align}
 & \mathbb{A}\equiv-\tfrac{1}{16}R_{\ \ \lambda\mu}^{\alpha\beta}R_{\ \ \ \rho\sigma}^{\lambda\mu}R_{\ \ \alpha\beta}^{\rho\sigma}=\mathbb{E}_{\ \beta}^{\alpha}\mathbb{E}_{\ \gamma}^{\beta}\mathbb{E}_{\ \alpha}^{\gamma}-3\mathbb{E}_{\ \beta}^{\alpha}\mathbb{H}_{\ \gamma}^{\beta}\mathbb{H}_{\ \alpha}^{\gamma}\ ,\label{eq:inv-C}\\
 & \mathbb{B}\equiv\tfrac{1}{16}\star\!R_{\ \ \lambda\mu}^{\alpha\beta}R_{\ \ \ \rho\sigma}^{\lambda\mu}R_{\ \ \alpha\beta}^{\rho\sigma}=\mathbb{H}_{\ \beta}^{\alpha}\mathbb{H}_{\ \gamma}^{\beta}\mathbb{H}_{\ \alpha}^{\gamma}-3\mathbb{E}_{\ \beta}^{\alpha}\mathbb{E}_{\ \gamma}^{\beta}\mathbb{H}_{\ \alpha}^{\gamma}\ ,\label{eq:inv-D}
\end{align}
which have no electromagnetic counterpart. {\em At any point these
four invariants may in principle take any value, independently of
each other}, and are all needed to determine whether $R_{\a\b\g\d}$
has a purely electric/magnetic character. Analogously to a \Maxwell
tensor, a non-zero vacuum Riemann tensor (or the spacetime) is called
{\em purely electric (purely magnetic)} at a point if there exists
an observer $\Oup$ measuring a vanishing gravitomagnetic (gravitoelectric)
tidal tensor: $\HH'_{\a\b}=0$ ($\EE'_{\a\b}=0$). By \eqref{eq:inv-A}-\eqref{eq:inv-D}
the existence of such an observer clearly requires $\mathbb{B}=0$
($\mathbb{A}=0$), besides $\innst{R}=0$ and $\inn{R}>0$ ($\inn{R}<0$),
but this is not sufficient.

To explain why, it is useful to define the complex tensor 
\begin{equation}
\mathbb{Q}_{\ \b}^{\a}\ \equiv\ \EE_{\ \b}^{\a}-i\HH_{\ \b}^{\a}\ .\label{eq:complexQab}
\end{equation}
This is a tensor which is spatial relative to the observer, $u_{\a}\QQ_{~\b}^{\a}=0$,
and consists of the sum of two symmetric spatial tensors, one real
and one purely imaginary, \emph{each of them} diagonalizable. Therefore,
the existence of observers $\Oup$ measuring $\mathbb{H}'_{\a\b}=0\,(\mathbb{E}'_{\a\b}=0)$
implies that the operator $\QQ'{}_{~\b}^{\a}$ has two properties:
it is diagonalizable and has real (purely imaginary) eigenvalues.
Now, both properties are independent of the observer, i.e., they are
shared by the respective tensors $\Qop$ measured by {\em arbitrary}
observers $\Ou$. To indicate the origin of this fact, we note that
$\QQ_{~\b}^{\a}$ can be viewed as a linear operator in (i.e., an
endomorphism of) the complexified rest space of $\Ou$. This is a
3-D complex vector space isomorphic to the space of those complex
anti-symmetric tensors $X_{\a\b}=X_{[\a\b]}$ satisfying $\star\!X_{\g\d}=i{X}_{\g\d}$
(so-called \emph{self-dual bivectors}, see e.g. \cite{StephaniExact});
by virtue of \eqref{eq:leftisright} and $R_{~\a\g\b}^{\g}=0$ the
tensor $-\frac{1}{2}R_{\a\b}{}^{\g\d}$ acts as a trace-free linear
operator on this space; moreover, all operators $\QQ_{~\b}^{\a}$
are {\em equivalent} to this observer-independent operator (see
(4.1)-(4.4) of \cite{StephaniExact}, or \cite{WylCosNat15}), meaning
that not only they have the same eigenvalues $\l_{k}\,(k=1,2,3)$,
i.e., the same characteristic polynomial 
\begin{equation}
c(x)=x^{3}-\frac{I}{2}x-\frac{\J}{3}=(x-\l_{1})(x-\l_{2})(x-\l_{3}),\label{eq:Qop-chareq}
\end{equation}
but also the same minimal polynomial $m(x)$, which in 3-D fully determines
the algebraic properties of operators.\footnote{\label{foot:minimal-poly} Recall that if a polynomial $p(x)$ annihilates
an endomorphism $L$ of a vector space (i.e., $p(L)=0$) it has all
eigenvalues of $L$ (i.e., roots of $c(x)$) as roots. The minimal
polynomial $m(x)$ is the (unique) annihilating polynomial of least
degree and leading coefficient 1, and {\em only} has the eigenvalues
as roots, possibly occurring with lower multiplicities than in $c(x)$;
$L$ is diagonalizable precisely if the multiplicities in $m(x)$
are 1 for all eigenvalues.} Here $I,\,\J$ are the complex invariants (e.g. \cite{McIntosh et al 1994,WyllePRD})
\begin{align}
I & =\tfrac{1}{8}\inn{R}-\tfrac{i}{8}\innst{R}=\l_{1}^{2}+\l_{2}^{2}+\l_{3}^{2}\nonumber \\
 & =\QQ_{~\b}^{\a}\QQ_{~\a}^{\b}=\EE_{~\b}^{\a}\EE_{~\a}^{\b}-\HH_{~\b}^{\a}\HH_{~\a}^{\b}-2i\EE_{~\b}^{\a}\HH_{~\a}^{\b}\ ;\label{eq:I-Q}\\
\J & \equiv\mathbb{A}+i\mathbb{B}=-\tfrac{1}{16}(R^{\a\b}{}_{\l\mu}-i\!\star\!R^{\a\b}{}_{\l\mu})R_{\ \ \rho\s}^{\lambda\mu}R_{\ \ \ \a\b}^{\rho\s}\nonumber \\
 & =\l_{1}^{3}+\l_{2}^{3}+\l_{3}^{3}=3\l_{1}\l_{2}\l_{3}=\QQ_{~\b}^{\a}\QQ_{~\g}^{\b}\QQ_{~\a}^{\g}\label{eq:J-Q}\\
 & =\mathbb{E}_{\ \beta}^{\alpha}\mathbb{E}_{\ \gamma}^{\beta}\mathbb{E}_{\ \alpha}^{\gamma}+i\mathbb{H}_{\ \beta}^{\alpha}\mathbb{H}_{\ \gamma}^{\beta}\mathbb{H}_{\ \alpha}^{\gamma}-3i\,\mathbb{E}_{\ \beta}^{\alpha}(\mathbb{E}_{\ \gamma}^{\beta}-i\mathbb{H}_{\ \gamma}^{\beta})\mathbb{H}_{\ \alpha}^{\gamma}.\nonumber 
\end{align}
One has $\l_{1}+\l_{2}+\l_{3}=\QQ_{~\a}^{\a}=0$, and the discriminant
of the cubic polynomial $c(x)$ equals (up to a factor 2) 
\begin{equation}
\Delta\equiv2(\l_{1}-\l_{2})^{2}(\l_{2}-\l_{3})^{2}(\l_{3}-\l_{1})^{2}=
I^{3}-6\J^{2}.\label{eq:Delta-def}
\end{equation}
The eigenvalue problem for $\Qop$ leads to the {\em Petrov classification}
of the vacuum Riemann tensor \cite{Petrov,StephaniExact}, which can
be formulated as follows: 
\begin{enumerate}
\item[{\bf (a)}] Petrov type I: this is the generic case where all eigenvalues differ
($\Delta\neq0$), and $m(x)=c(x)$. 
\item[{\bf (b)}] Petrov types D and II: both have $\Delta=0\neq I\J$, a double eigenvalue
$\l=-\J/I$ and a single eigenvalue $-2\l$, but $m(x)=(x+2\l)(x-\l)^{2}=c(x)$
for type II while $m(x)=(x+2\l)(x-\l)$ for type D; in terms of $\QQ_{\ \b}^{\a}$
and $h_{\b}^{\a}$ defined in \eqref{eq:SpaceProjector} this means
that 
\begin{equation}
{\cal F}({\bf \QQ})_{~\b}^{\a}\equiv\QQ_{~\g}^{\a}\QQ_{~\b}^{\g}+\lambda\QQ_{~\b}^{\a}-2\lambda^{2}h_{\b}^{\a}=0,\quad\lambda=-\frac{\J}{I}
\label{eq:distinctII-D}
\end{equation}
holds for type D (then implying $m(x)={\cal F}(x)$), while ${\cal F}({\bf \QQ})_{~\b}^{\a}\neq0$
for type II. 
\item[{\bf (c)}] Petrov types N, III and O: all have $I=\J=\Delta=0$ and a triple
eigenvalue $0$, but the respective minimal polynomials are $m(x)=x^{3}=c(x),\,x^{2}$
and $x$ (such that type O corresponds to $\mathbf{R}=0$). 
\end{enumerate}
Referring to footnote \ref{foot:minimal-poly}, diagonalizability
of the operators $\QQ_{\ \b}^{\a}$ precisely means that the Petrov
type is I, D or O (where in the last case $R_{\alpha\beta\gamma\delta}=0\Rightarrow\QQ_{\ \b}^{\a}=0$
for all observers, trivially). In particular, the characteristic property
\eqref{eq:distinctII-D} for type D ensures diagonalizability of $\QQ_{\ \b}^{\a}$
and distinguishes this type from the non-diagonal Petrov type II,
{\em a distinction that cannot be made in terms of invariants}
(which may be equal for both types). From the discussion above it
follows that a non-zero vacuum Riemann tensor that is purely electric
(purely magnetic) is of Petrov type I or D and the operators $\QQ_{\ \b}^{\a}$
have real (purely imaginary) eigenvalues. Now, it turns out that these
are not only \emph{necessary}, but also \emph{sufficient} conditions
for the existence of observers measuring a vanishing magnetic (electric)
tidal tensor: it is a well known property of Petrov types I and D
(see e.g. \cite{Pirani,StephaniExact,WylCosNat15,FerrSaezP2}) that
observers $\Oup$ always exist for which the operator ${\QQ'}_{\ \b}^{\a}$
allows for a basis of {\em real orthonormal} eigenvectors;\footnote{By \eqref{eq:complexQab} this means that ${\EE'}_{\ \b}^{\a}$ and
${\HH'}_{\ \b}^{\a}$ measured by these observers are simultaneously
diagonalizable, i.e., they commute (since they are symmetric), which
is equivalent to saying that the ``super-Poynting vector'' $\mathcal{P}'^{\alpha}$,
see (\ref{eq:super-Poynting}), vanishes for them.} in the case that the eigenvalues $\l_{k}$ are all real (purely imaginary),
it follows from the primed version of \eqref{eq:complexQab} that
such an observer $\Oup$ measures $\mathbb{H}'_{\alpha\beta}=0$ ($\mathbb{E}'_{\alpha\beta}=0$).
Hence we can say that a non-zero vacuum Riemann tensor {\em is purely
electric (purely magnetic) if and only if its Petrov type is I or
D and the eigenvalues $\l_{k}$ of the $\QQ_{\ \b}^{\a}$ operators
are real (purely imaginary)}. Furthermore, all eigenvalues being
real (purely imaginary) is equivalent to\footnote{That real (purely imaginary) eigenvalues imply \eqref{eq:PE/PM-Delta}
follows immediately from \eqref{eq:I-Q}-\eqref{eq:J-Q} and the first
expression for $\Delta$ in \eqref{eq:Delta-def}. Conversely, a cubic
polynomial $p(x)$ with real coefficients and vanishing quadratic
term has roots $-2a$ and $a\pm b$, with $a$ real and $b$ either
real or purely imaginary, and discriminant $D=2b^{2}(9a^{2}-b^{2})^{2}$
(cf.\ \eqref{eq:Delta-def}); hence $D\geq0$ implies $b$ real and
all roots are real; given \eqref{eq:PE/PM-Delta} one applies this
to $p(x)=c(x)$, with roots $\l_{k}$ and $D=\Delta$ ($p(x)=ic(ix)$,
with roots $-i\l_{k}$ and $D=-\Delta$) to see that the $\l_{k}$'s
are all real (purely imaginary). Compare to the more intricate reasoning
in \cite{McIntosh et al 1994,McIntosh2}.} 
\begin{equation}
I>0\ (<0)\ {\rm real},\quad\J\ {\rm real}\ ({\rm imaginary}),\quad\Delta\ge0\ (\leq0)\ \textrm{real}\,.\label{eq:PE/PM-Delta}
\end{equation}
Since $\Delta\neq0$ corresponds to Petrov type I in general, and
in view of \eqref{eq:I-Q}-\eqref{eq:J-Q} it follows that a non-zero
vacuum Riemann tensor {\em is purely electric (purely magnetic)
and of Petrov type I if and only if} 
\begin{equation}
\begin{aligned} & \innst{R}=\mathbb{B}=0\quad\textrm{and}\quad(\inn{R}/8)^{3}>6\mathbb{A}^{2}\geq0\\
 & (\innst{R}=\mathbb{A}=0\quad\textrm{and}\quad(\inn{R}/8)^{3}<-6\mathbb{B}^{2}\leq0)\ .
\end{aligned}
\label{eq:PEPM-I}
\end{equation}
As for the remaining possibility $\Delta=0\neq I\Leftrightarrow6\J^{2}=I^{3}\neq0$
the first part of \eqref{eq:PE/PM-Delta} implies the second part.
Contrary to the Petrov type I case the invariants are now insufficient
to formulate the purely electric or magnetic conditions: we also need
condition \eqref{eq:distinctII-D} to discriminate (allowed) Petrov
type D from (forbidden) Petrov type II. Note that the first part of
\eqref{eq:distinctII-D} {\em on itself} implies $I=6\l^{2},\,\J=-\l I$
by ${\cal F}({\QQ})_{~\a}^{\a}=\QQ_{~\a}^{\b}{\cal F}({\QQ})_{~\b}^{\a}=0$,
and thus the last part of \eqref{eq:distinctII-D}. It follows that
a non-zero vacuum Riemann tensor {\em is purely electric (purely
magnetic) and of Petrov type D if and only if $\inn{R}\neq0$ and}
\begin{equation}
\begin{aligned} & \QQ_{~\g}^{\a}\QQ_{~\b}^{\g}+\l\QQ_{~\b}^{\a}-2\l^{2}h_{\b}^{\a}=0\,\\
 & \textrm{with}\;\;\l=-{8\mathbb{A}}/{\inn{R}}\quad(\l=-{8i\mathbb{B}}/{\inn{R}})\,
\end{aligned}
\label{eq:PEPM-D}
\end{equation}
and $\QQ_{~\b}^{a},\,h_{\b}^{\a}$ relative to any observer $\Ou$;
compared to \eqref{eq:PEPM-I} one also has $\innst{R}=\mathbb{B}=0$
($\innst{R}=\mathbb{A}=0$) in this case but now $[\inn{R}/8]^{3}=6\mathbb{A}^{2}>0$
($[\inn{R}/8]^{3}=-6\mathbb{B}^{2}<0$). Alternatively, given the
first part of \eqref{eq:PE/PM-Delta} the second and third parts are
easily seen to be equivalent to either $\J=0$ or $\M\geq0$ real,
where for $\J\neq0$ the dimensionless invariant $\M$ is defined
by~\cite{McIntosh2} 
\begin{equation}
\M\equiv{\Delta}/{\J^{2}}={I^{3}}/{\J^{2}}-6\ .\label{eq:Mdef}
\end{equation}
Hence we find back the result~\cite{McIntosh et al 1994,FerrandoSaez2002}
that a non-zero vacuum Riemann tensor {\em is purely electric (purely
magnetic), i.e., an observer $\Oup$ exists for which $\mathbb{H}'_{\alpha\beta}$
($\mathbb{E}'_{\alpha\beta}$) vanishes, if and only if the following
conditions hold}: 
\begin{align}
 & \innst{R}=0\quad\textrm{and}\quad\inn{R}>0\,(<0)\,,\label{eq:grav-PEPM1}\\
 & \J=0\;\;\textrm{or}\;\;\{\J\neq0\,\,{\rm and}\,\,\M>0\,{\rm real}\}\;\;{\rm or}\;\;\eqref{eq:distinctII-D}\ ,\label{eq:grav-PEPMcond}
\end{align}
where the first two cases of \eqref{eq:grav-PEPMcond} automatically
give Petrov type I and \eqref{eq:distinctII-D} fully characterizes
Petrov type D. Using \eqref{eq:I-Q}-\eqref{eq:distinctII-D} the
conditions \eqref{eq:PEPM-I}-\eqref{eq:grav-PEPMcond} can be easily
tested by calculating $\QQ_{~\b}^{\a}$ relative to {\em any} observer
$\Ou$.

It follows that one can make formally similar statements to \textbf{(i)}-\textbf{(iii)}
of the electromagnetic case in Sec. \ref{sec:Electromagnetic-Scalar-Invariant},
replacing $\mathbf{F}$ by \textbf{$\mathbf{R}$} and adding the condition
(\ref{eq:grav-PEPMcond}) to \textbf{(ii)}: 
\begin{enumerate}
\item[\textbf{(i)}] $\innst{R}\neq0\,[\Leftrightarrow\EE^{\a\b}\HH_{\a\b}\neq0]\,\Rightarrow$
$\mathbb{E}_{\alpha\gamma}$ and $\mathbb{H}_{\alpha\gamma}$ are
both non-vanishing for all observers. 
\item[\textbf{(ii)}] $\innst{R}=0$, $\inn{R}>0\,(<0)$ \emph{and} (\ref{eq:grav-PEPMcond})
$\Rightarrow$ one can \textit{always} find an observer for which
$\mathbb{H}_{\alpha\gamma}$ ($\mathbb{E}_{\alpha\gamma}$) vanishes.
At the considered point the vacuum spacetime is classified as {\em
purely electric (magnetic)}. 
\item[\textbf{(iii)}] All invariants vanish: $\innst{R}=\inn{R}=\J=0$ $\Rightarrow$ either
$\mathbf{R}=0$ (Petrov type O) or the Petrov type is III or N and
$\mathbb{E}_{\alpha\gamma},\,\mathbb{H}_{\alpha\gamma}\ne0$ for all
observers\footnote{In direct analogy with electromagnetism (see footnote \ref{foot:pure radiation}),
the vanishing of all invariants has been proposed as a criterion (Bel's
second criterion) for ``pure'' gravitational radiation, see \cite{BelSecondCrit}
and also \cite{Zakharov} p. 53. Such \textcolor{black}{criterion
is based on ``super-energy'' \cite{belRadiation,SenovillaBigPaper,Maartens:1997fg,AlfonsoSE,FerrandoSaezSE,PaperAnalogies}.}}. 
\end{enumerate}
The implications in \textbf{(i)} and \textbf{(iii)} are obvious. A
vacuum Riemann tensor obeying \textbf{(i)} is either of Petrov type
I, II or D, while \textbf{(ii)} implies Petrov type I or D as explained
above. Notice that \textbf{(ii)} implies as well that $\mathbb{E}_{\alpha\beta}$
($\mathbb{H}_{\alpha\beta}$) is non-zero for all observers; in particular,
for a non-zero vacuum Riemann tensor at a point, there cannot, simultaneously,
exist observers for which $\HH_{\a\b}=0$ and observers for which
$\EE_{\a\b}=0$. 

The possibilities lying outside criteria \textbf{(i)}-\textbf{(iii)}
have no counterpart in the formal analogy with electromagnetism, and
they all preclude the existence of observers measuring a vanishing
$\mathbb{E}_{\alpha\beta}$ or $\mathbb{H}_{\alpha\beta}$. These
are:
\begin{enumerate}
\item[\textbf{(iv)}] $\innst{R}=0$, $\inn{R}\ne0$, $\J\ne0$ and either $\M<0$ real
or $\M$ non-real, which is a Petrov type I subcase. Examples of such
vacuum solutions are the Lewis metrics for the Lewis class \cite{SantosCQG95,Cilindros}
(or, equivalently, the van Stockum exterior solution for $aR>1/2$,
see \cite{Bonnor:1995zf}), describing a special class of the exterior
metrics produced by infinite rotating cylinders. 
\item[\textbf{(v)}] $\innst{R}=0$, $\inn{R}\ne0$, $\J\ne0$ and $\M=0$ but without
(\ref{eq:distinctII-D}) holding, corresponding to Petrov type II
{[}see point (b) above{]}; an example is the limiting case $aR=1/2$
of the van Stockum exterior solution \cite{Bonnor:1995zf}. 
\item[\textbf{(vi)}] $\innst{R}=\inn{R}=0$ and $\J\ne0$ (corresponding to $I=0\Leftrightarrow\M=-6$),
which is a Petrov type I subcase whose only known vacuum solution
is Petrov's homogeneous metric~\cite{Petrovmetric,StephaniExact},
having a constant $\J$ and a simply-transitive maximal group $G_{4}$
of motions. 
\end{enumerate}
Finally, it is worth mentioning that no vacuum spacetimes are known
for which the Riemann tensor is purely magnetic \emph{in an open 4-D
region} (i.e., where $\mathbb{E}_{\alpha\beta}=0$ with respect to
some observer {\em congruence}). It has therefore been conjectured
that no such spacetimes exist (see e.g. \cite{Barnes_Magnetic})\footnote{The conjecture has been proven for Petrov type D~\cite{Hall} and
for type I with $\J=0$~\cite{Brans}; it has further been shown
that, in any vacuum spacetime, $\mathbb{E}_{\alpha\beta}\ne0$ with
respect to \emph{any} observer congruence that is either shear-free
\cite{Barry}, non-rotating \cite{VandenBergh:2002fb} or geodesic
\cite{VBergh2003}; see \cite{WyllePRD} for a complete survey.}. However, a vacuum Riemann tensor can be purely magnetic in 3-D hypersurfaces
(exemplified in Fig. \ref{fig:KerrInvariants}) or lower-dimensional
sets.

\subsection{Observers measuring no gravitomagnetic/gravitoelectric tidal tensor
in vacuum\label{sub:grav-PEPM-obs}}

As explained above, the existence of observers for which the gravitomagnetic
(gravitoelectric) tidal tensor vanishes at a point of a non-flat vacuum
spacetime requires the Petrov type to be either I or D. Their appearances
for both Petrov types are well-known~\cite{Pirani,belRadiation,FerrandoSaez2002,WylCosNat15}.
In the Petrov type I case there is a unique such observer. A vacuum
Riemann tensor of type D, on the other hand, shows strong analogy
with a \Maxwell tensor: it has exactly two {\em principal null
directions (PNDs)} which are spanned by those null vectors $k^{\alpha}$
that satisfy 
\begin{equation}
k^{[\a}R_{\ \g\d\e}^{\b]}k^{\g}k^{\e}=0\ ;\label{Weyl-PND}
\end{equation}
these null vectors generate the {\em timelike principal plane},
and analogously to the situation in electromagnetism, the observers
measuring no $\mathbb{H}_{\alpha\beta}$ ($\mathbb{E}_{\alpha\beta}$)
are precisely those with a 4-velocity in this plane, see Fig. \ref{fig:PNDblade}.

\begin{figure}
\includegraphics[width=1\columnwidth]{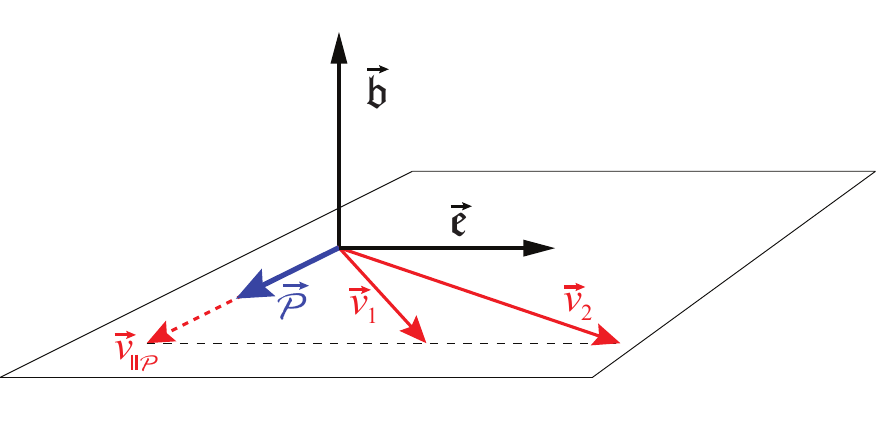}

\protect\protect\protect\protect\caption{\label{fig:BoostGrav}Boosted observers that measure $\mathbb{H}'_{\alpha\beta}=0$
($\mathbb{E}'_{\alpha\beta}=0$) in Petrov type D spacetimes. Their
velocities $\vec{v}_{i}$ have a component $\vec{v}_{\parallel\mathcal{P}}$
parallel to the super-Poynting vector $\vec{\mathcal{P}}$ given by
Eq. (\ref{eq:vGrav}), and an arbitrary component parallel to the
vector $\vec{\eb}$ defined in Eq. (\ref{eq:evec}). Notice the analogy
with Fig. \ref{EBfig}.}
\end{figure}

As proved in general in the companion paper \cite{WylCosNat15}, the
following algorithm, specified to the purely electric (purely magnetic)
case, gives the 4-velocity of the observers $\Oup$ for which $\mathbb{H}'_{\alpha\beta}=0$
($\mathbb{E}'_{\alpha\beta}=0$) in terms of the tidal tensors measured
by an arbitrary observer $\Ou$: 
\begin{enumerate}
\item For each non-degenerate eigenvalue $\l_{k}$ of $\Qop$ (take $k=1$
for type D, while $k=1,2,3$ for type I) construct a vector $\v_{k}^{\a}$
in the corresponding eigenspace as follows, where $\spacev^{\a}$
is any spatial vector in the rest space of $\Ou$ such that $\v_{k}^{\a}\neq0$:
\begin{align}
 & \textrm{type D:}\ \ \v_{1}^{\a}\equiv\QQ_{~\b}^{\a}\spacev^{\b}-\l\spacev^{\a},\quad\l=-8\J/(\inn{R})\ ;\label{eq:wvector}\\
 & \textrm{type I:}\ \v_{k}^{\a}\equiv(\QQ^{\a}{}_{\g}\QQ^{\g}{}_{\b}
+\l_{k}\QQ^{\a}{}_{\b})\spacev^{\b}+(\l_{k}^{2}-\tfrac{1}{16}\inn{R})\spacev^{\a},\nonumber \\
 & \l_{k}=\a\,\sqrt{\frac{|\inn{R}|}{12}}\cos\left(\frac{\arccos\Theta_{\a}}{3}-k\frac{2\pi}{3}\right),\quad k=1,2,3,\nonumber \\
 & \Theta_{\a}\equiv\frac{\textrm{sgn}(\a\J)}{\sqrt{1+\M/6}},\quad\a=\begin{cases}
1, & \mathbb{H}'_{\alpha\beta}=0,\\
i, & \mathbb{E}'_{\alpha\beta}=0.
\end{cases}\label{eq:L-alpha}
\end{align}

\item \label{enu:Point2}The complex vectors $\v_{k}^{\a}$ are orthogonal
to $u^{\a}$, non-null, and mutually orthogonal in the type I case;
normalize them to unit vectors $\v_{k}^{\a}/\sqrt{\v_{k}^{\sigma}(\v_{k})_{\sigma}}\equiv\eb_{k}^{\a}-i\,\bb_{k}^{\a}$,
such that $\eb_{k}^{\a}(\bb_{l})_{\a}+\eb_{l}^{\a}(\bb_{k})_{\a}=0$
and $\eb_{k}^{\a}(\eb_{l})_{\a}-\bb_{k}^{\a}(\bb_{l})_{\a}=\delta_{kl}$. 
\item In the type D case, the observers measuring no $\mathbb{H}'_{\alpha\beta}$
($\mathbb{E}'_{\alpha\beta}$) are those with 4-velocity of the form
\begin{align}
 & u'^{\alpha}=C\frac{t^{\alpha}}{\sqrt{-t^{\a}t_{\a}}}+D\frac{\eb^{\alpha}}{\sqrt{\eb^{\a}\eb_{\a}}},\quad D^{2}-C^{2}=-1,\label{eq:u'grav}\\
 & t^{\alpha}\equiv u^{\alpha}+\frac{\epsilon_{\ \sigma\tau\beta}^{\alpha}\eb^{\sigma}\bb^{\tau}u^{\beta}}{\eb_{\nu}\eb^{\nu}}\,,\label{eq:t-grav}\\
 & \eb^{\alpha}\equiv\eb_{1}^{\alpha}=\Re\left(\frac{\v_{1}^{\a}}{\sqrt{\v_{1}^{\sigma}(\v_{1})_{\sigma}}}\right)\,.\label{eq:evec}
\end{align}
In the type I case, the unique observer $\Oup$ has 4-velocity $u'^{\a}=\tI^{\a}/\sqrt{-\tI^{\a}{\tI}_{\alpha}}$,
where 
\begin{equation}
\tI^{\a}=\left(\sum_{k=1}^{3}\eb_{k}^{\b}(\eb_{k})_{\b}-1\right)u^{\a}+\eta_{~\b\g\d}^{\a}u^{\d}\sum_{k=1}^{3}\eb_{k}^{\b}\bb_{k}^{\g}\ .\label{eq:u'-grav-I}
\end{equation}

\end{enumerate}
In the Petrov type D case, the expressions above can also be used
to compute the PNDs, generated by the null vectors 
\begin{equation}
k_{\pm}^{\a}=\frac{t^{\a}}{\sqrt{-2t^{\a}t_{\a}}}\pm\frac{\eb^{\a}}{\sqrt{2\eb^{\a}\eb_{\a}}}\ .\label{PNDs}
\end{equation}
Notice the formal analogy between (\ref{eq:u'grav})-(\ref{eq:t-grav})
and the electromagnetic expressions (\ref{eq:t})-(\ref{eq:u'}) which
yield the 4-velocities of the observers for which the magnetic field
vanishes in the purely electric case. An alternative expression for
$t^{\a}$ is~\cite{WylCosNat15} 
\begin{align}
t^{\alpha} & =u^{\alpha}+\frac{\e^{\a\b\g\d}\mathbb{E}_{\b\mu}\mathbb{H}_{~\g}^{\mu}u_{\d}}{3\xi\AD(\AD+1)}\ ;\nonumber \\
\AD & \equiv\sqrt{\frac{2}{3}\frac{\EE^{\a\b}\EE_{\a\b}+\HH^{\a\b}\HH_{\a\b}}{|\EE^{\a\b}\EE_{\a\b}-\HH^{\a\b}\HH_{\a\b}|}+\frac{1}{3}}\ ;\label{eq:tgrav2}\\
\xi & =\frac{1}{4}|\EE^{\a\b}\EE_{\a\b}-\HH^{\a\b}\HH_{\a\b}|=\frac{|\inn{R}|}{32}\ ,\nonumber 
\end{align}
where we recognize the ``super-Poynting'' vector (see e.g.\ \cite{WylCosNat15,belRadiation,SenovillaBigPaper,Maartens:1997fg,AlfonsoSE,FerrandoSaezSE,PaperAnalogies},
and compare to \eqref{eq:poynting}) 
\begin{equation}
\mathcal{P}^{\alpha}\equiv\frac{1}{2}\e^{\a\b\g\d}{\EE}_{\b\mu}{\HH}_{~\g}^{\mu}u_{\d},\quad{\rm or}\quad\vec{\mathcal{P}}\equiv\frac{1}{2}\overleftrightarrow{\mathbb{E}}\times\overleftrightarrow{\mathbb{H}}\ ,\label{eq:super-Poynting}
\end{equation}
the second expression holding in dyadic notation and in the rest frame
of $\Ou$. From Eqs. (\ref{eq:u_u'}), (\ref{eq:u'grav}), and (\ref{eq:tgrav2})
(noticing that $\gamma=C/\sqrt{-t^{\a}t_{\a}}$), it follows that
the observers $\Oup$ for which $\mathbb{H}'_{\alpha\beta}=0$ ($\mathbb{E}'_{\alpha\beta}=0$)
must move, relative to $\Ou$, with a velocity $v^{\alpha}=v_{\parallel\mathcal{P}}^{\alpha}+v_{\parallel\eb}^{\alpha}$
that has a component $v_{\parallel\mathcal{P}}^{\alpha}$ parallel
to $\mathcal{P}^{\alpha}$ given by 
\begin{equation}
v_{\parallel\mathcal{P}}^{\alpha}=\frac{\e^{\a\b\g\d}\mathbb{E}_{\b\mu}\mathbb{H}_{~\g}^{\mu}u_{\d}}{3\xi\AD(\AD+1)},\quad\textrm{i.e.,}\quad\vec{v}_{\parallel\mathcal{P}}=\frac{\overleftrightarrow{\mathbb{E}}\times\overleftrightarrow{\mathbb{H}}}{3\xi\AD(\AD+1)},\label{eq:vGrav}
\end{equation}
and an arbitrary component $v_{\parallel\eb}^{\alpha}$ parallel to
$\eb^{\alpha}$, see Fig \ref{fig:BoostGrav}. Notice the similarity
with the electromagnetic counterparts in Eqs. (\ref{eq:explicitv})-(\ref{eq:t})
and Fig.\ \ref{EBfig}. These however hold only in the purely electric
case (for the observers measuring $B^{\alpha}=0$); in order to obtain
the velocities of the observers measuring $E^{\alpha}=0$ (in purely
magnetic fields), one needs to replace $E^{\nu}E_{\nu}$ by $B^{\nu}B_{\nu}$
in the denominators, and switch $\vec{E}$ and $\vec{B}$ in Fig.\ \ref{EBfig}.
For electromagnetic expressions encompassing both the purely electric
and magnetic cases {[}hence the closest analogues of Eqs.\ (\ref{eq:tgrav2})-(\ref{eq:vGrav}){]},
see the companion paper \cite{WylCosNat15}. 

\subsection{General Riemann tensor\label{sub:Rie-gen}}

In the presence of sources, the Riemann tensor does not obey \eqref{eq:leftisright}.
Generically it has 20 independent components in any frame, and relative
to an arbitrary observer $\Ou$ it may be completely characterized
by the three spatial tensors~\cite{Beldecomp} 
\begin{equation}
\begin{aligned} & \mathbb{E}_{\alpha\beta}\equiv R_{\alpha\mu\beta\nu}u^{\mu}u^{\nu}\ ,\quad\mathbb{H}_{\alpha\beta}\equiv\star\!R_{\alpha\mu\beta\nu}u^{\mu}u^{\nu}\ ,\\
 & \mathbb{F}_{\alpha\beta}\equiv\star\!R\!\star_{\alpha\mu\beta\nu}u^{\mu}u^{\nu}\ ,
\end{aligned}
\label{eq:BelTensors}
\end{equation}
according to the formula~\cite{PaperAnalogies,AlfonsoSE} 
\begin{equation}
\begin{aligned}R_{\ \ \gamma\delta}^{\alpha\beta} & =4\mathbb{E}_{\ \ [\gamma}^{[\alpha}u_{\delta]}u^{\beta]}+\epsilon^{\alpha\beta\phi\psi}u_{\psi}\epsilon_{\ \ \gamma\delta}^{\mu\nu}u_{\nu}\mathbb{F}_{\phi\mu}\\
 & +2\left\{ \epsilon_{\ \ \gamma\delta}^{\mu\chi}u_{\chi}\mathbb{H}_{\mu}^{\ [\beta}u^{\alpha]}+\epsilon^{\mu\alpha\beta\chi}u_{\chi}\mathbb{H}_{\mu[\delta}u_{\gamma]}\right\} .
\end{aligned}
\label{Bel}
\end{equation}
The tensors $\mathbb{E}_{\alpha\beta}$ and $\mathbb{F}_{\alpha\beta}$
are symmetric and spatial relative to $u^{\a}$, thus having 6 independent
components each; $\mathbb{H}_{\alpha\beta}$ is spatial and traceless,
possessing 8 independent components; $\mathbb{F}_{\alpha\beta}$ has
no electromagnetic analogue.

From the Riemann tensor one can generically construct 14 algebraically
independent scalar invariants \cite{StephaniExact,Zakharov}. In particular,
the Kretschmann and Chern-Pontryagin scalars are given in terms of
the tensors \eqref{eq:BelTensors} by 
\begin{align}
 & \frac{1}{4}\inn{R}=\mathbb{E}^{\alpha\gamma}\mathbb{E}_{\alpha\gamma}+\mathbb{F}^{\alpha\gamma}\mathbb{F}_{\alpha\gamma}-2\mathbb{H}^{\alpha\gamma}\mathbb{H}_{\alpha\gamma}\ ,\label{eq:Kretsch}\\
 & \frac{1}{8}\innst{R}=(\mathbb{E}^{\alpha\gamma}-\mathbb{F}^{\alpha\gamma})\mathbb{H}_{\alpha\gamma}\ ,\label{eq:Pontry}
\end{align}
The Riemann tensor may be decomposed as follows: 
\begin{equation}
R^{\a\b}{}_{\g\d}=C^{\a\b}{}_{\g\d}+2\d_{[\g}^{[\a}R^{\b]}{}_{\d]}-\frac{1}{3}R\,\d_{[\g}^{\a}\d_{\d]}^{\b}\ .\label{Riemann decomp-R}
\end{equation}
Here $R_{\a\b}\equiv R_{~\a\g\b}^{\g}$ is the Ricci tensor, $R\equiv R_{~\a}^{\a}$
the Ricci scalar, and $C_{\a\b\g\d}\equiv\mathbf{C}$ the {\em Weyl
tensor}, which has the same symmetries as the Riemann tensor, obeys
\eqref{eq:leftisright} with ${\bf R}$ replaced by ${\bf C}$, and
is moreover trace-free: $C_{~\a\g\b}^{\g}=0$. One defines the electric
and magnetic parts of the Weyl tensor relative to an observer $\Ou$
by 
\begin{equation}
\mathcal{E}_{\alpha\beta}\equiv C_{\alpha\gamma\beta\d}u^{\gamma}u^{\d};\qquad\mathcal{H}_{\alpha\beta}\equiv\star C_{\alpha\gamma\beta\d}u^{\gamma}u^{\d}\ .\label{eq:def-EH-Weyl}
\end{equation}
These tensors are symmetric, spatial and traceless, and as in \eqref{eq:complexQab}
can be assembled into the complex tensor
\[
{\cal Q}_{\a\b}\equiv\mathcal{E}_{\alpha\beta}-i\mathcal{H}_{\alpha\beta}.
\]
By \eqref{Riemann decomp-R} the relation with the tensors \eqref{eq:BelTensors}
is 
\begin{align}
 & \mathbb{E}_{\alpha\beta}=\mathcal{E}_{\alpha\beta}+\left(\frac{R}{6}+\frac{R_{\gamma\delta}u^{\gamma}u^{\delta}}{2}\right)h_{\a\b}-\frac{1}{2}h_{\a}^{\g}h_{\b}^{\d}R_{\g\d},\label{ERiemann-EWeyl}\\
 & \mathbb{H}_{\alpha\beta}=\mathcal{H}_{\alpha\beta}+\frac{1}{2}\epsilon_{\alpha\beta\sigma\gamma}R^{\sigma\tau}u_{\tau}u^{\gamma}\ ,\label{HRiemann-HWeyl}\\
 & \mathbb{F}_{\alpha\beta}=-\mathcal{E}_{\alpha\beta}+\left(\frac{R}{3}+\frac{R_{\gamma\delta}u^{\gamma}u^{\delta}}{2}\right)h_{\a\b}-\frac{1}{2}h_{\a}^{\g}h_{\b}^{\d}R_{\g\d},\label{FRiemann-EWeyl}
\end{align}
with $h_{\b}^{\a}$ as defined in \eqref{eq:SpaceProjector}. 

In the vacuum case $R_{\a\b}=0$ of the previous subsections the Riemann
tensor equals the Weyl tensor. For a non-vacuum Riemann tensor, {\em
everything in the previous subsections holds for the Weyl tensor but
not (necessarily) for the Riemann tensor itself}, i.e., everything
holds if one replaces $R_{\a\b\g\d},\,\EE_{\a\b},\HH_{\a\b},\QQ_{\a\b}$
by $C_{\a\b\g\d},\,{\cal E}_{\a\b},\,{\cal H}_{\a\b},\,{\cal Q}_{\a\b}$
in the definitions and results of Secs.~\ref{sub:Rie-vac} and \ref{sub:grav-PEPM-obs},
referred to as their {\em Weyl generalizations}, and the scalars
$\l_{k}\,(k=1,2,3)$ are the common eigenvalues of the operators $-\tfrac{1}{2}C_{\a\b}{}^{\g\d}$
and ${\cal Q}_{~\b}^{\a}$ for any observer $\Ou$. In particular,
a non-zero Weyl tensor is called {\em purely electric (purely magnetic)}
if there exists an observer $\Oup$ for which the magnetic (electric)
part of the Weyl tensor vanishes; this happens precisely when the
Petrov type is I or D and %
the Weyl generalization of, respectively, \eqref{eq:PEPM-I} or \eqref{eq:PEPM-D}
{[}or, equivalently, of \eqref{eq:grav-PEPM1}-\eqref{eq:grav-PEPMcond}{]}
holds. The observers $\Oup$ for which ${\cal H}'_{\a\b}=0$ (${\cal E}'_{\a\b}=0$)
can be obtained from the Weyl generalization of the algorithm in Sec.~\ref{sub:grav-PEPM-obs}. 

Regarding the Riemann tensor itself, one can study the vanishing of
the gravitoelectric or gravitomagnetic tidal tensor relative to some
observer $\Oup$; one must however be careful with introducing the
terminology `purely electric' or `purely magnetic' here, because for
a non-vacuum Riemann tensor the conditions $\EE'_{\a\b}=0$ and $\HH'_{\a\b}=0$
may hold simultaneously, even for the same observer $\Oup$~\cite{Barry}\footnote{This happens precisely when $R_{\a\b\g\d}u'^{\d}=0$ and is exemplified
by Einstein's static universe metric, or more generally by 1+3 decomposable
spacetimes with $u'^{\alpha}$ orthogonal to the 3-D spacelike hypersurfaces
\cite{HallPEPM}.}. We will call the Riemann tensor and the spacetime {\em purely
electric (purely magnetic)} at a point if an observer $\Oup$ exists
for which $\HH'_{\a\b}=0$ ($\EE'_{\a\b}=0$) and moreover $\EE_{\a\b}\neq0$
($\HH_{\a\b}\neq0$) for {\em all} observers $\Oup$ (compare to
\cite{MagneticCurvatures,Barry,HerOrtWyll13}). 

The frame projections of conditions $\EE'_{\a\b}\equiv R_{\alpha\mu\beta\nu}u'^{\mu}u'^{\nu}=0$
or $\HH'_{\a\b}\equiv\star\!R_{\alpha\mu\beta\nu}u'^{\mu}u'^{\nu}=0$
form an overdetermined system of 6, resp.~8 homogeneous quadratic
equations in the components of $u'^{\a}$, which should be augmented
with the component form of the inhomogeneous quadratic condition $g_{\a\b}u'^{\a}u'^{\b}=-1$.
As for $\mathbb{E}'_{\a\b}=0$ no clear-cut general criterion is known
for when the resulting system allows for solutions (however, see \cite{Barry,BeemHarris,HerOrtWyll13}
for some partial results). As for $\mathbb{H}'_{\a\b}=0$, on the
other hand, Eq.~\eqref{HRiemann-HWeyl} gives the decomposition of
$\HH_{\a\b}$ into symmetric and antisymmetric parts, $\HH_{(\a\b)}={\cal H}_{\a\b}$
and $\HH_{[\a\b]}=\frac{1}{2}\epsilon_{\alpha\beta\sigma\gamma}R^{\sigma\tau}u_{\tau}u^{\gamma}$,
where the latter is equivalent to $\epsilon^{\alpha\beta\sigma\gamma}\HH_{[\sigma\gamma]}=2u^{[\a}R^{\b]}{}_{\gamma}u^{\gamma}$.%
{} Hence 
\begin{equation}
\HH'_{\a\b}=0\quad\Leftrightarrow\quad\begin{cases}
{\cal H}'_{\a\b}=0\ , & ({\rm i})\\
{u'}^{[\a}R^{\b]}{}_{\g}u'^{\g}=0\ . & ({\rm ii})
\end{cases}\label{eq:extr-gravmag-curv}
\end{equation}
Using the projector orthogonal to $u'^{\a}$, ${h'}_{\b}^{\a}=\delta_{\b}^{\a}+{u'}^{\a}{u'}_{\b}$,
we can still write 
\begin{align}
 & {u'}^{[\a}R^{\b]}{}_{\g}u'^{\g}=0\quad\Leftrightarrow\quad{h'}_{\b}^{\a}R_{~\g}^{\b}{u'}^{\g}=-8\pi{\mathcal{J}'}^{\alpha}=0\ .\label{A2}
\end{align}
where ${\mathcal{J}'}^{\alpha}\equiv-{h'}_{\b}^{\a}T^{\beta\gamma}{u'}_{\gamma}$
is the spatial mass/energy current density as measured by an observer
of 4-velocity $u^{\alpha}$, and in the second equality we used Einstein's
field equations $R_{\mu\nu}=8\pi(T_{\mu\nu}-\frac{1}{2}g_{\mu\nu}T_{\,\,\,\alpha}^{\alpha})+\Lambda g_{\mu\nu}$.
Equations \eqref{eq:extr-gravmag-curv} tell us that {\em an observer
$\Oup$ measures a zero gravitomagnetic tidal tensor if and only if
the magnetic part of the Weyl tensor relative to it vanishes and its
4-velocity $u'^{\a}$ is a Ricci eigenvector}~\cite{Trumper,HerOrtWyll13}.
Or, equivalently, if both the magnetic part of the Weyl tensor and
the mass-energy currents relative to it vanish. 

Define the traceless Ricci tensor as $S_{\a\b}\equiv R_{\a\b}-\tfrac{1}{4}Rg_{\a\b}$,
and its associated invariants~\cite{CampWain77,JolyMacC90}: 
\begin{align}
 & I_{6}\equiv S_{~\b}^{\a}S_{~\a}^{\b}\,,\quad I_{7}\equiv S_{~\b}^{\a}S_{~\g}^{\b}S_{~\a}^{\g}\,,\quad I_{8}\equiv S_{~\b}^{\a}S_{~\g}^{\b}S_{~\d}^{\g}S_{~\a}^{\d},\nonumber \\
 & \IS\equiv7I_{6}{}^{2}-12I_{8},\quad\JS\equiv36I_{6}I_{8}-17I_{6}{}^{3}-12I_{7}{}^{2}.\label{Ric-invars}
\end{align}
A timelike eigenvector ${u'}^{\alpha}$ of the Ricci tensor $R_{\alpha\beta}$
(or, equivalently, of $S_{\alpha\beta}$, both regarded as linear
operators in tangent space), exists if and only if $S_{~\b}^{\a}$
is diagonalizable with real eigenvalues\footnote{In general, eigenspaces of $S_{~\b}^{\a}$ corresponding to different
eigenvalues are orthogonal {[}if $S_{~\b}^{\a}v^{\b}=\l v^{\a}$ and
$S_{~\b}^{\a}w^{\b}=\mu w^{\a}$ with $\l\neq\mu$ then $S_{\a\b}v^{\a}w^{\b}=\l v_{\b}w^{\b}=\mu v_{\a}w^{\a}$
and so $v^{\a}w_{\a}=0${]}. If $S_{~\b}^{\a}$ is diagonalizable
with real eigenvalues then tangent space is an orthogonal direct sum
of the real $S_{~\b}^{\a}$-eigenspaces, so one of these is timelike
and thus contains (all) timelike eigenvectors. Conversely, if a timelike
eigenvector exists then its orthogonal complement is a real Euclidean
space and the restriction of $S_{~\b}^{\a}$ to this space is a symmetric
endomorphism; hence this restriction, and thus $S_{~\b}^{\a}$ itself,
is diagonalizable with real eigenvalues.}, which, as shown in Appendix \ref{app:extr-gravmag-curv}, happens
precisely in one of the following cases%

(a) $S_{\alpha\beta}=0$;

(b) $I_{6}\neq0$ and 
\begin{equation}
S_{~\b}^{\a}S_{~\g}^{\b}+2\lambda S_{~\g}^{\a}-3\lambda^{2}\delta_{~\g}^{\a}=0\label{eq:pf-charact}
\end{equation}
~~~with $\l\equiv-{I_{7}}/{(2I_{6})}$;

(c) $I_{6}\neq0$ and $S_{~\b}^{\a}S_{~\g}^{\b}=\frac{1}{4}I_{6}\delta_{~\g}^{\a}$;

(d) $\JS\neq-2I_{6}\IS$ and
\[
(S_{~\b}^{\a}-\l\delta_{~\b}^{\a})(S_{~\zeta}^{\b}S_{~\g}^{\zeta}+2\l S_{~\g}^{\b}+(3\l^{2}-I_{6}/2)\delta_{~\g}^{\b})=0
\]
~~~with $\lambda\equiv{-I_{7}\IS}/{(\JS+2I_{6}\IS)}$ and $I_{6}/2>2\lambda^{2}$;

(e) $(\IS)^{3}>(\JS)^{2}$.\\
Note that this property can be expressed in terms of scalar invariants
of the Ricci tensor only in case (e) but not in the more degenerate
cases (a)-(d). {[}This somewhat parallels the situation of condition
\eqref{eq:PEPM-I} for Weyl-Petrov type I vs.~%
\eqref{eq:PEPM-D} for type D and $C_{\a\b\g\d}=0$ for type O{]}.
For instance, the invariant relations $\IS=\JS=0<I_{6}$ identify
(b) among the subcases above, but Ricci tensors satisfying these relations
{\em and not} \eqref{eq:pf-charact} do exist.

Introducing also the Weyl generalizations of \eqref{eq:inv-A}-\eqref{eq:inv-D},
\begin{align*}
 & \tfrac{1}{8}\inn{C}\equiv\tfrac{1}{8}C^{\a\b\g\d}C_{\a\b\g\d}={\cal E}^{\a\b}{\cal E}_{\a\b}-{\cal H}^{\a\b}{\cal H}_{\a\b}\ ,\\
 & \tfrac{1}{16}\innst{C}\equiv\tfrac{1}{16}\star\!C^{\a\b\g\d}C_{\a\b\g\d}={\cal E}^{\a\b}{\cal H}_{\a\b}\ ,\\
 & \AC\equiv-\tfrac{1}{16}C_{\ \ \lambda\mu}^{\alpha\beta}C_{\ \ \ \rho\sigma}^{\lambda\mu}C_{\ \ \alpha\beta}^{\rho\sigma}={\cal E}_{\ \beta}^{\alpha}{\cal E}_{\ \gamma}^{\beta}{\cal E}_{\ \alpha}^{\gamma}-3{\cal E}_{\ \beta}^{\alpha}{\cal H}_{\ \gamma}^{\beta}{\cal H}_{\ \alpha}^{\gamma},\\
 & \BC\equiv\tfrac{1}{16}\star\!C_{\ \ \lambda\mu}^{\alpha\beta}C_{\ \ \ \rho\sigma}^{\lambda\mu}C_{\ \ \alpha\beta}^{\rho\sigma}={\cal H}_{\ \beta}^{\alpha}{\cal H}_{\ \gamma}^{\beta}{\cal H}_{\ \alpha}^{\gamma}-3{\cal E}_{\ \beta}^{\alpha}{\cal E}_{\ \gamma}^{\beta}{\cal H}_{\ \alpha}^{\gamma},
\end{align*}
equations \eqref{eq:extr-gravmag-curv} then imply the following new
criterion, providing the necessary and sufficient conditions for the
existence of observers measuring $\HH'_{\a\b}=0$.

\emph{Criterion:} observers $\Oup$ measuring a vanishing gravitomagnetic
tidal tensor ($\HH'_{\a\b}=0$) exist in three Petrov types of spacetime,
and precisely under the following conditions.

\subsubsection{Petrov type O\label{sub:Petrov-type-O}}

In this case $C_{\a\b\g\d}=0$, thus the existence of an observer
for which $\mathbb{H}'_{\alpha\beta}=0$ reduces to the existence
of a time-like eigenvector of the Ricci tensor, corresponding to cases
(a)-(e) above. The explicit expressions for the 4-velocities of such
observers are given, respectively, in items (a)-(e) of Appendix \ref{app:extr-gravmag-curv}
(to which we refer for further details).

\subsubsection{Petrov type D\label{sub:Petrov-type-D}}

First the Weyl tensor needs to be purely electric. A Weyl tensor is
purely electric and of type D if and only if the corresponding generalization
of \eqref{eq:PEPM-D}, 
\begin{equation}
\QC_{~\g}^{\a}\QC_{~\b}^{\g}+\l\QC_{\b}^{\a}-2\l^{2}h_{\b}^{\a}=0\,,\quad\l=-\frac{8\AC}{\inn{C}}\,,\label{eq:Weyl-PE-D}
\end{equation}
is satisfied. Condition (\ref{eq:extr-gravmag-curv}i) then holds
for an observer $\Oup$ if and only if $u'^{\a}$ lies in the timelike
Weyl principal plane $\Sigma$, spanned by null vectors $k^{\a}$
and $l^{\a}$, with $k^{\a}l_{\a}=-1$. These are given by the Weyl
generalization of \eqref{PNDs} with \eqref{eq:wvector}, \eqref{eq:evec}
and \eqref{eq:tgrav2}, identifying $k^{\a}=k_{+}^{\a},\;l^{\a}=k_{-}^{\a}$.
Such a vector $u'^{\a}$ can be parametrized by 
\begin{equation}
u'^{\a}=\frac{1}{\sqrt{2q}}(qk^{\a}+l^{\a}),\qquad q>0,\label{up-para}
\end{equation}
{[}where $q=(C+D)^{2}$%
{} compared to \eqref{eq:u'grav}{]} and we need to find when \eqref{A2}
holds for some $q$. Note that for each $q>0$ the vector $\x'^{\a}=qk^{\a}-l^{\a}$
spans the spacelike direction in $\Sigma$ orthogonal to $u'^{\a}$.
Thus condition \eqref{A2} geometrically means that the vector $R_{~\b}^{\a}u'^{\b}$
lies in $\Sigma$ and is orthogonal to $\x'^{\a}$, which is respectively
expressed by $\e_{\a\b\g\d}k^{\b}l^{\g}R_{~\zeta}^{\d}u^{\zeta}=0$
and $\x'^{\a}R_{\a\b}u'^{\b}=0$. With the definitions 
\begin{align*}
 & \K^{\a}\equiv\e_{\a\b\g\d}k^{\b}l^{\g}R_{~\zeta}^{\d}k^{\zeta},\quad\Rkk\equiv R_{\a\b}k^{\a}k^{\b}\,;\\
 & \L^{\a}\equiv\e_{\a\b\g\d}k^{\b}l^{\g}R_{~\zeta}^{\d}l^{\zeta},\quad\Rll\equiv R_{\a\b}l^{\a}l^{\b}\,,
\end{align*}
these conditions are equivalent to 
\begin{align}
\L^{\a}=-q\K^{\a},\qquad\Rll=q^{2}\Rkk,\qquad q>0.
\end{align}
Note that the vectors $\K^{\a}$ and $\L^{\a}$ are orthogonal to
$\Sigma$, and are thus spacelike or zero. It follows that an observer
exists at $p$ for which $\mathbb{H}'_{\a\b}=0$ if and only if either\\
 \vspace{-0.2cm}

(2-a) $\K^{\a}=\L^{\a}=0$, $\Rkk=\Rll=0$;\; or

(2-b) $\K^{\a}=\L^{\a}=0$, $\Rkk\Rll>0$;\; or

(2-c) $\K^{[\a}\L^{\b]}=0$, $\K^{\a}\L_{\a}<0$, $\Rkk=\Rll\left(\frac{\K^{\a}\K_{\a}}{K^{\b}L_{\b}}\right)^{2}$\\
 \vspace{-0.2cm}

\noindent is satisfied on top of \eqref{eq:Weyl-PE-D}. The characteristic
condition $\K^{\a}=\L^{\a}=0$ in (2-a) and (2-b) means that the restriction
of the Ricci operator $R_{~\b}^{\a}$ to $\Sigma$ is an endomorphism;
if moreover $\Rkk=\Rll=0$ then $k^{\a}$ and $l^{\a}$ are both eigenvectors
with the same eigenvalue $-R_{\a\b}k^{\a}l^{\b}$ and we have (2-a)
{[}else (2-b){]}, in which case $\Sigma$ is a timelike eigenplane.
The observers $\Oup$ for which (\ref{eq:extr-gravmag-curv}) holds
are those with 4-velocity \eqref{up-para} and the following values
of $q>0$:
\begin{enumerate}
\item[(2-a)] any $q>0$ (all observers with 4-velocity in $\Sigma$). Examples:
all ``doubly aligned'' electrovacuum spacetimes which are (at the
given point) Weyl purely electric%
.
\item[(2-b)] $q=\sqrt{\Rll/\Rkk}$ (unique observer). Examples: G\"{o}del universe
(and all other ``aligned'' perfect fluid spacetimes), Som-Raychaudhuri
\emph{uniform} metrics.
\item[(2-c)] $q=-\K^{\a}\L_{\a}/\K^{\b}\K_{\b}$ (unique observer). Examples:
special types of ``non-aligned'' electrovacuum spacetimes which
are (at the given point) Weyl purely electric.
\end{enumerate}
The above mentioned examples are discussed in Sec. \ref{sec:Non-vacuum-examples}.
The `aligned' perfect fluids mentioned in case (2-b) consist of Petrov
D fluids for which the fluid's 4-velocity belongs to the time-like
principal plane $\Sigma$ \cite{WyllPhD}. They include all spherically
or plane symmetric perfect fluid (including dust) models \cite{StephaniExact}.

\subsubsection{Petrov type I\label{sub:Petrov-type-I}}

A Weyl tensor is purely electric and of Petrov type I if and only
if the corresponding generalization of \eqref{eq:PEPM-I},
\begin{align}
\innst{C} & =\BC=0\,,\quad(\inn{C}/8)^{3}>6\AC^{2}\label{eq:Weyl-PE-I}\\
\Leftrightarrow\{\innst{C} & =0,\,\inn{C}>0\}\ \ \mbox{and}\ \ \{\J_{C}=0\;\textrm{or}\;\M_{C}>0\}\nonumber 
\end{align}
is satisfied. In this case there is at each point a \emph{unique}
observer $\Oup$ for which (\ref{eq:extr-gravmag-curv}i) holds, with
4-velocity $u'^{\a}\propto\tI^{\a}$ given by the Weyl generalization
of \eqref{eq:u'-grav-I}. Hence, an observer measuring a vanishing
gravitomagnetic tidal tensor $\mathbb{H}'_{\a\b}$ exists if and only
if the condition 
\begin{equation}
\tI^{[\a}R_{\;\;\g}^{\b]}\tI^{\g}=0\label{tI-cond}
\end{equation}
is satisfied on top of \eqref{eq:Weyl-PE-I}, in which case \eqref{eq:extr-gravmag-curv}
holds for the unique observer $\Oup$ of 4-velocity $u'^{\a}\propto\tI^{\a}$. 

Among the spacetimes verifying these conditions are all Petrov type
I (locally or globally) static spacetimes: by definition (see e.g.
\cite{Cilindros}), they admit a hypersurface orthogonal time-like
Killing vector field; the congruence of observers tangent to such
field is vorticity-free, i.e., they measure no gravitomagnetic field
$H^{\alpha}=2\omega^{\alpha}=0$, cf. Eq. (\ref{eq:GEMfields}). Hence,
by (\ref{Hij}), $\mathbb{H}_{\alpha\beta}=0$ for all such observers.
Examples of non-vacuum solutions of this kind are the locally static
cylindrically symmetric Einstein-Maxwell solutions discussed in Sec.
4 of \cite{BerghWilsCQG1985}, or Sec. 3 of \cite{MacCallumEMCylinders}.
Non-static examples are the ``gravito-electric'' dust models in
\cite{VdBWyll1}.

An example of the more common situation that one condition but not
the other is satisfied is the van Stockum rotating cylinder, discussed
in Sec. \ref{sub:Van-Stockum-cylinder}.%

\subsubsection{What can be said based only on the curvature invariants}

It follows from \eqref{eq:extr-gravmag-curv} that any invariant condition
which ensures that ${\cal H}_{\a\b}\neq0$ for all observers (i.e.,
that the Weyl tensor is non-zero and {\em not} purely electric)
also ensures that $\HH_{\a\b}\neq0$ for all observers. In line with
\eqref{eq:Weyl-PE-I}, condition \eqref{eq:Weyl-PE-D} implies ${\innst{C}}=\BC=0$
and $(\inn{C}/8)^{3}=6\AC^{2}>0$; hence $\innst{C}\neq0$ or $\BC\neq0$
or $(\inn{C}/8)^{3}<6\AC^{2}$ (or, less stringently, $\inn{C}<0$)
all imply that $\HH_{\a\b}\neq0$ for all observers. In particular,
one obtains from (\ref{eq:Pontry}) and (\ref{ERiemann-EWeyl})-(\ref{FRiemann-EWeyl})%
, that 
\begin{equation}
\mathbf{\star R}\cdot\mathbf{R}=\mathbf{\star C}\cdot\mathbf{C}=16\mathcal{E}^{\alpha\beta}\mathcal{H}_{\alpha\beta}\ ,\label{starRR}
\end{equation}
and so {\em a non-zero Chern-Pontryagin scalar, $\mathbf{\star R}\cdot\mathbf{R}\ne0$,
implies $\mathcal{H}_{\alpha\beta}\ne0$ and thus $\mathbb{H}_{\alpha\beta}\ne0$
for all observers}.

Moreover, if there is a non-zero cosmological constant $\Lambda$
but no sources ($T_{\alpha\beta}=0,\,R_{\a\b}=\Lambda g_{\a\b},\,R=4\Lambda\neq0$)
then Eqs.~(\ref{ERiemann-EWeyl})-(\ref{FRiemann-EWeyl}) reduce
to 
\begin{align}
 & \mathbb{E}_{\alpha\beta}=\mathcal{E}_{\alpha\beta}-\frac{\Lambda}{3}(g_{\alpha\beta}+u_{\alpha}u_{\beta})=-\mathbb{F}_{\alpha\beta},\label{eq:ELambda}\\
 & \mathbb{H}_{\alpha\beta}=\mathcal{H}_{\alpha\beta}\quad(\Rightarrow\mathbb{H}_{[\alpha\beta]}=0)\ .\label{eq:HLambda}
\end{align}
From (\ref{eq:ELambda}) one has $\mathbb{E}_{\ \alpha}^{\alpha}=-\Lambda$,
and ${\cal E}_{\a\b}$ is simply obtained as the tracefree part of
$\mathbb{E}_{\a\b}$. It follows that the Riemann tensor is never
purely magnetic, i.e., $\mathbb{E}_{\alpha\beta}\ne0$ for all observers,
and is purely electric if and only if the Weyl tensor is zero or purely
electric, where the last case is equivalent to either \eqref{eq:Weyl-PE-D}
(not totally based on curvature invariants) or \eqref{eq:Weyl-PE-I}.%

\section{Interpretation of the invariant structure of the relevant electromagnetic
setups\label{sec:Interpretation_Invariants_EM}}

We are especially interested in understanding physically the invariant
$\vec{E}\cdot\vec{B}$, and why the magnetic field vanishes for some
observers in certain setups. Consider an arbitrary distribution of
charges and currents, and an arbitrary congruence of observers $\Ou$
of 4-velocity $u^{\alpha}$. The projections parallel and orthogonal
to $u^{\alpha}$ of the Maxwell field equations $F_{\ \ ;\beta}^{\alpha\beta}=4\pi j^{\alpha}$
and $\star F_{\ \ ;\beta}^{\alpha\beta}=0$, respectively, yield the
source equations for the magnetic field, which, in an orthonormal
frame ``adapted'' to the observers $\Ou$, read (see \cite{PaperAnalogies},
Sec. 3.4.1 for details) 
\begin{eqnarray}
\nabla^{\perp}\times\vec{B} & = & \dot{\vec{E}}-\vec{a}\times\vec{B}+4\pi\vec{j}-\sigma_{\ }^{\hat{\imath}\hat{\jmath}}E_{\hat{\jmath}}\vec{e}_{\hat{\imath}}+\frac{2}{3}\theta\vec{E}\;;\label{eq:curlBtetrad}\\
\nabla^{\perp}\cdot\vec{B} & = & -2\vec{\omega}\cdot\vec{E}\;,\label{eq:DivBtetrad}
\end{eqnarray}
where 
\begin{align}
 & a^{\alpha}=u_{\ ;\beta}^{\alpha}u^{\beta}\ ;\qquad\omega^{\alpha}=\frac{1}{2}\epsilon^{\alpha\beta\gamma\delta}u_{\gamma;\beta}u_{\delta}\ ;\nonumber \\
 & \sigma^{\alpha\beta}=h_{\lambda}^{\alpha}h_{\tau}^{\beta}u^{(\lambda;\tau)}-\frac{1}{3}u_{\ ;\tau}^{\tau}\ ;\qquad\theta=u_{\ ;\alpha}^{\alpha}\label{eq:Kinematics}
\end{align}
are, respectively, the acceleration, vorticity, shear, and expansion
scalar of the observer congruence; $h_{\ \beta}^{\alpha}$ is the
spatial projector defined in \eqref{eq:SpaceProjector}, $\nabla^{\perp}$
is the spatial projection of the Levi-Civita covariant derivative
$\nabla$, 
\begin{equation}
\nabla_{\gamma}^{\perp}X^{\alpha_{1}...\alpha_{n}}=h_{\beta_{1}}^{\alpha_{1}}...h_{\beta_{n}}^{\alpha_{n}}\nabla_{\gamma}X^{\beta_{1}...\beta_{n}}\ ,\label{eq:nabla_perp}
\end{equation}
and dot denotes the ordinary time derivative along the observer's
worldline, $\dot{X}^{\hat{\alpha}_{1}...\hat{\alpha}_{n}}\equiv u^{\hat{\beta}}\partial_{\hat{\beta}}X^{\hat{\alpha}_{1}...\hat{\alpha}_{n}}$.
Hats in the indices (e.g. $\hat{\imath}$) denote tetrad components.
In \emph{an} \emph{inertial frame}, all the kinematical quantities
\eqref{eq:Kinematics} vanish, $\nabla_{i}^{\perp}=\nabla_{i}$ and
Eqs. \eqref{eq:curlBtetrad}-\eqref{eq:DivBtetrad} take the well
known form 
\begin{equation}
\nabla\times\vec{B}=\dot{\vec{E}}+4\pi\vec{j}\quad({\rm i}),\qquad\nabla\cdot\vec{B}=0\quad({\rm ii)}.\label{eq:BeqsInertial}
\end{equation}
Based on these equations we make the following observations: 
\begin{enumerate}
\item \label{enu:Empoint1}If $\dot{\vec{E}}+4\pi\vec{j}\ne\vec{0}$ in
an \emph{inertial} frame, then, according to Eq. (\ref{eq:BeqsInertial}i),
$\vec{B}$ cannot vanish in that frame on any \emph{spatial} 3-D open
region (only on 2-surfaces or lower-dimensional sets). 
\item \label{enu:EMpoint2}If there exists an \emph{inertial} frame where
$\dot{\vec{E}}+4\pi\vec{j}=\vec{0}$ \emph{everywhere}, and no fields
are present other than those arising from the sources, then $\vec{B}=0$
\emph{globally} in that frame. This implies $\vec{E}\cdot\vec{B}=0$
everywhere. Example, consider a system of $N$ point charges: if an
\emph{inertial} frame exists where they are all at rest, then $\vec{j}=\dot{\vec{E}}=\vec{0}$
and so $\vec{B}=0\Rightarrow\vec{E}\cdot\vec{B}=0$ everywhere (cf.
Eq.~\eqref{eq:Bsuperposition}). 
\item \label{enu:EMpoint3}Observation \ref{enu:EMpoint2} is guaranteed
\emph{only when} one is dealing with an inertial frame; for in an
arbitrary frame, as can be seen from Eqs.~\eqref{eq:curlBtetrad}-\eqref{eq:DivBtetrad},
the vorticity and shear/expansion of the observer congruence contribute
as sources for $\vec{B}$. Examples: spinning charged body, or a system
of point charges in rigid rotational motion (e.g. the rotating pair
of charges in Fig. \ref{fig:Two}); the bodies are at rest with respect
to the co-rotating frame, and so $\vec{j}=\dot{\vec{E}}=\vec{0}$
everywhere in this frame. In spite of that, generically $\nabla^{\perp}\cdot\vec{B}\ne0\Rightarrow\vec{B}\ne\vec{0}$,
cf.~Eq.~\eqref{eq:DivBtetrad}, and \emph{also} $\vec{E}\cdot\vec{B}\ne0$
generically. 
\item \label{enu:EMpoint4}The converse of \ref{enu:EMpoint2} is not true:
when there is no inertial frame where $\dot{\vec{E}}+4\pi\vec{j}=\vec{0}$
everywhere, this does not necessarily mean that $\vec{E}\cdot\vec{B}\ne0$.
The magnetic field can still vanish at some region (3-D or lower-dimensional)
with respect to inertial or non-inertial frames. Even within a region
where $\dot{\vec{E}}+4\pi\vec{j}\ne\vec{0}$, it can vanish with respect
to inertial frames at spatial 2-surfaces or lower-dimensional sets
(cf. point \ref{enu:Empoint1}), or, for non-inertial frames, even
on 3-D spatial regions. In fact, from Eq.~\eqref{eq:curlBtetrad}
we see that, in a region where $\dot{\vec{E}}+4\pi\vec{j}\ne0$, we
can still have $\nabla^{\perp}\times\vec{B}=0$ for a non-inertial
frame, which is compatible with a vanishing $\vec{B}$. These situations
will be exemplified in Secs.~\ref{sub:Two-charges,-planar}-\ref{sub:Further-examples:-the Cylinder}. 
\end{enumerate}
Point \ref{enu:Empoint1} is just the statement that $\nabla\times\vec{B}\ne0$
implies that $\vec{B}$ cannot vanish on spatial open sets. To prove
point \ref{enu:EMpoint2}, one notes that, when $\vec{B}$ is well
defined in the \emph{whole }space, $\nabla\times\vec{B}=0$ means
that $\vec{B}=\nabla\psi$, for some scalar function $\psi$; the
equation $\nabla\cdot\vec{B}=\nabla^{2}\psi=0$ then implies (via
Green's theorem) that if, at infinity, $\vec{B}=0$ (no sources at
infinity), then $\vec{B}=0$ everywhere.\\

\emph{Systems of N point charges.---} Systems of point particles are
of special interest herein as they may be cast as the building blocks
of the classical systems we will study. In the case of gravitational
systems, they are studied in the framework of the first order post-Newtonian
approximation (1PN); in electromagnetism we shall use an analogous
approximation that we dub, following \cite{WillBook}, the first ``post-Coulombian''
(1PC) approximation. It can be stated as follows: one scales, by some
small dimensionless parameter $\epsilon$, 
\begin{equation}
\frac{q}{m}\phi\sim\epsilon^{2};\qquad v\lesssim\epsilon;\qquad v_{{\rm s}}\lesssim\epsilon\ ,\label{eq:approxScheme}
\end{equation}
where $\phi$ is the Coulomb potential, $q$ and $m$ are the charge
and mass of a test particle, $v$ and $v_{{\rm s}}$ are the velocities
of the test particle and of the sources. Time derivatives increase
the degree of smallness of a quantity by a factor $\epsilon$; for
example, $\partial\phi/\partial t\sim\phi v_{{\rm s}}\sim\epsilon\phi$.
This is thus both a weak field and slow motion assumption. The approximation
consists of keeping terms up to $\Os{4}$ in the equations of motion;
take the case of the Lorentz force, 
\[
\vec{a}=\frac{q}{m}\left(\vec{E}+\vec{v}\times\vec{B}\right)=\frac{q}{m}\left(-\nabla\phi-\partial_{t}\vec{A}+\vec{v}\times(\nabla\times\vec{A})\right)\ ;
\]
in order to know $\vec{a}$ to order $\O{4}{}$, one needs to know
$q\phi/m$ to order $\Os{4}$ and $q\vec{A}/m$ to order $\Os{3}$.
For a system of $N$ point charges, this amounts to considering a
4-potential\footnote{These expressions follow from Eqs. (2.73)-(2.74) of \cite{WillBook},
for the case of zero gravitational field, setting $\mu_{0}=\varepsilon_{0}=1$
therein. They could also be obtained from the exact Liénard-Wiechert
retarded potentials Eqs. (14.8) of \cite{Jackson} (superimposing
the potentials of single moving particles given therein, since electromagnetism
is linear), expanding them to 1PC order, and noting that to 1PC order
the instantaneous relative position $\vec{r}_{{\rm a}}$ is related
by a quadratic extrapolation to the retarded relative position $\vec{R}_{{\rm a}}$,
$\vec{r}_{{\rm a}}\simeq\vec{R}_{{\rm a}}-\vec{v}R_{{\rm a}}-\frac{1}{2}\vec{a}_{{\rm a}}R_{{\rm a}}^{2}$.} $A^{\alpha}=(A^{0},\vec{A})$ whose components read, in an \emph{inertial}
frame, 
\begin{equation}
A^{0}=\sum_{{\rm a}}Q_{{\rm a}}\left[\frac{1}{r_{{\rm a}}}\left(1+\frac{v_{{\rm a}}^{2}}{2}-\frac{1}{2}\vec{r}_{{\rm a}}\cdot\vec{a}_{{\rm a}}\right)-\frac{(\vec{r}_{{\rm a}}\cdot\vec{v}_{{\rm a}})^{2}}{2r_{{\rm a}}^{3}}\right]\ ;\label{eq:A0_EM}
\end{equation}
\begin{equation}
\vec{A}=\frac{\sum_{{\rm a}}Q_{{\rm a}}\vec{v}_{{\rm a}}}{r_{{\rm a}}}\ ,\label{eq:A_EM}
\end{equation}
where $Q_{{\rm a}}$ is the charge of particle ``${\rm a}$'', $\vec{r}_{{\rm a}}\equiv\vec{x}-\vec{x}_{{\rm a}}$,
$\vec{x}$ is the point of observation, $\vec{x}_{{\rm a}}$ is the
instantaneous position of particle ``${\rm a}$'', $\vec{v}_{{\rm a}}=\partial\vec{x}_{{\rm a}}/\partial t$
its velocity and $\vec{a}_{{\rm a}}=\partial\vec{v}_{{\rm a}}/\partial t$
its acceleration; for a system of interacting bodies \emph{with no
external forces}, to 1PC accuracy, $\vec{a}_{{\rm a}}$ is to be taken
in Eq. \eqref{eq:A0_EM} as the acceleration caused by the Coulomb
field produced by the other charges, i.e., $\vec{a}_{{\rm a}}=(Q_{{\rm a}}/m_{{\rm a}})\sum_{{\rm b}\ne{\rm a}}Q_{{\rm b}}\vec{r}_{{\rm a}{\rm b}}/r_{{\rm a}{\rm b}}^{3}$,
with $\vec{r}_{{\rm a{\rm b}}}\equiv\vec{x}_{{\rm a}}-\vec{x}_{{\rm b}}$.
The 1PC electric and magnetic fields, $\vec{E}=-\nabla A^{0}-\partial_{t}\vec{A}$
and $\vec{B}=\nabla\times\vec{A}$, follow as 
\begin{eqnarray}
\vec{E} & = & \sum_{{\rm a}}Q_{{\rm a}}(1+\varphi_{{\rm a}})\frac{\vec{r}_{{\rm a}}}{r_{{\rm a}}^{3}}-\frac{1}{2}\sum_{{\rm a}}Q_{{\rm a}}\frac{\vec{a}_{{\rm a}}}{r_{{\rm a}}}\ ;\label{eq:E_Nbody}\\
\vec{B} & = & \sum_{{\rm a}}\frac{Q_{{\rm a}}}{r_{{\rm a}}^{3}}\vec{v}_{{\rm a}}\times\vec{r}_{{\rm a}}=\sum_{{\rm a}}\vec{v}_{{\rm a}}\times\left[\vec{E}_{{\rm a}}\right]_{{\rm C}}\ ,\label{eq:Bsuperposition}
\end{eqnarray}
where 
\[
\varphi_{{\rm a}}\equiv\frac{v_{{\rm a}}^{2}}{2}-\frac{1}{2}(\vec{r}_{{\rm a}}\cdot\vec{a}_{{\rm a}})-\frac{3}{2}\frac{(\vec{r}_{{\rm a}}\cdot\vec{v}_{{\rm a}})^{2}}{r_{{\rm a}}^{2}}
\]
and $\left[\vec{E}_{{\rm a}}\right]_{{\rm C}}=Q_{{\rm a}}\vec{r}_{{\rm a}}/r_{{\rm a}}^{3}$
denotes the Coulomb (i.e., 0PC) electric field of particle ``a''.

We shall next discuss the electromagnetic invariants and the fields
measured by different observers in setups which may be cast as the
analogues of the gravitational systems of interest, and where the
observations \ref{enu:Empoint1}-\ref{enu:EMpoint4} above will be
exemplified.

\subsection{One single point charge\label{sub:EMOne-single-point}}

In the inertial rest frame of the charge, one has $A^{\alpha}=(\frac{Q}{r},0,0,0)$,
$\vec{E}=(Q/r^{2})\vec{e}_{r}$, $\vec{B}=0$. The two scalar invariants
of $F_{\alpha\beta}$ are 
\begin{equation}
\vec{E}^{2}-\vec{B}^{2}=\frac{Q^{2}}{r^{4}}>0\ ,\qquad\vec{E}\cdot\vec{B}=0\ \ \mbox{(everywhere)}\ ,\label{eq:InvariantsSingleCharge}
\end{equation}
telling us that $F_{\alpha\beta}$ is purely electric (everywhere),
i.e., everywhere there are observers for which $B^{\alpha}=0$. Those
are the observers at rest in the inertial rest frame of the source
(\emph{``static''} observers), and also observers in purely radial
motion, since, as we have seen in Sec. \ref{sub:Observers-with-no},
the component $\vec{v}_{\parallel E}$ along $\vec{E}$ of the velocity
of the observers measuring no magnetic field is arbitrary. In other
words, such observers have a 4-velocity of the form\footnote{This can be explicitly checked by noticing that the magnetic field
$B^{'\alpha}=\star F^{\alpha\beta}u'_{\beta}$ as measured by an arbitrary
observer of 4-velocity $u'^{\alpha}=(u'^{t},u'^{r},u'^{\theta},u'^{\phi})$
has, as only non-vanishing components \cite{PaperGyros}, $B'^{\theta}=-Qu'^{\phi}\sin\theta/r^{2}$,
$B'^{\phi}=Qu'^{\theta}\csc\theta/r^{2}$.} $u^{\alpha}=(u^{0},u^{r},0,0)$.

In order to understand the invariant structure \eqref{eq:InvariantsSingleCharge},
let $\mathcal{S}$ and $\mathcal{S}'$ be, respectively, the inertial
rest frame of the point charge, and an inertial frame moving relative
to it with some velocity $\vec{v}$ (non-parallel to $\vec{E}$).
In $\mathcal{S}$, $\vec{B}=0$ \emph{globally}, which implies $\vec{E}\cdot\vec{B}=0$
\emph{everywhere}. Observers $\Op$ at rest in $\mathcal{S}'$, in
turn, measure a non-zero magnetic field $\vec{B}'$; but it is such
that it is always orthogonal to $\vec{E}'$, ensuring $\vec{E}'\cdot\vec{B}'=0$,
as we shall now explicitly show. By equation (\ref{Bvec'}) {[}or
its covariant form \eqref{b2}{]}, 
\begin{equation}
\vec{B}'=-\gamma\vec{v}\times\vec{E}\ ,\qquad\left[B'^{\alpha}=-\epsilon^{\alpha\beta\gamma\delta}E_{\gamma}u_{\delta}u'_{\beta}\right]\ .\label{BSingleCharge}
\end{equation}
Thus, $\vec{B}'$ is perpendicular to the electric field $\vec{E}$
measured in the charge's rest frame and to the velocity $\vec{v}$;
hence it is also perpendicular to $\vec{E}'$, as is seen from \eqref{Evec'}:
\begin{equation}
\vec{E}'=\gamma\vec{E}-\frac{\gamma^{2}}{\gamma+1}\vec{v}\left(\vec{v}\cdot\vec{E}\right)\ .\label{eq:EsingleCharge}
\end{equation}

\subsection{System of two point charges\label{sub:Two-charges,-planar}}

We shall now consider two moving charged particles with charges $Q_{1}$,
$Q_{2}$ of the same sign. If they move with different velocities
with respect to some inertial frame, then there is no inertial frame
where they are \emph{both} at rest. From Eq. (\ref{eq:BeqsInertial}i)
we see that, by contrast with the example in the previous section,
in this case the magnetic field cannot vanish \emph{globally} in an
inertial frame. Let us see how this reflects in the invariants. By
\eqref{eq:E_Nbody} the electric field $\vec{E}$, at an arbitrary
point $P$ with coordinates $\vec{x}$, is 
\begin{eqnarray}
\vec{E} & = & \frac{Q_{1}}{r_{1}^{3}}(1+\varphi_{1})\vec{r}_{1}+\frac{Q_{2}}{r_{2}^{3}}(1+\varphi_{2})\vec{r}_{2}\nonumber \\
 &  & -\frac{1}{2}Q_{1}\frac{\vec{a}_{1}}{r_{1}}-\frac{1}{2}Q_{2}\frac{\vec{a}_{2}}{r_{2}}\ ,
\end{eqnarray}
where $\vec{r}_{1}=\vec{x}-\vec{x}_{1}$, $\vec{r}_{2}=\vec{x}-\vec{x}_{1}$.
By \eqref{eq:Bsuperposition} the magnetic field is 
\begin{equation}
\vec{B}=\frac{Q_{1}}{r_{1}^{3}}\vec{v}_{1}\times\vec{r}_{1}+\frac{Q_{2}}{r_{2}^{3}}\vec{v}_{2}\times\vec{r}_{2}\ .\label{eq:BTwoCharges}
\end{equation}
At an arbitrary point the invariant $\vec{E}\cdot\vec{B}$ is thus

\begin{align}
 & \vec{E}\cdot\vec{B}=\frac{Q_{1}}{r_{1}^{3}}(\vec{v}_{1}\times\vec{r}_{1})\cdot\left[\frac{Q_{2}}{r_{2}^{3}}(1+\varphi_{2})\vec{r}_{2}-\frac{1}{2}Q_{1}\frac{\vec{a}_{1}}{r_{1}}-\frac{1}{2}Q_{2}\frac{\vec{a}_{2}}{r_{2}}\right]\nonumber \\
 & +\frac{Q_{2}}{r_{2}^{3}}(\vec{v}_{2}\times\vec{r}_{2})\cdot\left[\frac{Q_{1}}{r_{1}^{3}}(1+\varphi_{1})\vec{r}_{1}-\frac{1}{2}Q_{1}\frac{\vec{a}_{1}}{r_{1}}-\frac{1}{2}Q_{2}\frac{\vec{a}_{2}}{r_{2}}\right]\label{eq:E.BTwoCharges}
\end{align}
which is generically non-vanishing. To lowest order, 
\begin{equation}
\vec{E}\cdot\vec{B}\simeq\frac{Q_{1}Q_{2}}{r_{1}^{3}r_{2}^{3}}\left[(\vec{v}_{1}\times\vec{r}_{1})\cdot\vec{r}_{2}+(\vec{v}_{2}\times\vec{r}_{2})\cdot\vec{r}_{1}\right]\ .\label{eq:EBTwoChargeslowest}
\end{equation}
As for the invariant $\vec{E}^{2}-\vec{B}^{2}$, there is a region
between the two charges where $\vec{B}^{2}>\vec{E}^{2}$ (magnetic
dominance), around the point where $\vec{E}\simeq\frac{Q_{1}}{r_{1}^{3}}\vec{r}_{1}+\frac{Q_{2}}{r_{2}^{3}}\vec{r}_{2}=0$;
elsewhere $\vec{E}^{2}\geq\vec{B}^{2}$, and henceforth we restrict
attention to this region of electric dominance.\\

\subsubsection*{Coplanar motion\label{sub:Coplanar-motion}}

Take now the case when $\vec{v}_{1}$, $\vec{v}_{2}$ and the position
vectors of the bodies are coplanar (i.e., the two bodies move in the
same plane). It follows that $\vec{E}\cdot\vec{B}$ vanishes in the
plane of motion, and is generically non-zero outside that plane. It
is easy to see from Eq. (\ref{eq:E.BTwoCharges}) that in the plane
of motion $\vec{E}\cdot\vec{B}=0$: taking the point of observation
$P$ to lie on that plane, then the $\vec{r}_{{\rm a}}$, $\vec{v}_{{\rm a}}$
and $\vec{a}_{{\rm a}}$ all lie on that plane; hence $(\vec{v}_{{\rm a}}\times\vec{r}_{{\rm a}})\cdot\vec{r}_{{\rm b}}=(\vec{v}_{{\rm a}}\times\vec{r}_{{\rm a}})\cdot\vec{a}_{{\rm b}}=0$.
This means that, in this plane, \textit{\emph{there are observers
for which the magnetic field vanishes}}.

\begin{figure}[H]
\includegraphics[height=2in]{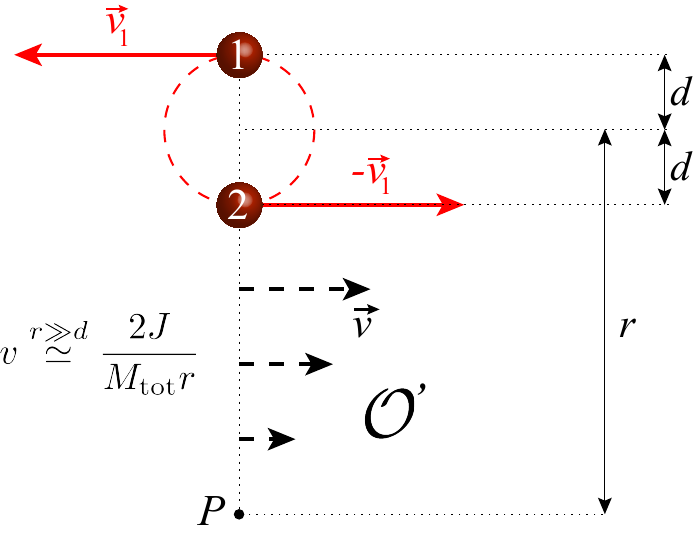}

\protect\protect\protect\protect\caption{\label{fig:Two}Two equal charges in antipodal circular motion ---
observers $\Op$ for which the magnetic field vanishes (represented
only along the axis passing through the particles); their velocity
$\vec{v}$ is depicted by black dashed arrows.}
\end{figure}

We will investigate such observers in the simple example in Fig. \ref{fig:Two},
that will prove enlightening for the next section: two particles,
with equal charge $Q$, in circular motion of radius $d$ and in antipodal
positions (e.g., with some rod holding them), so that their velocities
are equal in magnitude but opposite in direction: $\vec{v}_{1}=-\vec{v}_{2}$.

In this special case $\vec{a}_{1}=a(\vec{r}_{1}-\vec{r}_{2})/2d$,
$\vec{a}_{2}=a(\vec{r}_{2}-\vec{r}_{1})/2d$ and $(\vec{v}_{2}\times\vec{r}_{2})\cdot\vec{r}_{1}=(\vec{v}_{1}\times\vec{r}_{1})\cdot\vec{r}_{2}$,
such that Eq.\ (\ref{eq:E.BTwoCharges}) simplifies to 
\begin{eqnarray}
\vec{E}\cdot\vec{B} & = & Q^{2}\left[\frac{2+\varphi_{2}+\varphi_{1}}{r_{1}^{3}r_{2}^{3}}+\frac{a}{4d}\left(\frac{1}{r_{1}}-\frac{1}{r_{2}}\right)\left(\frac{1}{r_{1}^{3}}-\frac{1}{r_{2}^{3}}\right)\right]\nonumber \\
 &  & (\vec{v}_{1}\times\vec{r}_{1})\cdot\vec{r}_{2}\ ,\label{eq:EB2Equalcharges}
\end{eqnarray}
which has the structure 
\[
\begin{cases}
\vec{E}\cdot\vec{B}=0\ \mbox{in the plane of motion,}\\
\vec{E}\cdot\vec{B}\ne0\ \mbox{elsewhere}.
\end{cases}
\]
That $\vec{E}\cdot\vec{B}\ne0$ at any point outside the plane of
motion (denote it by $\Sigma$) can be seen as follows. First note
that the first line of \eqref{eq:EB2Equalcharges} cannot be zero,
since by the 1PC assumptions all terms must be much smaller than the
first one ($2Q^{2}/r_{1}^{3}r_{2}^{3}$). Then note that $(\vec{v}_{1}\times\vec{r}_{1})\cdot\vec{r_{2}}=(\vec{v}_{1}\times[\vec{r}_{1}-\vec{r}_{2}])\cdot\vec{r_{2}}$.
Since $\vec{r_{1}}-\vec{r}_{2}\in\Sigma$, the vector $\vec{v}_{1}\times[\vec{r}_{1}-\vec{r}_{2}]$
is orthogonal to $\Sigma$; hence $\vec{E}\cdot\vec{B}=0$ only if
$\vec{r}_{2}$ lies on $\Sigma$, which is possible only if the point
of observation $P\in\Sigma$; at any point $P$ outside $\Sigma$,
$\vec{E}\cdot\vec{B}\ne0$. This means that \emph{outside }$\Sigma$
the magnetic field is \emph{non-vanishing for all observers}, and
that \emph{in }$\Sigma$\emph{ there are observers for which} $\vec{B}'=0$.
From Eq.~(\ref{eq:explicitv}), such observers must have a velocity
whose component orthogonal to $\vec{E}$, $\vec{v}_{\perp E}=\vec{v}_{\parallel p}$,
reads 
\begin{equation}
\vec{v}_{\perp E}=\frac{r_{2}^{4}\vec{v}_{1\perp r_{1}}-r_{1}^{4}\vec{v}_{1\perp r_{2}}+r_{1}r_{2}\left[(\vec{r}_{1}\cdot\vec{v}_{1})\vec{r}_{2}-(\vec{r}_{2}\cdot\vec{v}_{1})\vec{r}_{1}\right]}{r_{1}^{4}+r_{2}^{4}+2r_{1}r_{2}\vec{r}_{1}\cdot\vec{r_{2}}}\ ,\label{eq:vnoB2chargesGen}
\end{equation}
where $\vec{v}_{1\perp r_{1}}$ and $\vec{v}_{1\perp r_{2}}$ are
the components of $\vec{v}_{1}$ orthogonal to $\vec{r}_{1}$ and
$\vec{r}_{2}$, respectively. We first notice that $\vec{v}_{\perp E}$,
and, therefore, $\vec{v}$ (since $\vec{E}$ at any point of $\Sigma$
lies on $\Sigma$, except at the middle point $\vec{r}_{1}=-\vec{r}_{2}$
where $\vec{E}=0$) lies on the plane of motion $\Sigma$. The reason
why $\vec{B}'$ vanishes for these observers is especially easy to
understand along the axis passing through the two particles, see Fig.
\ref{fig:Two}. First note that, clearly, the magnetic field at $P$,
as measured by the static observer $\OO$, is non-vanishing, because
although the magnetic field produced by particle 1 acts in opposite
direction to the magnetic field from particle 2, the latter is closer
to $P$ so that the two fields do not cancel out. By choosing an observer
$\Op$ moving with 3-velocity $\vec{v}$ in the same direction as
particle 2, one is decreasing particle 2's velocity, and, at the same
time, increasing particle 1's velocity relative to the observer's
inertial rest frame. That means \emph{decreasing} the magnetic field
$\vec{B}'_{2}$ generated by particle 2 and \emph{increasing} the
magnetic field $\vec{B}'_{1}$ generated by particle 1, so that eventually
one can make the (total) magnetic field $\vec{B}'=\vec{B}'_{1}+\vec{B}'_{2}$
vanish. Along the axis (with $r_{1}=r+d$, $r_{2}=r-d$, cf. Fig.
\ref{fig:Two}), the observers $\Op$ for which $\vec{B}'=0$ have
velocities 
\begin{equation}
\vec{v}=\vec{v}_{1}\frac{-2dr}{(r^{2}+d^{2})}\ \ \Rightarrow\ \ v\ \stackrel{r\gg d}{\simeq}\ \frac{2v_{1}d}{r}=\frac{2J}{M_{{\rm tot}}r}\label{eq:vnoBtwocharges}
\end{equation}
where $M_{{\rm tot}}=M_{1}+M_{2}=2M_{1}$ is the system's total mass,
and we noted that $M_{{\rm tot}}v_{1}d=J$ is the system's angular
momentum as measured in the center of mass frame.

\subsection{Spinning spherical charge\label{sub:A-spinning-spherical}}

\begin{figure}[H]
\includegraphics[width=0.98\columnwidth]{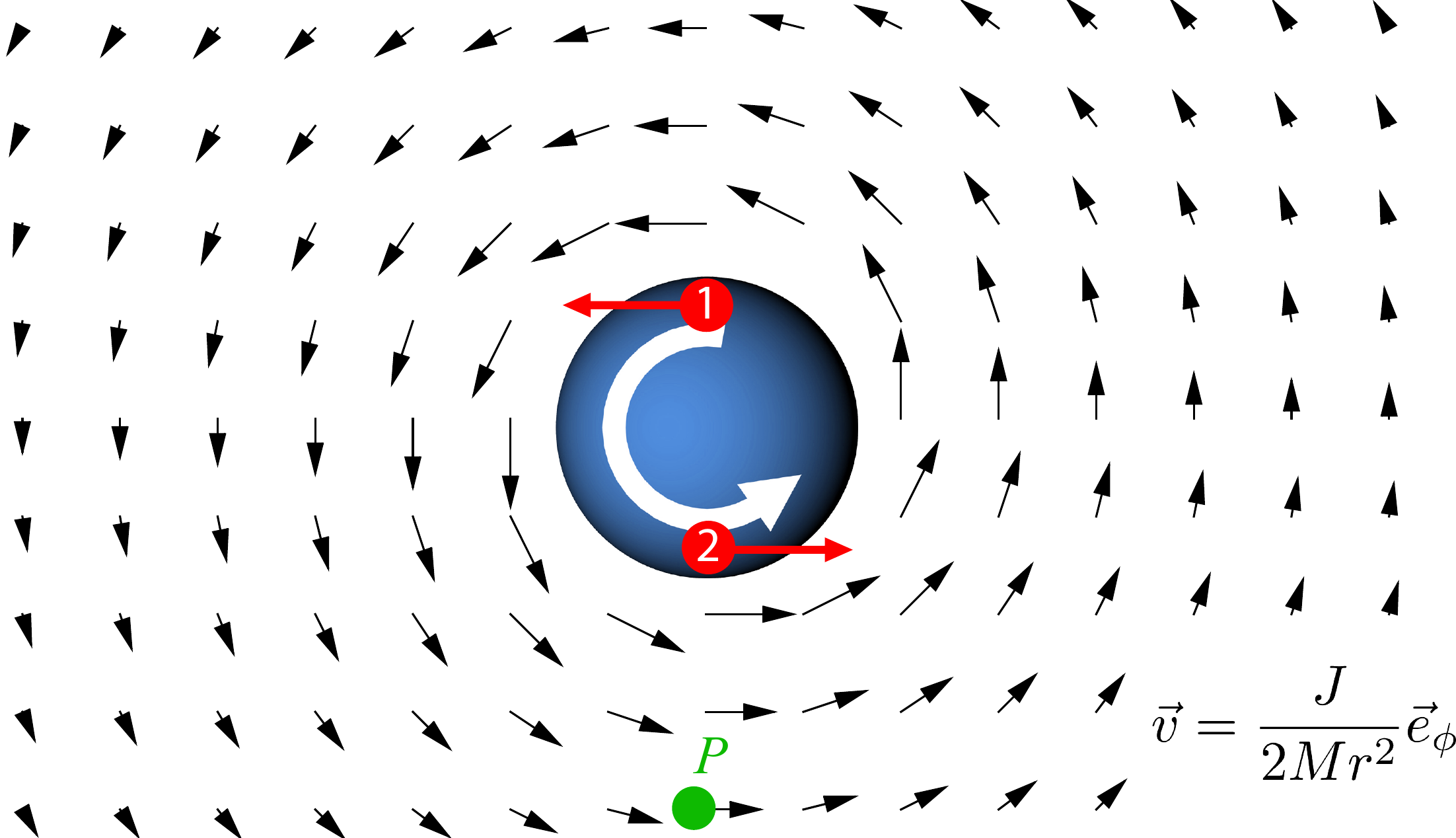}

\protect\protect\protect\protect\caption{\label{fig:SchargeDots}Observers for which the magnetic field vanishes.
(Note that these are \textit{not} observers ``co-rotating'' with
the same angular velocity of the spinning body). $\vec{e}_{\phi}$
is the coordinate basis vector $\vec{e}_{\phi}\equiv\vec{\partial}_{\phi}=r\vec{e}_{\hat{\phi}}$.
Observer $\Op$ at point $P$ must have a velocity that decreases
the magnetic field generated by the charge elements of the closer
hemisphere (e.g., charge element $2$), and increases the magnetic
field produced by the charge elements of the opposite hemisphere (e.g.,
charge element $1$), such that they eventually cancel out.}
\end{figure}

Consider a spinning charged spherical body with mass $M$, angular
momentum $\vec{J}=J\vec{e}_{z}$, charge $Q$ and dipole moment $\vec{\mu}_{{\rm s}}=(Q/2M)J\vec{e}_{z}$.
The electric and magnetic fields produced are, as measured by the
rest observers, 
\begin{equation}
\vec{E}=\frac{Q}{r^{2}}\vec{e}_{r}\ ,\qquad\vec{B}=\frac{2\mu_{{\rm s}}\cos\theta}{r^{3}}\vec{e}_{r}+\frac{\mu_{{\rm s}}\sin\theta}{r^{4}}\vec{e}_{\theta}\ ,\label{EMsphere}
\end{equation}
where $\vec{e}_{i}\equiv\vec{\partial}_{i}$ denote \emph{coordinate}
basis vectors. The invariants are given by\footnote{The first inequality always holds assuming the classical gyromagnetic
ratio $\mu_{{\rm s}}/J=Q/2M$, corresponding to a classical source
where the charge and mass are identically distributed. In that case
\[
\frac{\mu_{{\rm s}}^{2}}{r^{6}}=\frac{Q^{2}}{4r^{6}}\frac{J^{2}}{M^{2}}\le\frac{1}{4}\frac{Q^{2}}{r^{4}}\frac{R^{2}}{r^{2}}
\]
where $R$ is the body's radius and we have used the fact that, in
order for the dominant energy condition to be obeyed, $R\ge J/M$,
see \cite{CostaNatario2014}. Since $r>R$ at any point exterior to
the particle, we have $\mu_{{\rm s}}^{2}/r^{6}<Q^{2}/r^{4}\Rightarrow\vec{E}^{2}-\vec{B}^{2}>0$. }: 
\begin{equation}
\left\{ \begin{array}{l}
{\displaystyle \vec{E}^{2}-\vec{B}^{2}=\frac{Q^{2}}{r^{4}}-\frac{\mu_{{\rm s}}^{2}(5+3\cos2\theta)}{2r^{6}}>0\ ,}\\
\\
\vec{E}\cdot\vec{B}={\displaystyle \frac{2\mu_{{\rm s}}Q\cos\theta}{r^{5}}\,\ \ \ \ (\vec{E}\cdot\vec{B}=0\mbox{ in the plane }\theta=\pi/2)\ .}
\end{array}\right.\label{eq:InvariantsSpinningCharge}
\end{equation}
Since $\vec{E}\cdot\vec{B}=0$ in the equatorial plane ($\theta=\pi/2$),
observers $\mathcal{O}'$ exist in this plane for which $\vec{B}'=0$.
From Eq. \eqref{eq:explicitv}, the velocity of those observers is
such that its component $\vec{v}_{\perp E}=\vec{v}_{\parallel p}$
orthogonal to $\vec{E}$ is 
\[
\vec{v}_{\perp E}=\frac{\vec{E}\times\vec{B}}{\vec{E}^{2}}=\frac{\mu_{{\rm s}}}{Qr^{2}}\vec{e}_{\phi}\ ,
\]
no restriction being imposed on the (radial) component $\vec{v}_{\parallel E}=v^{r}\vec{e}_{r}$
parallel to $\vec{E}$ (apart from the normalization condition $u'^{\alpha}u'_{\alpha}=-1$).
That is, observers moving in the equatorial plane ($v^{\theta}=0$)
with angular velocity 
\begin{equation}
\frac{d\phi}{dt}=\frac{u'^{\phi}}{u'^{t}}=\frac{\mu_{{\rm s}}}{Qr^{2}}=\frac{J}{2Mr^{2}}\label{eq:EMvplot}
\end{equation}
measure a vanishing magnetic field. One might check that these are
indeed the only observers for which $B^{'\alpha}=0$ by computing
explicitly $B^{'\alpha}$ for an arbitrary 4-velocity $u'^{\alpha}=(u'^{t},u'^{r},u'^{\theta},u'^{\phi})$,
as done in \cite{PaperGyros}. If we take the \emph{special case}
$v^{r}=u'^{r}=0$, we obtain the velocity field $\vec{v}=J/(2Mr^{2})\vec{e}_{\phi}$
depicted in Fig. \ref{fig:SchargeDots}. 

The vanishing of \textbf{$\vec{B}'$} for such observers can be understood
in the same spirit as in the case of the two point charges in coplanar
motion of Fig. \ref{fig:Two}. A rotating charged body may be decomposed
in arbitrarily small charge elements in translation; and its electromagnetic
field \eqref{EMsphere} cast as a superposition of the field produced
by each such elements. In particular, to 1PC order, we may write for
$\vec{B}$ (cf. Eq. \eqref{eq:Bsuperposition}) 
\begin{equation}
\vec{B}(\vec{r})=\int_{{\rm body}}\rho_{{\rm c}}\frac{\vec{v}_{{\rm c}}(\vec{x}')\times(\vec{r}-\vec{x}')}{|\vec{r}-\vec{x}'|^{3}}d^{3}\vec{x}'\ ,\label{eq:Bintegral}
\end{equation}
where $\vec{v}_{{\rm c}}(\vec{x}')$ is the velocity of the charge
element $\rho_{{\rm c}}d^{3}\vec{x}'$ at the point $\vec{x}'$. Consider
the situation in Fig. \ref{fig:SchargeDots}. Relative to an observer
at rest at a point $P$ of the equatorial plane, the charge elements
in the closer hemisphere (e.g., charge element $2$) move in opposite
direction to the ones in the opposite hemisphere (e.g., charge element
$1$), so their contributions $\vec{v}_{{\rm c}}(\vec{x}')\times(\vec{r}-\vec{x}')$
to the integral \eqref{eq:Bintegral} have opposite signs. The net
field $\vec{B}$ is different from zero because one hemisphere is
closer than the other (leading to a dipole field). The observers $\Op$
for which $\vec{B}'=0$ at $P$ must move in the same sense as the
rotational motion of the body, thereby decreasing the relative velocity
of the charge elements in the closer hemisphere (decreasing the magnitude
of their magnetic field), and increasing the relative velocity of
the elements in the farther hemisphere (increasing the magnitude of
their magnetic field), with a suitable velocity $\vec{v}$ such that
the fields from the two hemispheres cancel out.

Notice the similarity with the result obtained in Sec. \ref{sub:Coplanar-motion}:
the magnitude of the velocity field in Fig. \ref{fig:SchargeDots}
is $v=J/2Mr$, which, up to a factor of four, matches the asymptotic
behavior of the field \eqref{eq:vnoBtwocharges} depicted in Fig.
\ref{fig:Two}.

Using Eq. (\ref{Bvec'}), and noting that $\vec{B}\cdot\vec{v}=0$
to obtain $\vec{B}'=\gamma\vec{B}-\gamma\vec{v}\times\vec{E}$, one
can also interpret the vanishing of $\vec{B}'$ for the observers
in Fig. \ref{fig:SchargeDots} as a cancellation between the magnetic
field $\gamma\vec{B}$ arising from the rotational motion of the source
and the magnetic field $-\gamma\vec{v}\times\vec{E}$ arising from
the translational motion of the source relative to the observer. The
fact that such cancellation may occur only in the equatorial plane
is easy to see noting that since the translational magnetic field
$-\gamma\vec{v}\times\vec{E}$ is orthogonal to $\vec{E}$, it can
kill the rotational field only if $\vec{B}$ is also orthogonal to
$\vec{E}$ which, for this setup, happens only in the equatorial plane.

Finally, notice that the observers in Fig. \ref{fig:SchargeDots}
exemplify one case of point \ref{enu:EMpoint4} of Sec. \ref{sec:Interpretation_Invariants_EM}:
there is no \emph{inertial} frame where the different charge elements
are at rest\footnote{In a co-rotating frame the whole spinning body is at rest; but such
frame consists of a congruence of observers all having\emph{ different}
4-velocities $U^{\alpha}$, thus different inertial rest frames, whilst
having the same angular velocity. Actually, no single point in the
body is at rest with respect to the inertial frame of a co-rotating
observer if the latter lies outside the body. $\vec{B}$ does \emph{not}
vanish in the co-rotating frame, even though the body is at rest therein
(so $\vec{j}=\dot{\vec{E}}=0$); taking the perspective of such frame,
this is justified with the fact that the vorticity of the observer
congruence contributes as a source for $\vec{B}$, cf. Eqs. \eqref{eq:curlBtetrad}-\eqref{eq:DivBtetrad}.}, i.e., where $\vec{j}=0$ everywhere; moreover, as in the two-body
system of Sec. \ref{sub:Two-charges,-planar}, the magnetic field
does not \emph{globally} vanish in any inertial frame. Yet, in a spatial
2-surface, there are still observers measuring no magnetic field,
only they do not form an inertial frame (take e.g. the congruence
with $d\phi/dt$ given by \eqref{eq:EMvplot}, and $u'^{\theta}=u'^{r}=0$;
such congruence is accelerated and shears\footnote{\label{fn:Kinematics_obs_noB}For the congruence $u'^{\alpha}=u'^{0}(1,0,0,d\phi/dt)$,
with $u'^{0}=1/\sqrt{1-(d\phi/dt)^{2}g_{\phi\phi}}$, the non-vanishing
components of the acceleration and shear are, in the equatorial plane,
$a^{r}=a^{2}/(a^{2}r-4r^{3})$, $\sigma_{r\phi}=\sigma_{\phi r}=-4ar^{2}/(4r^{2}-a^{2})^{3/2}$,
$\sigma_{rt}=\sigma_{tr}=2a^{2}/(4r^{2}-a^{2})^{3/2}$, where $a\equiv J/M$.
The vorticity and expansion vanish in that plane.}).

\subsection{Further examples --- infinite rotating cylinder\label{sub:Further-examples:-the Cylinder}}

Here we consider a simple physical system that exemplifies the remaining
cases mentioned in point~\ref{enu:EMpoint4} of Sec.~\ref{sec:Interpretation_Invariants_EM}.
Consider the electromagnetic field produced by a uniform, rotating,
and infinitely long cylinder of radius $R$ and charge density $\rho_{{\rm c}}$.
The electric and magnetic fields, as measured by static observers,
read, in cylindrical coordinates $(r,\phi,z)$, 
\begin{align*}
r<R: & \quad\vec{E}=2\pi\rho_{{\rm c}}r\vec{e}_{r}\ ;\qquad\vec{B}=2\pi\rho_{{\rm c}}\Omega(R^{2}-r^{2})\vec{e}_{z}\ ;\\
r\ge R: & \quad\vec{E}=\frac{2\pi\rho_{{\rm c}}R^{2}}{r}\vec{e}_{r}\ ;\qquad\vec{B}=\vec{0}\ .
\end{align*}
It follows that $\vec{E}\cdot\vec{B}=0$ everywhere, and $\vec{E}^{2}-\vec{B}^{2}>0$
($<0$) for $r>r_{{\rm c}}$($<r_{{\rm c}}$), where the critical
radius $r_{{\rm c}}^{2}=R^{2}+(1-\sqrt{1+4R^{2}\Omega^{2}})/(2\Omega^{2})$
defines the boundary between the \emph{purely} electric/magnetic regions,
and lies inside the cylinder ($r_{{\rm c}}<R$). The magnetic field
$\vec{B}$ vanishes at every point outside the cylinder in the inertial
frame of the static observers; this exemplifies one of the situations
in point \ref{enu:EMpoint4} of Sec. \ref{sec:Interpretation_Invariants_EM}:
even when there is no inertial frame where the currents are zero everywhere,
still\textbf{ $\vec{B}$} can vanish in a 3-D region relative to an
inertial frame. Inside the cylinder, for $r>r_{{\rm c}}$ (purely
electric region), it vanishes for certain observers. Such observers
have a velocity whose component orthogonal to $\vec{E}$ is obtained
from Eq. \eqref{eq:explicitv}, 
\[
\vec{v}_{\perp E}=\frac{\vec{E}\times\vec{B}}{\vec{E}^{2}}=\frac{\Omega(R^{2}-r^{2})}{r}\vec{e}_{r}\times\vec{e}_{z}=\Omega\left[1-\frac{R^{2}}{r^{2}}\right]\vec{e}_{\phi}\ ;
\]
i.e., observers with angular velocity $d\phi/dt=\Omega\left[1-R^{2}/r^{2}\right]$
(in the sense {\em opposite} to the cylinder's rotation). Such
observer congruences are not inertial, as they are shearing, rotating
and accelerating. This exemplifies another situation in point \ref{enu:EMpoint4}
of Sec. \ref{sec:Interpretation_Invariants_EM}: with respect to non-inertial
frames, even in a 3-D region where $\vec{j}\ne0$, one can have $\vec{B}=\vec{0}$.

\section{Interpretation of the invariant structure for the relevant astrophysical
setups\label{sec:Interpretation-of-the-Gravitational}}

In the gravitational case we are interested in understanding the curvature
invariants of the gravitational fields of current experimental interest,
in particular the Chern-Pontryagin invariant $\star\mathbf{R}\cdot\mathbf{R}$
and its vanishing in some setups.

From the differential Bianchi identities\textbf{ $R_{\alpha\beta[\gamma\delta;\mu]}=0$},
written in terms of the electric part $\mathcal{E}_{\alpha\beta}=C_{\alpha\gamma\beta\sigma}u^{\gamma}u_{\textrm{ }}^{\sigma}$
and magnetic part $\mathcal{H}_{\alpha\beta}=\star C_{\alpha\gamma\beta\sigma}u^{\gamma}u_{\textrm{ }}^{\sigma}$
of the Weyl tensor with respect to the observers $\Ou$, see \eqref{eq:def-EH-Weyl},
one obtains the source equations for $\mathcal{H}_{\alpha\beta}$
\cite{EMMbook}, which read\footnote{To obtain Eqs. \eqref{eq:CurlHij}-\eqref{eq:DivHij} from (6.34)-(6.35)
of \cite{EMMbook}, one notes that the ``dot'' derivative in \cite{EMMbook},
Eq. (4.6) therein, denotes $\nabla_{\mathbf{u}}=u^{\alpha}\nabla_{\alpha}$
(not an ordinary time derivative, as it does in the present paper);
that for a spatial, traceless, and symmetric 2-tensor, $\nabla_{\mathbf{u}}\mathcal{E}^{\langle\hat{\imath}\hat{\jmath}\rangle}=\nabla_{\mathbf{u}}^{\perp}\mathcal{E}^{\hat{\imath}\hat{\jmath}}=\dot{\mathcal{E}}^{\hat{\imath}\hat{\jmath}}+2\Gamma_{\hat{0}\hat{k}}^{(\hat{\imath}}\mathcal{E}^{\hat{\jmath})\hat{k}}=\dot{\mathcal{E}}^{\hat{\imath}\hat{\jmath}}+2\omega_{\ \hat{k}}^{(\hat{\imath}}\mathcal{E}^{\hat{\jmath})\hat{k}}$
(see the connection coefficients in \cite{PaperAnalogies}, with $\vec{\Omega}=\vec{\omega}$);
and that $\omega_{\gamma\langle\alpha}\mathcal{E}_{\beta\rangle}^{\ \gamma}=\omega_{\gamma(\alpha}\mathcal{E}_{\beta)}^{\ \gamma}$
and $\omega_{\ \langle\alpha}^{\gamma}\pi_{\beta\rangle\gamma}=\omega_{\gamma(\alpha}\pi_{\beta)}^{\ \gamma}$.}, in an orthonormal frame ``adapted'' \cite{PaperAnalogies} to
the observers,

\begin{align}
\begin{array}{c}
{\displaystyle {\rm curl}\mathcal{H}_{\hat{\imath}\hat{\jmath}}}\end{array} & =\dot{\mathcal{E}}_{\hat{\imath}\hat{\jmath}}+\omega^{\hat{m}}\epsilon_{\hat{m}\hat{k}(\hat{\imath}}\mathcal{E}_{\hat{\jmath})}^{\ \hat{k}}-3\sigma_{\hat{k}\langle\hat{\imath}}\mathcal{E}_{\jmath\rangle}^{\ \hat{k}}-2a^{\hat{k}}\epsilon_{\hat{k}\hat{m}(\hat{\imath}}\mathcal{H}_{\hat{\jmath})}^{\ \hat{m}}\nonumber \\
 & +\mathcal{E}_{\hat{\imath}\hat{\jmath}}\theta+4\pi\left[(\rho+p)\sigma_{\hat{\imath}\hat{\jmath}}+\nabla_{\langle\hat{\imath}}^{\perp}\mathcal{J}_{\hat{\jmath}\rangle}+2a_{\langle\hat{\imath}}\mathcal{J}_{\hat{\jmath}\rangle}\right.\nonumber \\
 & \left.+\dot{\pi}_{\hat{\imath}\hat{\jmath}}+\omega^{\hat{m}}\epsilon_{\hat{m}\hat{k}(\hat{\imath}}\pi_{\hat{\jmath})}^{\ \hat{k}}+\frac{\theta}{3}\pi_{\hat{\imath}\hat{\jmath}}+\sigma_{\ \langle\hat{\imath}}^{\hat{k}}\pi_{\hat{\jmath}\rangle\hat{k}}\right]\ ;\label{eq:CurlHij}\\
\nabla_{\hat{\jmath}}^{\perp}\mathcal{H}_{\ \hat{\imath}}^{\hat{\jmath}} & =-3\omega^{\hat{\jmath}}\mathcal{E}_{\hat{\imath}\hat{\jmath}}-\epsilon_{\hat{\imath}\hat{\jmath}\hat{k}}\sigma_{\ \hat{m}}^{\hat{\jmath}}\mathcal{E}^{\hat{k}\hat{m}}-4\pi\left[2(\rho+p)\omega_{\hat{\imath}}\right.\nonumber \\
 & \left.-\epsilon_{\hat{\imath}\hat{\jmath}\hat{k}}\sigma_{\ \hat{m}}^{\hat{\jmath}}\pi^{\hat{m}\hat{k}}-\pi_{\hat{\imath}\hat{\jmath}}\omega^{\hat{\jmath}}+(\nabla^{\perp}\times\vec{\mathcal{J}})_{\hat{\imath}}\right]\ ,\label{eq:DivHij}
\end{align}
where ${\rm curl}A_{\alpha\beta}\equiv-\epsilon_{\ \ \ (\alpha}^{\gamma\mu\nu}A_{\beta)\nu;\mu}u_{\gamma}$
and the index notation $\langle\mu\nu\rangle$ stands for the spatially
projected, symmetric and trace-free part of a rank two tensor: 
\[
A_{\langle\mu\nu\rangle}\equiv h_{(\mu}^{\alpha}h_{\nu)}^{\beta}A_{\alpha\beta}-\frac{1}{3}h_{\mu\nu}h_{\alpha\beta}A^{\alpha\beta}\ ,
\]
with $h_{\ \beta}^{\alpha}$ defined in Eq.~\eqref{eq:SpaceProjector}.
In these equations $\rho\equiv T^{\alpha\beta}u_{\alpha}u_{\beta}$
is the mass/energy density, $\mathcal{J}^{\alpha}\equiv-h_{\beta}^{\alpha}T^{\beta\gamma}u_{\gamma}$
is the spatial mass/energy current density as measured by an observer
of 4-velocity $u^{\alpha}$, $p=T^{\alpha\beta}h_{\alpha\beta}/3$
is the pressure, and $\pi^{\alpha\beta}\equiv T^{\langle\alpha\beta\rangle}$
is the \emph{traceless} spatial projection of $T^{\alpha\beta}$ with
respect to $u^{\alpha}$ (i.e., the traceless stress tensor, cf. \cite{EMMbook}
Eq. (5.9)). The tensors $\mathcal{E}_{\alpha\beta}$ and $\mathcal{H}_{\alpha\beta}$
are related to the electric and magnetic parts of the Riemann tensor,
$\mathbb{E}_{\alpha\beta}$ and $\mathbb{H}_{\alpha\beta}$, by Eqs.
\eqref{ERiemann-EWeyl}-\eqref{HRiemann-HWeyl}. We note in particular
that 
\begin{equation}
\mathbb{H}_{\alpha\beta}=\mathcal{H}_{\alpha\beta}-4\pi\epsilon_{\alpha\beta\sigma\gamma}\mathcal{J}^{\sigma}u^{\gamma}\ .\label{eq:H_HWeyl2}
\end{equation}
Equations \eqref{eq:CurlHij}-\eqref{eq:DivHij} exhibit formal similarities
with Maxwell's equations \eqref{eq:curlBtetrad}-\eqref{eq:DivBtetrad}.

The gravitational fields of the astrophysical setups of interest are
considered in the literature at post-Newtonian accuracy. Such approximation
may be cast as follows (see e.g. \cite{Gravitation,WillBook,DSX,Kaplan,JantzenThomas}).
One scales 
\[
U\sim\epsilon^{2}\ ;\qquad v\lesssim\epsilon\ ;\qquad v_{{\rm s}}\lesssim\epsilon\ ,
\]
where $U$ is the Newtonian potential and $v_{s},v$ are the velocities
of the sources and of the test particle. The first post-Newtonian
order (1PN) consists of keeping terms up to $\mathcal{O}(\epsilon^{4})\equiv\Os{4}$
in the equations of motion (see e.g. \cite{WillBook} Sec. 4.1 (b)).
This amounts to considering a metric of the form \cite{DSX,SoffelKlioner,Kaplan}
\begin{align}
g_{00} & =-1+2w-2w^{2}+\Os{6}\ ;\nonumber \\
g_{i0} & =\mathcal{A}_{i}+\Os{5}\ ;\qquad g_{ij}=\delta_{ij}\left(1+2U\right)+\Os{4}\ ,\label{eq:PNmetric}
\end{align}
where $\vec{\mathcal{A}}$ is the ``gravitomagnetic vector potential''
and the scalar $w$ consists of the sum of $U$ plus \emph{non-linear}
terms of order $\epsilon^{4}$, $w=U+\Os{4}$. The electric and magnetic
parts of the Riemann tensor, as measured by an observer at rest ($u^{i}=0$)
in the coordinate system of (\ref{eq:PNmetric}), are, using the 1PN
Christoffel symbols in e.g. Eqs. (8.15) of \cite{WillPoissonBook},
\begin{eqnarray}
\mathbb{E}_{ij} & = & -w_{,ij}+\dot{\mathcal{A}}_{(i,j)}+3U_{,i}U_{,j}\nonumber \\
 &  & -\delta_{ij}(\ddot{U}+(\nabla U)^{2})+\ensuremath{\O{6}{2}}\ ;\label{eq:EijPN}\\
\mathbb{H}_{ij} & = & -\frac{1}{2}\epsilon_{i}^{\ lk}\mathcal{A}_{k,lj}-\epsilon_{ij}^{\ \ k}\dot{U}_{,k}+\ensuremath{\O{5}{2}}\ ;\label{eq:HijPN}
\end{eqnarray}
$\mathbb{E}_{\alpha0}=\mathbb{E}_{0\alpha}=\mathbb{H}_{\alpha0}=\mathbb{H}_{0\alpha}=0$.
It is useful to note that, \emph{to the accuracy at hand,} one may
substitute into Eq.~\eqref{eq:HijPN}: 
\begin{equation}
\mathcal{\vec{A}}(x)=-4\int\frac{\vec{\mathcal{J}}}{|\vec{x}-\vec{x}'|}d^{3}\vec{x}'\ ;\qquad\dot{U}(x)=-\int\frac{\nabla\cdot\vec{\mathcal{J}}}{|\vec{x}-\vec{x}'|}d^{3}\vec{x}'\ .\label{eq:A,U_PerfectFluid}
\end{equation}
In $\dot{U}$ we used the relation (see e.g. \cite{Gravitation})
$\partial\rho/\partial t=-\nabla\cdot\vec{\mathcal{J}}+O(\rho_{,j}\epsilon^{3})$,
which is an approximation (accurate enough \emph{for Eq. \eqref{eq:HijPN}})
to the conservation equation $T_{\ ;\beta}^{0\beta}=0$. We may also
re-write Eqs. \eqref{eq:CurlHij}-\eqref{eq:DivHij} to 1PN order,
\begin{eqnarray}
\begin{array}{c}
{\displaystyle {\rm curl}\mathcal{H}_{ij}}\end{array} & = & [\dot{\mathcal{E}}_{ij}]_{{\rm N}}+4\pi\mathcal{J}_{\langle i,j\rangle}+\ensuremath{\O{5}{3}}\ ;\label{eq:curlHijPN}\\
\mathcal{H}_{\ i,j}^{j} & = & -4\pi(\nabla\times\vec{\mathcal{J}})_{i}+\ensuremath{\O{5}{3}},\label{eq:DivHijPN}
\end{eqnarray}
where $[\mathcal{E}_{ij}]_{{\rm N}}$ is the \emph{traceless} Newtonian
(i.e., 0PN) tidal tensor, $[\mathcal{E}_{ij}]_{{\rm N}}=-U_{,ij}+\delta_{ij}U_{\ ,k}^{k}/3$,
and we noted that the shear of a 1PN frame vanishes, $\sigma_{ij}=(\Gamma_{0j}^{i}+\Gamma_{0i}^{j})/2-\dot{U}\delta_{ij}=0$,
and that $\vec{\omega}=\nabla\times\vec{\mathcal{A}}/2\sim\ensuremath{\O{3}{}}$,
$\theta=3\dot{U}\sim\ensuremath{\O{3}{}}$, $\mathcal{J}^{i}\sim O(\rho\epsilon)$,
$\pi_{ij}\sim O(\rho\epsilon^{2})\sim p$, and $\rho\sim\ensuremath{\O{2}{2}}$
(via $\nabla^{2}U\simeq-4\pi\rho$). It is important in this context
to notice that the neglect of the terms involving contractions of
the tensors $\mathcal{E}_{\alpha\beta}$ or $\mathcal{H}_{\alpha\beta}$
with the kinematical quantities \eqref{eq:Kinematics} embodies a
restriction on the type of reference frame (for, e.g., if one chooses
an accelerated or rapidly rotating frame, even for weak sources or
in the far field regime, one could not neglect the terms involving
the vorticity and acceleration); it is reasonable in post-Newtonian
frames \cite{WillNordvedt1972,Gravitation,DSX} (such as the one associated
to the coordinate system of the metric \eqref{eq:PNmetric}), because
they are as close as possible to inertial frames.

Eqs.~\eqref{eq:curlHijPN}-\eqref{eq:DivHijPN}, together with \eqref{eq:HijPN}-\eqref{eq:A,U_PerfectFluid},
allow one to draw conclusions to some extent analogous to points \ref{enu:Empoint1}-\ref{enu:EMpoint4}
of Sec. \ref{sec:Interpretation_Invariants_EM}, using PN frames instead
of inertial frames: 
\begin{enumerate}
\item \label{enu:Gravpoint1}If, in a PN frame, the right-hand side of Eqs.~\eqref{eq:curlHijPN}
or \eqref{eq:DivHijPN} is non-zero, then $\mathcal{H}_{ij}\ne0\Rightarrow\mathbb{H}_{ij}\ne0$
\emph{in that frame} (it can be zero only on 2-surfaces or lower-dimensional
sets). 
\item \label{enu:Gravpoint2}If there exists a PN frame where $\vec{\mathcal{J}}=0$
everywhere then $\mathbb{H}_{ij}=\mathcal{H}_{ij}=0$ everywhere,
and so $\star\mathbf{R}\cdot\mathbf{R}=0$ everywhere. Example: system
of $N$ point masses; if there exists a PN frame where they are all
at rest then $\mathbb{H}_{ij}=\mathcal{H}_{ij}=0$ and $\star\mathbf{R}\cdot\mathbf{R}=0$
everywhere, cf.~Eq.~\eqref{eq:Hij_Nbody} below. 
\item \label{enu:Gravpoint3}Observation \ref{enu:Gravpoint2} is guaranteed
\emph{only }for PN frames, where \eqref{eq:HijPN} holds. For arbitrary
frames, the vorticity and shear/expansion of the observer congruence
can be arbitrarily large; the terms involving them in \eqref{eq:CurlHij}-\eqref{eq:DivHij}
can no longer be neglected, and act as sources for $\mathcal{H}_{\alpha\beta}$.
Example: gravitational field generated by a spinning body; in the
frame co-rotating with the body there are no mass-currents ($\vec{\mathcal{J}}=0$),
but, in spite of that, ${\rm curl}\mathcal{H}_{\hat{\imath}\hat{\jmath}}\ne0$
and $\nabla_{\hat{\jmath}}^{\perp}\mathcal{H}_{\ \hat{\imath}}^{\hat{\jmath}}\ne0$
generically, implying $\mathcal{H}_{\alpha\beta}=\mathbb{H}_{\alpha\beta}\ne0$
and \emph{also} $\star\mathbf{R}\cdot\mathbf{R}\ne0$ generically. 
\item \label{enu:Gravpoint4}The converse of \ref{enu:Gravpoint2} is not
true: when there is no PN frame where $\vec{\mathcal{J}}=0$ everywhere,
that does not necessarily mean that $\star\mathbf{R}\cdot\mathbf{R}\ne0$
or that $\mathbb{H}_{ij}$ cannot vanish in some region with respect
to some frame. Examples: the 2-body system of Sec.~\ref{sub:Two-masses,-planar},
or the spinning body of Sec. \ref{sub:Kerr}: although there are no
PN frames where both bodies are/the whole body is at rest, still $\star\mathbf{R}\cdot\mathbf{R}=0$
in the orbital/equatorial plane, and $\mathbb{H}_{\alpha\beta}=\mathcal{H}_{\alpha\beta}=0$
with respect to certain observer congruences that \emph{do} \emph{not}
correspond to PN frames. 
\end{enumerate}
Point \ref{enu:Gravpoint1} is the statement that either $\nabla_{j}\mathcal{H}_{\ i}^{j}\ne0$
or ${\displaystyle {\rm curl}\mathcal{H}_{ij}}\ne0$ imply that $\mathcal{H}_{ij}$
cannot vanish in an open 3-D spatial set. Point \ref{enu:Gravpoint2}
follows directly from Eqs.~\eqref{eq:HijPN}-\eqref{eq:A,U_PerfectFluid};
one may check also that it is consistent with Eqs.~\eqref{eq:curlHijPN}-\eqref{eq:DivHijPN}
by noting that, when $\vec{\mathcal{J}}=0$ everywhere \emph{in a
PN frame}, $[\dot{\mathcal{E}}_{ij}]_{{\rm N}}=0$, cf.~Eqs.~\eqref{eq:A,U_PerfectFluid},
so that indeed ${\rm curl}\mathcal{H}_{ij}=\partial_{j}\mathcal{H}_{\ i}^{j}=0$.

Regarding point \ref{enu:Gravpoint3} one observes that, for arbitrary
frames the different terms involving $\sigma_{\alpha\beta}$, $\theta$,
$\omega^{\alpha}$ and $a^{\alpha}$ in Eqs. \eqref{eq:CurlHij}-\eqref{eq:DivHij}
cannot in general be neglected. It follows that when $\vec{\mathcal{J}}=0$,
one can still have ${\rm curl}\mathcal{H}_{ij}\ne0$ and/or $\nabla_{\hat{\jmath}}^{\perp}\mathcal{H}_{\ \hat{\imath}}^{\hat{\jmath}}\ne0$,
implying $\mathcal{H}_{\alpha\beta}\ne0\Rightarrow\mathbb{H}_{\alpha\beta}\ne0$
in any open 3-D spatial set. In general this will also imply $\star\mathbf{R}\cdot\mathbf{R}\neq0$.
In the case of the stationary gravitational field of a spinning body,
from the point of view of the frame rigidly co-rotating\footnote{The associated coordinate system is obtained from the Boyer Lindquist
coordinates by the simple transformation $\phi'=\phi+\Omega t$, $\Omega\equiv$
body's angular velocity.} with it there are no mass-currents, $\vec{\mathcal{J}}=0$; moreover
$\sigma_{\alpha\beta}=\theta=0$ (since the frame is rigid), $\dot{\mathcal{E}}_{\hat{\imath}\hat{\jmath}}=\dot{\pi}_{\hat{\imath}\hat{\jmath}}=0$
(since the setup is stationary in this frame), and, outside the body,
$\rho=p=\pi_{ij}=0$. But still ${\rm curl}\mathcal{H}_{\hat{\imath}\hat{\jmath}}=\omega^{\hat{m}}\epsilon_{\hat{l}\hat{k}(\hat{\imath}}\mathcal{E}_{\hat{\jmath})}^{\ \hat{k}}-2a^{\hat{k}}\epsilon_{\hat{k}\hat{m}(\hat{\imath}}\mathcal{H}_{\hat{\jmath})}^{\ \hat{m}}$
and $\nabla_{\hat{\jmath}}^{\perp}\mathcal{H}_{\ \hat{\imath}}^{\hat{\jmath}}=-3\omega^{\hat{\jmath}}\mathcal{E}_{\hat{\imath}\hat{\jmath}}$,
which are generically non-zero, implying $\mathcal{H}_{\alpha\beta}=\mathbb{H}_{\alpha\beta}\ne0$
and also $\star\mathbf{R}\cdot\mathbf{R}\ne0$ generically, as we
shall see explicitly in Sec.~\ref{sub:Kerr} (Eq. (\ref{eq:InvKerrPN})
therein).

Regarding point \ref{enu:Gravpoint4} it is also worth mentioning
that it is possible, even in a region where $\vec{\mathcal{J}}\ne0$,
to have $\star\mathbf{C}\cdot\mathbf{C}=0$ and $\mathcal{H}_{\alpha\beta}=0$
with respect to some observers; an example is the Van Stockum interior
solution, corresponding to an infinitely long and rigidly rotating
cylinder of dust; it is shown in \cite{Bonnor_Stockum} that there
is a region within the cylinder (the inner cylinder $r<(2a)^{-1}$,
in the notation therein) where there are observers for whom $\mathcal{H}_{\alpha\beta}=0$,
in analogy with the situation for the magnetic field within a rotating
charged cylinder discussed in Sec.~\ref{sub:Further-examples:-the Cylinder}.
This is consistent with Eqs.~\eqref{eq:CurlHij}-\eqref{eq:DivHij},
as in a region where $\vec{\mathcal{J}}\ne0$ we can still have (depending
on the kinematical quantities of the chosen frame) $\begin{array}{c}
{\displaystyle {\rm curl}\mathcal{H}_{\hat{\imath}\hat{\jmath}}}\end{array}=\nabla_{\hat{\jmath}}^{\perp}\mathcal{H}_{\ \hat{\imath}}^{\hat{\jmath}}=0$. The same does not apply however to $\mathbb{H}_{\alpha\beta}$,
which is \emph{always} non-zero when $\vec{\mathcal{J}}\ne0$ by virtue
of Eq. (\ref{eq:H_HWeyl2}).\\

\emph{Systems of N point masses.---} Systems of point masses are of
special interest in this work; for such systems the metric potentials
read, in the \emph{harmonic gauge} (e.g. \cite{SoffelKlioner,DSX,Kaplan,WillPoissonBook}),
\begin{align}
w & =\sum_{{\rm a}}\frac{M_{{\rm a}}}{r_{{\rm a}}}\left(1+2v_{{\rm a}}^{2}-\sum_{{\rm b\ne a}}\frac{M_{{\rm b}}}{r_{{\rm a}{\rm b}}}-\frac{1}{2}\vec{r}_{{\rm a}}\cdot\vec{a}_{{\rm a}}-\frac{(\vec{r}_{{\rm a}}\cdot\vec{v}_{{\rm a}})^{2}}{2r_{{\rm a}}^{2}}\right)\nonumber \\
\vec{\mathcal{A}} & =-4\sum_{{\rm a}}\frac{M_{{\rm a}}}{r_{{\rm a}}}\vec{v}_{{\rm a}}\ ;\qquad U=\sum_{{\rm a}}\frac{M_{{\rm a}}}{r_{{\rm a}}}\ ,\label{eq:PNPot_Nbodies}
\end{align}
where $M_{{\rm a}}$ is the mass of particle ``a'', $\vec{r}_{{\rm a}}$,
$\vec{r}_{{\rm a{\rm b}}}$ and $\vec{v}_{{\rm a}}$ are defined in
Sec. \ref{sec:Interpretation_Invariants_EM} (after Eq. \eqref{eq:A_EM}),
and $\vec{a}_{{\rm a}}=\partial\vec{v}_{{\rm a}}/\partial t$ is the
\emph{coordinate} acceleration. For a system of gravitating bodies
(with no external forces), to 1PN accuracy, $\vec{a}_{{\rm a}}$ is
to be taken above as the Newtonian field caused by the other bodies,
i.e., $\vec{a}_{{\rm a}}=-\sum_{{\rm b}\ne{\rm a}}M_{{\rm b}}\vec{r}_{{\rm a}{\rm b}}/r_{{\rm a}{\rm b}}^{3}$.

Observe that $\mathcal{A}_{i}$ and $U$, and hence $\mathbb{H}_{ij}$,
are \emph{linear}, i.e., they are a superposition of the contribution
of each source \cite{DSX}, just like the electromagnetic potentials
and fields. One can write \eqref{eq:HijPN} explicitly in the suggestive
forms 
\begin{eqnarray}
\mathbb{H}_{ij} & = & 6\sum_{{\rm a}}\frac{M_{{\rm a}}}{r_{{\rm a}}^{5}}(\vec{r}_{{\rm a}}\times\vec{v}_{{\rm a}})_{(i}(r_{{\rm a}})_{j)}\label{eq:Hij_Nbody}\\
\Leftrightarrow\overleftrightarrow{\mathbb{H}} & = & \sum_{{\rm a}}\left(\vec{v}_{{\rm a}}\times\left[\overleftrightarrow{\mathbb{E}_{{\rm a}}}\right]_{{\rm N}}-\left[\overleftrightarrow{\mathbb{E}_{{\rm a}}}\right]_{{\rm N}}\times\vec{v}_{{\rm a}}\right)\ ,
\end{eqnarray}
where $\left[\mathbb{E}_{{\rm a}}^{ij}\right]_{{\rm N}}$ denotes
the Newtonian tidal tensor of particle ``a'', and we used the dyadic
notation in point \ref{enu:Diadic-notation} of Sec. \ref{sub:Notation-and-conventions}.
Notice the formal analogy with the post-Coulombian expression for
the magnetic field \eqref{eq:Bsuperposition}. This will allow us
to understand the structure of the curvature invariants of the relevant
gravitational setups by a reasoning analogous to that in the corresponding
electromagnetic setups.

\subsection{One single point mass\label{sub:One-single-point-mass}}

Drawing a parallel with Sec. \ref{sec:Interpretation_Invariants_EM},
we will start by studying the invariants of the gravitational field
produced by a single point mass. This is the gravitational field effectively
involved in the translational form of gravitomagnetism detected in
the observations of the binary system\footnote{Even though the binary system is a two-body system, the effect being
measured is the influence of the translational gravitomagnetic field
produced by one body (playing the role of the source) on the motion
of the other body. Hence, in what pertains to this effect, the system
may effectively be regarded as a one-body (the source) system, the
other body being the test particle.} PSR 1913 +16 (the Hulse-Taylor binary pulsar) \cite{Nordtvedt1988}.
It also describes the relevant contribution to the geodetic precession
measured in different systems: the precession of the Earth-Moon system
along its orbit around the Sun, detected in the analysis of Lunar
Laser Ranging (LLR) data \cite{BertottiCiufoliniBender1987,Shapiro_et_alPRL1988,WilliamsNewhallDickey1996},
the geodetic precession of the gyroscopes in the Gravity Probe-B \cite{GPB},
and the precession of the pulsar's spin vector in the binary systems
PSR J0737\textminus 3039A/B \cite{BretonStairsScience2008} and PSR
B1534+12 \cite{StairsThorsettArzoumanianPRL2004}.

The metric is described by the Schwarzschild solution, which reads,
in Schwarzschild coordinates, 
\[
ds^{2}=-\left(1-\frac{2M}{r}\right)dt^{2}+\frac{dr^{2}}{\left(1-\frac{2M}{r}\right)}+r^{2}d\Omega^{2}\ .
\]
The observers at rest in this coordinate system are the Killing or
\emph{``static''} observers $\mathbf{u}\propto\partial/\partial t$,
which may be thought of as rigidly attached to the asymptotic inertial
rest frame of the source.

This spacetime is of Petrov type D (everywhere); thus the third condition
in (\ref{eq:grav-PEPMcond}) is satisfied everywhere, and one only
has to worry about the quadratic invariants, which have the structure:
\[
\left\{ \begin{array}{l}
{\displaystyle \frac{\inn{R}}{8}}=\mathbb{E}^{\alpha\gamma}\mathbb{E}_{\alpha\gamma}-\mathbb{H}^{\alpha\gamma}\mathbb{H}_{\alpha\gamma}={\displaystyle \frac{6M^{2}}{r^{6}}>0}\\
\\
{\displaystyle \frac{\innst{R}}{16}}=\mathbb{E}^{\alpha\gamma}\mathbb{H}_{\alpha\gamma}=0\ \ \mbox{(everywhere)}
\end{array}\right.\ .
\]
Thus this is (everywhere) a purely electric spacetime, i.e., everywhere
there are observers for which $\mathbb{H}'_{\alpha\beta}=0$, see
Sec. \ref{sub:Rie-vac}. Their 4-velocities $u'^{\alpha}$ are obtained
from Eqs. (\ref{eq:u'grav}), (\ref{eq:wvector}), (\ref{eq:evec}),
(\ref{eq:tgrav2}), and (\ref{eq:super-Poynting}). We have, for the
auxiliary quantities involved%
, $\mathcal{P}^{\alpha}=0\Rightarrow t^{\alpha}=u^{\alpha}$, $\J=\mathbb{A}=-6M^{3}/r^{9}$,
$\lambda=M/r^{3}$, $\eb^{\alpha}=-\sqrt{1-2M/r}\delta_{r}^{\alpha}$;
therefore, by (\ref{eq:u'grav})%
, 
\begin{equation}
u'^{\alpha}=(u'^{0},u'^{r},0,0)\ ,\label{eq:4velSchw}
\end{equation}
with $u'^{r}$ arbitrary (under the normalization condition $u'^{\alpha}u'_{\alpha}=-1$).
That is, observers which are either \emph{static} or moving radially,
in analogy with the situation in the analogous electromagnetic system
of Sec. \ref{sub:EMOne-single-point}. One might check these results
by computing explicitly $\mathbb{H}'_{\alpha\beta}$ for an arbitrary
$u'^{\alpha}$, as done in \cite{PaperGyros}. Note that the fact
that $\mathbb{H}'_{\alpha\beta}=0$ for the static observers means
that it \emph{globally vanishes in a rigid frame.}

One can also get intuition on why $\mathbb{E}^{\alpha\gamma}\mathbb{H}_{\alpha\gamma}$
remains zero for any observer from arguments analogous to those that
explain why $E^{\alpha}B_{\alpha}=0$ for the point charge; namely
the formal similarity between the transformation laws. Let $\Ou$
and $\Oup$ be, respectively, a static observer and an observer moving
relative to it with some velocity $\vec{v}$. For $\Ou$, $\mathbb{H}_{\alpha\beta}=0\Rightarrow\mathbb{E}^{\alpha\gamma}\mathbb{H}_{\alpha\gamma}=0$.
The moving observer $\Oup$ will in turn measure a non-vanishing gravitomagnetic
tidal tensor, $\mathbb{H}'_{\alpha\beta}\ne0$, but it will be such
that it is always ``orthogonal'' to the gravitoelectric tidal tensor
$\mathbb{E}'_{\alpha\beta}$, in analogy with the situation for the
magnetic field (\ref{BSingleCharge}). In order to see this, first
observe, from the decompositions \eqref{eq:Rie-decomp}-\eqref{eq:Riest-decomp},
that with respect to the congruence of static observers ($u^{\alpha}=u^{0}\delta_{0}^{\alpha}$),
the Riemann tensor and its dual are completely described by the electric
part: 
\begin{eqnarray*}
R_{\alpha\beta}^{\ \ \ \gamma\delta} & = & 4\left\{ 2u_{[\alpha}u^{[\gamma}+g_{[\alpha}^{\,\,\,\,[\gamma}\right\} \mathbb{E}_{\beta]}^{\,\,\,\delta]}\ ;\\
\star R_{\alpha\beta}^{\ \ \ \gamma\delta} & = & 2\epsilon_{\alpha\beta\lambda\tau}\mathbb{E}^{\lambda[\gamma}u^{\delta]}u^{\tau}+2\epsilon^{\lambda\tau\gamma\delta}\mathbb{E}_{\lambda[\alpha}u_{\beta]}u_{\tau}\ .
\end{eqnarray*}
Hence, to linear order, the space components of $\mathbb{E}'_{\alpha\beta}\equiv R_{\alpha\mu\beta\nu}u'^{\mu}u'^{\nu}$
and $\mathbb{H}'_{\alpha\beta}\equiv\star R_{\alpha\mu\beta\nu}u'^{\mu}u'^{\nu}$
read (using dyadic notation, see point \ref{enu:Diadic-notation}
of Sec. \ref{sub:Notation-and-conventions}), 
\begin{equation}
\overleftrightarrow{\mathbb{E}}'\simeq\overleftrightarrow{\mathbb{E}};\qquad\overleftrightarrow{\mathbb{H}}'\simeq\overleftrightarrow{\mathbb{E}}\times\vec{v}-\vec{v}\times\overleftrightarrow{\mathbb{E}}\ ,\label{eq:Transform_E_H}
\end{equation}
which have formal similarities with Eqs. \eqref{BSingleCharge}-\eqref{eq:EsingleCharge},
and, together with $\mathbb{H}'_{0\alpha}=\mathbb{H}'_{\alpha0}=0$,
lead immediately to $\mathbb{E'}^{\alpha\beta}\mathbb{H}'_{\alpha\beta}=\mathbb{E'}^{ij}\mathbb{H}'_{ij}=0$.
The verification using the exact expressions for $\mathbb{E}'_{\alpha\beta}$,
$\mathbb{H}'_{\alpha\beta}$ is also straightforward.

\subsection{Two bodies, coplanar motion --- the Earth-Sun system\label{sub:Two-masses,-planar}}

We consider here the gravitational field generated by two bodies ---
the Earth and the Sun --- orbiting each other, whose (translational)
gravitomagnetic effects are implied in \cite{Nordvedt1973,Nordtvedt2003,SoffelKlioner}.
The metric, accurate to first post-Newtonian order, is, cf. Eqs. \eqref{eq:PNmetric}
and \eqref{eq:PNPot_Nbodies}, 
\begin{eqnarray}
g_{00} & = & -1+2\frac{M_{\oplus}}{r_{\oplus}}+4\frac{M_{\oplus}v_{\oplus}^{2}}{r_{\oplus}}-\frac{M_{\oplus}(\vec{v}_{\oplus}\cdot\vec{r}_{\oplus})^{2}}{r_{\oplus}^{3}}-\frac{2M_{\oplus}M_{\odot}}{r_{\odot\oplus}r_{\oplus}}\nonumber \\
 &  & -\frac{M_{\oplus}}{r_{\oplus}}\vec{r}_{\oplus}\cdot\vec{a}_{\oplus}+\odot\leftrightarrow\oplus-2\left(\frac{M_{\odot}}{r_{\odot}}+\frac{M_{\oplus}}{r_{\oplus}}\right)^{2}\ ;\label{eq:g00TwoMass}\\
\ g_{0i} & = & 4\left[\frac{M_{\odot}(v_{\odot})_{i}}{r_{\odot}}+\frac{M_{\oplus}(v_{\oplus})_{i}}{r_{\oplus}}\right]\ ;\label{eq:g01TwoMass}\\
g_{ij} & = & \left[1+2\left(\frac{M_{\odot}}{r_{\odot}}+\frac{M_{\oplus}}{r_{\oplus}}\right)\right]\delta_{ij}\ ,\label{eq:gijTwoMass}
\end{eqnarray}
where $\odot\equiv$ Sun, $\oplus\equiv$ Earth. Since $\mathbb{E}_{\alpha\beta}\sim\O{2}{2}$
and $\mathbb{H}_{\alpha\beta}\sim\O{3}{2}$, cf.~Eqs.~\eqref{eq:EijPN}-\eqref{eq:HijPN},
then, generically $\mathbb{E}^{\alpha\gamma}\mathbb{E}_{\alpha\gamma}>\mathbb{H}^{\alpha\gamma}\mathbb{H}_{\alpha\gamma}$,
i.e., $\mathbf{R}\cdot\mathbf{R}>0$. In this respect we note that
the region of magnetic dominance that exists between the two charges
in the electromagnetic system of Sec. \ref{sub:Two-charges,-planar}
(around the point where $\vec{E}=0$), has no counterpart in the present
gravitational system, because here we are dealing with \emph{tidal
tensors}, not vector fields, and these add differently. Namely, in
the region corresponding to that where, in the electromagnetic system,
the electric fields cancel out, the tidal tensors $\mathbb{E}_{\alpha\beta}$
of each body \emph{add up} instead (in electromagnetism this is analogous
instead to the situation with the electric \emph{tidal tensor}, as
defined in \cite{CHPRD,PaperAnalogies}). As for the Chern-Pontryagin
invariant $\star\mathbf{R}\cdot\mathbf{R}=16\mathbb{E}^{\alpha\beta}\mathbb{H}_{\alpha\beta}$,
one has, to lowest order, cf.~Eqs.~\eqref{eq:EijPN}-\eqref{eq:HijPN},
\eqref{eq:PNPot_Nbodies}-\eqref{eq:Hij_Nbody}, 
\begin{figure}
\includegraphics[width=1\columnwidth]{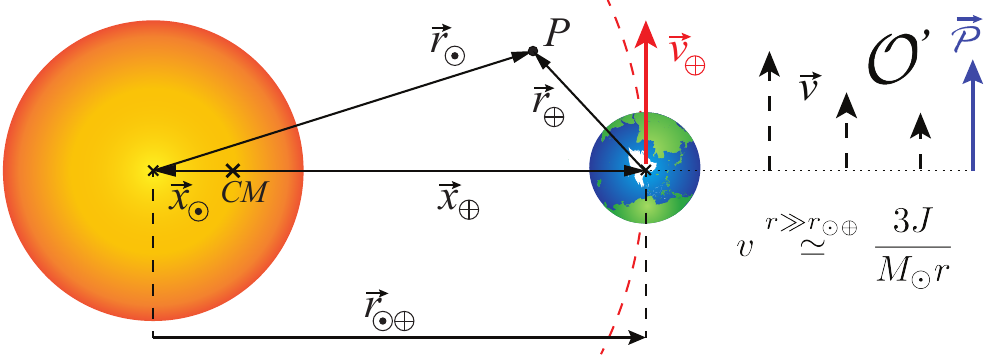}

\protect\protect\protect\protect\caption{\label{fig:Earth-Sun}The Earth-Sun system. Observers $\Oup$ for
which $\mathbb{H}'_{\alpha\beta}=0$ (represented only along the Earth-Sun
axis); their velocity $\vec{v}$ is depicted by dashed arrows. }
\end{figure}

\begin{align}
 & \mathbb{E}^{\alpha\beta}\mathbb{H}_{\alpha\beta}=U^{,ij}\left(\frac{1}{2}\epsilon_{i}^{\ lk}\mathcal{A}_{k,lj}+\epsilon_{ij}^{\ \ k}\dot{U}_{,k}\right)+\O{7}{2}\nonumber \\
 & {\displaystyle =6\left(\sum_{{\rm a=\oplus,\odot}}\frac{M_{{\rm a}}}{r_{{\rm a}}}\right)^{,ij}\sum_{{\rm a=\oplus,\odot}}\frac{M_{{\rm a}}}{r_{{\rm a}}^{5}}(\vec{r}_{{\rm a}}\times\vec{v}_{{\rm a}})_{(i}(r_{{\rm a}})_{j)}}\nonumber \\
 & =\frac{18M_{\oplus}M_{\odot}}{r_{\oplus}^{5}r_{\odot}^{5}}(\vec{r}_{\oplus}\cdot\vec{r}_{\odot})\left[(\vec{v}_{\oplus}\times\vec{r}_{\oplus})\cdot\vec{r}_{\odot}+(\vec{v}_{\odot}\times\vec{r}_{\odot})\cdot\vec{r}_{\oplus}\right]\ ,\label{eq:RRstar2body}
\end{align}
which agrees\footnote{To obtain Eq. (7) of \cite{Ciufolini LLR} from Eq. (\ref{eq:RRstar2body})
above, one makes $\vec{v}_{\odot}=0$ (as in \cite{Ciufolini LLR}
an heliocentric reference frame is used), and notes that $\vec{r}_{\odot}$
and $\vec{r}_{\oplus}$ read, in the notation therein, respectively,
$\vec{x}_{M}$ and $\vec{r}_{M\oplus}$. } with Eq. (7) of \cite{Ciufolini LLR}. This invariant (namely the
expressions between square brackets) exhibits formal similarities
with the electromagnetic invariant \eqref{eq:EBTwoChargeslowest};
it has the structure 
\[
\mathbb{E}^{\alpha\gamma}\mathbb{H}_{\alpha\gamma}\begin{cases}
=0\ \ \mbox{in the orbital plane;}\\
\ne0\:\;\mbox{generically.}
\end{cases}
\]
That $\mathbb{E}^{\alpha\beta}\mathbb{H}_{\alpha\beta}=0$ in the
orbital plane $\Sigma$ can be seen observing that, when the point
of observation $P$ lies on $\Sigma$, then $\{\vec{r}_{\odot},\vec{r}_{\oplus}\}\in\Sigma$;
and since also $\{\vec{v}_{\odot},\vec{v}_{\oplus}\}\in\Sigma$ (always),
it follows that $(\vec{v}_{\oplus}\times\vec{r}_{\oplus})\cdot\vec{r}_{\odot}=(\vec{v}_{\odot}\times\vec{r}_{\odot})\cdot\vec{r}_{\oplus}=0$,
implying $\mathbb{E}^{\alpha\beta}\mathbb{H}_{\alpha\beta}=0$. This
structure is analogous to that of \eqref{eq:EBTwoChargeslowest} for
coplanar motion, except that here the factor $(\vec{r}_{\oplus}\cdot\vec{r}_{\odot})$,
which has no electromagnetic counterpart, introduces an additional
1-D region where $\mathbb{E}^{\alpha\beta}\mathbb{H}_{\alpha\beta}=0$
(the circle determined by $\vec{r}_{\oplus}\perp\vec{r}_{\odot}$).
The existence of observers for which $\mathbb{H}_{\alpha\beta}=0$
on $\Sigma$ can also be understood in analogy with the electromagnetic
apparatus of Sec. \ref{sub:Two-charges,-planar}. Let us compute their
velocities. The gravitomagnetic tidal tensor measured by an observer
$\Op$ moving with velocity $\vec{v}$ with respect to the chosen
PN frame is, cf. Eq. (\ref{eq:Hij_Nbody}),

\begin{equation}
\mathbb{H}'_{ij}=6\left[\frac{M_{\oplus}}{r_{\oplus}^{5}}(\vec{r}_{\oplus}\times[\vec{v}_{\oplus}-\vec{v}])_{(i}(r_{\oplus})_{j)}+\oplus\leftrightarrow\odot\right]\ .\label{eq:Hij'}
\end{equation}
Choose, for convenience, the PN frame comoving with the center of
mass (CM) of the Earth-Sun system (which is close to the ``barycentric''
reference frame considered in e.g. \cite{NordtvedtPRL1988,SoffelKlioner,Nordtvedt2003},
or to the heliocentric system in \cite{Ciufolini LLR}), and take
$z=0$ to be the orbital plane $\Sigma$. Firstly one observes that,
in order for $\mathbb{H}'_{ij}=0$, the observer's velocity $\vec{v}$
must be parallel to $\Sigma$, except at some special points on $\Sigma$,\footnote{If $\vec{v}$ has a component orthogonal to $\Sigma$, then, taking
the $x$-axis along the Earth-Sun axis, such that $r_{\odot}^{y}=r_{\oplus}^{y}$
and $r_{\odot}^{x}=r_{\oplus}^{x}+r_{\odot\oplus}$, one obtains from
the conditions $\mathbb{H}'_{xy}=\mathbb{H}'_{xx}=0$ a system of
two equations in two unknowns $r_{\oplus}^{x}$ and $r_{\oplus}^{y}$,
namely $M_{\oplus}((r_{\oplus}^{x})^{2}-(r_{\oplus}^{y})^{2})/r_{\oplus}^{5}+M_{\odot}((r_{\odot}^{x})^{2}-(r_{\oplus}^{y})^{2})/r_{\odot}^{5}=0$
and $M_{\oplus}r_{\oplus}^{x}/r_{\oplus}^{5}+M_{\odot}r_{\odot}^{x}/r_{\odot}^{5}=0$.
With $M_{\odot}\gg M_{\oplus}$ this system has two solutions, corresponding
to two points in $\Sigma$ very close to the Earth and off the Earth-Sun
axis. An observer $\Op$ at those points measures $\mathbb{H}'_{\alpha\beta}=0$
precisely if the component of its velocity parallel to $\Sigma$ equals
\eqref{eq:v-gen}, the component orthogonal to $\Sigma$ being arbitrary.} in analogy with the electromagnetic problem in Sec. \ref{sub:Two-charges,-planar}
(coplanar motion). This implies $\vec{r}_{\oplus}\times[\vec{v}_{\oplus}-\vec{v}]=(\vec{r}_{\oplus}\times[\vec{v}_{\oplus}-\vec{v}])^{z}\vec{e}_{z}$,
and thus trivially $\mathbb{H}'_{ii}=0$, $\mathbb{H}'_{xy}=\mathbb{H}'_{yx}=0$.
The only surviving components are $\mathbb{H}'_{xz}=\mathbb{H}'_{zx}$
and $\mathbb{H}'_{yz}=\mathbb{H}'_{zy}$, whose vanishing amounts
to the conditions 
\begin{equation}
\vec{v}\times\vec{a}^{(i)}=\vec{b}^{(i)},\qquad i=x,\,y,\label{eq:H'zi}
\end{equation}
where the vectors $\vec{a}^{(i)}$ and $\vec{b}^{(i)}$ are defined
by 
\begin{align*}
 & \vec{a}^{(i)}\equiv\frac{r_{\oplus}^{i}M_{\oplus}}{r_{\oplus}^{5}}\vec{r}_{\oplus}+\frac{r_{\odot}^{i}M_{\odot}}{r_{\odot}^{5}}\vec{r}_{\odot},\\
 & \vec{b}^{(i)}\equiv\frac{r_{\oplus}^{i}M_{\oplus}}{r_{\oplus}^{5}}\vec{v}_{\oplus}\times\vec{r}_{\oplus}+\frac{r_{\odot}^{i}M_{\odot}}{r_{\odot}^{5}}\vec{v}_{\odot}\times\vec{r}_{\odot}\ .
\end{align*}
Since $(\vec{r}_{\oplus}\times\vec{r}_{\odot})=(\vec{r}_{\oplus}\times\vec{r}_{\odot})^{z}\vec{e}_{z}$,
we have 
\begin{equation}
\vec{a}^{(x)}\times\vec{a}^{(y)}=\frac{M_{\oplus}M_{\odot}}{r_{\oplus}^{5}r_{\odot}^{5}}\|\vec{r}_{\oplus}\times\vec{r}_{\odot}\|^{2}\vec{e}_{z}\ ,\label{eq:axay}
\end{equation}
and the solution of \eqref{eq:H'zi} splits into two cases:

{\em 1. Observer off Earth-Sun axis ($\vec{r}_{\oplus}\times\vec{r}_{\odot}\neq0$)}.
By \eqref{eq:axay} $\vec{a}^{(x)}$ and $\vec{a}^{(y)}$ span in
this case the orbital plane, therefore one may write $\vec{v}=\lambda\vec{a}^{(x)}+\mu\vec{a}^{(y)}$;
substituting into \eqref{eq:H'zi} and using \eqref{eq:axay} readily
gives the unique solution 
\begin{equation}
\vec{v}=\frac{r_{\oplus}^{5}r_{\odot}^{5}}{M_{\oplus}M_{\odot}||\vec{r}_{\oplus}\times\vec{r}_{\odot}||^{2}}\left((b^{(y)})^{z}\vec{a}^{(x)}-(b^{(x)})^{z}\vec{a}^{(y)}\right),\label{eq:v-gen}
\end{equation}
where $(b^{(i)})^{z}=\vec{b}^{(i)}\cdot\vec{e}_{z}$. Hence, at each
point of $\Sigma$ off the Earth-Sun axis, there is a \emph{unique}
observer moving parallel to $\Sigma$ for which $\mathbb{H}'_{\alpha\beta}=0$.
This is in contrast with the electromagnetic analogue in Sec. \ref{sub:Two-charges,-planar}
(coplanar motion), where the velocity of the observers measuring $\vec{B}'=0$
had an arbitrary component along the electric field (hence there was
an infinite number of such observers at each point). One can say that
the situation is similar (in this respect) to purely electric exact
solutions of the general Petrov type I.

{\em 2. Observer on Earth-Sun axis ($\vec{r}_{\oplus}\times\vec{r}_{\odot}=0$)},
depicted in Fig. \ref{fig:Earth-Sun}. In this case $\vec{a}^{(x)}/r_{\oplus}^{x}=\vec{a}^{(y)}/r_{\oplus}^{y}\equiv\vec{V}$
and $\vec{b}^{(x)}/r_{\oplus}^{x}=\vec{b}^{(y)}/r_{\oplus}^{y}\equiv\vec{W}$,
such that \eqref{eq:H'zi} reduces to the single equation $\vec{v}\times\vec{V}=\vec{W}$.
Clearly, the component $\vec{v}_{\parallel V}$ of $\vec{v}$ parallel
to $\vec{V}$ is arbitrary; for the orthogonal component one obtains
(taking a cross product with $\vec{V}$): 
\begin{equation}
\vec{v}_{\perp V}\equiv\vec{v}_{\perp r_{\odot\oplus}}=\frac{\vec{V}\times\vec{W}}{V^{2}}\simeq\frac{M_{\oplus}(r_{\odot}^{3}-r_{\oplus}^{3})}{r_{\odot}^{3}M_{\oplus}+M_{\odot}r_{\oplus}^{3}}\vec{v}_{\oplus}\ ;\label{eq:vNoH2BodyLine}
\end{equation}
where we noted that $\vec{V}$ is parallel to the Earth-Sun axis ($\vec{V}\parallel\vec{r}_{\odot\oplus}$,
where $\vec{r}_{\odot\oplus}\equiv\vec{r}_{\odot}-\vec{r}_{\oplus}=\vec{x}_{\oplus}-\vec{x}_{\odot}$),
and the last approximate equality follows from the fact that, along
the axis, $\vec{V}\cdot\vec{v}_{\oplus}=\vec{V}\cdot\vec{v}_{\odot}=0$,
and that $M_{\oplus}\vec{v}_{\oplus}\simeq-M_{\odot}\vec{v}_{\odot}$
(since the system's momentum vanishes in the CM frame). We note moreover
that, along the axis, the super-Poynting vector as measured by the
rest observers, $\mathcal{P}^{i}=\epsilon_{\ jk}^{i}\mathbb{E}^{jl}\mathbb{H}_{l}^{\ k}$,
reads 
\begin{align*}
\vec{\mathcal{P}} & =9\left[\frac{M_{\oplus}^{2}}{r_{\oplus}^{6}}\vec{v}_{\oplus}+\frac{M_{\odot}^{2}}{r_{\odot}^{6}}\vec{v}_{\odot}+\frac{M_{{\rm \odot}}M_{\oplus}}{r_{\odot}^{3}r_{\oplus}^{3}}(\vec{v}_{\odot}+\vec{v}_{\oplus})\right]\\
 & \simeq9M_{\oplus}\left[\frac{M_{\oplus}}{r_{\oplus}^{6}}-\frac{M_{\odot}}{r_{\odot}^{6}}+\frac{M_{{\rm \odot}}}{r_{\odot}^{3}r_{\oplus}^{3}}-\frac{M_{\oplus}}{r_{\odot}^{3}r_{\oplus}^{3}}\right]\vec{v}_{\oplus}\ ,
\end{align*}
which is parallel to $\vec{v}_{\perp r_{\odot\oplus}}$. This means
that $\vec{v}_{\perp r_{\odot\oplus}}$ is in fact the component of
$\vec{v}$ parallel to $\vec{\mathcal{P}}$; therefore, along the
axis, the situation is similar to a purely electric Petrov type D
exact solution (and to the electromagnetic case): at each point a
class of observers exists for which $\mathbb{H}_{\alpha\beta}=0$;
such observers have a velocity consisting of a component $\vec{v}_{\parallel\mathcal{P}}=\vec{v}_{\perp r_{\odot\oplus}}$
along $\vec{\mathcal{P}}$ fixed by Eq. (\ref{eq:vNoH2BodyLine}),
plus an arbitrary component $\vec{v}_{\parallel r_{\oplus\odot}}$
parallel to the Earth-Sun axis.

Using $M_{\odot}\gg M_{\oplus}$ we have $x_{\odot}\approx0$; considering
moreover an observation point much farther than the Earth-Sun distance
($r\gg r_{\odot\oplus}$), as depicted in Fig. \ref{fig:Earth-Sun},
we obtain, in the special case where $\vec{v}$ has no component along
the axis ($\vec{v}=\vec{v}_{\perp r_{\odot\oplus}}=\vec{v}_{\parallel\mathcal{P}}$),
the limit 
\begin{equation}
v\ \stackrel{r\gg r_{\odot\oplus}}{\simeq}\ \frac{3M_{\oplus}v_{\oplus}r_{\odot\oplus}}{M_{\odot}r}=\frac{3J}{M_{\odot}r}\ ,\label{eq:vNoH2BodyLimit}
\end{equation}
where we noted that, to lowest order (which is the accuracy needed
for $\vec{v}$ in Eq. (\ref{eq:Hij'})), $J=M_{\oplus}v_{\oplus}r_{\odot\oplus}$
is the system's angular momentum as measured in the center of mass
PN frame. This is analogous to the situation in the electromagnetic
problem of Sec. \ref{sub:Coplanar-motion}, and the velocity field
\eqref{eq:vnoBtwocharges}. The analogy can be made even closer by
considering the gravitational counterpart of the system in Fig. \ref{fig:Two},
i.e., two particles with the same mass $M_{1}=M_{2}=M_{{\rm tot}}/2$
and with velocities $\vec{v}_{1}$ and $-\vec{v}_{1}$, orbiting each
other (no ``rod'' is necessary in this case) in a circular motion
of radius $d$. The velocities of the observers $\Op$ (at points
$P$ along the axis) for which $\mathbb{H}'_{ij}=0$ are obtained
from \eqref{eq:vNoH2BodyLine} setting $M_{\oplus}=M_{\odot}=M_{{\rm tot}}/2$,
$|\vec{x}_{\oplus}|=|\vec{x}_{\odot}|=d$, $v_{\oplus}=v_{1}$, leading
to 
\begin{equation}
v=v_{1}\frac{(3dr^{2}+d^{3})}{r^{3}+3d^{2}r}\ \stackrel{r\gg d}{\simeq}\ \frac{3v_{1}d}{r}=\frac{3J}{M_{{\rm tot}}r}\ ,\label{eq:v_field_2equalMasses}
\end{equation}
where, again, we identified $J=v_{1}M_{{\rm tot}}d$. This is similar
(up to a factor $3/2$), for large $r$, to the velocity \eqref{eq:vnoBtwocharges}
for which $\vec{B}'=0$ in the electromagnetic system. The reason
why $\mathbb{H}'_{ij}=0$ for these observers (and not for others)
can also be understood by a reasoning analogous to the one we made
at the end of Sec. \ref{sub:Coplanar-motion}: $\mathbb{H}_{ij}$
is the superposition of the individual gravitomagnetic tidal tensors
produced by each body, cf. Eq. \eqref{eq:Hij'}, which, for the setup
analogous to Fig. \ref{sub:Coplanar-motion}, has non-vanishing components
$\mathbb{H}_{zy}=\mathbb{H}_{yz}=(\mathbb{H}_{1})_{zy}+(\mathbb{H}_{2})_{zy}$.
The contributions $(\mathbb{H}_{1})_{zy}$ and $(\mathbb{H}_{2})_{zy}$
have opposite signs since $v_{1}^{x}=-v_{2}^{x}$. Thus, for an observer
at rest in the system's CM frame ($v=0$), $|(\mathbb{H}_{2})_{zy}|>|(\mathbb{H}_{1})_{zy}|$,
since body 2 is closer to the observer. Increasing the observer's
velocity $v$ (in the sense of the orbital motion) means decreasing
$|(\mathbb{H}'_{2})_{zy}|$ whilst increasing $|(\mathbb{H}'_{1})_{zy}|$,
so that they eventually cancel out, $\mathbb{H}'_{ij}=0$. These similarities
with electromagnetism can be traced back to the facts that, to first
post-Newtonian order, $\mathbb{H}_{ij}$, Eq. \eqref{eq:Hij_Nbody},
is linear (so a superposition principle applies just like in electromagnetism),
and has a dependence on the velocities of the sources (and transformation
laws under a change of PN frame, Eq. \eqref{eq:Transform_E_H}) that
are, to some extent, also analogous to their electromagnetic counterparts.

Finally, we note that this application exemplifies point \ref{enu:Gravpoint4}
of Sec. \ref{sec:Interpretation-of-the-Gravitational}: although there
is no PN frame where \emph{both} bodies are at rest, still $\star\mathbf{R}\cdot\mathbf{R}=0$
in a 2-D region (the orbital plane), where $\mathbb{H}'_{\alpha\beta}=0$
for certain families of observers (which do not form PN frames).

\subsection{The gravitational field of a spinning body\label{sub:Kerr}}

The gravitational field of a compact, spinning body of mass $M$ and
angular momentum $J$, whose center of mass is at rest in the given
PN frame is, to 1PN order, obtained by substituting $w=U=M/r$, $\vec{\mathcal{A}}=2\vec{r}\times\vec{J}/r^{3}$
into Eqs. \eqref{eq:PNmetric}, see e.g. \cite{Kaplan,WillPoissonBook}.
This coincides with the 1PN limit of the Kerr solution (in isotropic
coordinates), which is the field we shall consider here, since this
is an exact solution well suited to our methods. Its well-known form
in Boyer-Lindquist coordinates is 
\begin{eqnarray*}
ds^{2} & = & -\frac{\Delta}{\Sigma}\left(dt-a\sin^{2}\theta d\phi\right)^{2}+\frac{\Sigma}{\Delta}dr^{2}+\Sigma d\theta^{2}\\
 &  & +\frac{\sin^{2}\theta}{\Sigma}\left(adt-(r^{2}+a^{2})d\phi\right)^{2}\ ,
\end{eqnarray*}
where 
\[
\Delta\equiv r^{2}-2Mr+a^{2}\ ,\ \ \Sigma\equiv r^{2}+a^{2}\cos^{2}\theta\ ,\ \ a\equiv\frac{J}{M}\ .
\]
This spacetime is of Petrov type D, so the third condition in \eqref{eq:grav-PEPMcond}
is satisfied everywhere; hence it suffices to study the quadratic
invariants, which read \cite{SemerakDeformedBlackholes,LakeKerrInv,cherubini:02}
\begin{figure}
\includegraphics[width=1\columnwidth]{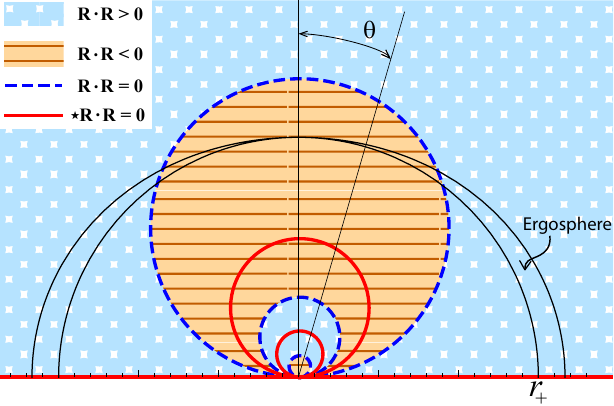}

\protect\protect\protect\protect\caption{\label{fig:KerrInvariants}Structure of the quadratic invariants $\mathbf{R}\cdot\mathbf{R}$
and $\star\mathbf{R}\cdot\mathbf{R}$ in the Kerr spacetime, in Boyer-Lindquist
coordinates. Only the half-plane $\theta\le\pi/2$ is represented.
Regions of electric dominance ($\mathbf{R}\cdot\mathbf{R}>0$) are
colored in dotted blue, regions of magnetic dominance ($\mathbf{R}\cdot\mathbf{R}<0$)
are colored in stripped yellow. Dashed blue circles represent the
zeros of $\mathbf{R}\cdot\mathbf{R}$, solid red lines (two circles
plus the $\theta=\pi/2$ axis) represent the zeros of $\star\mathbf{R}\cdot\mathbf{R}$.}
\end{figure}

\begin{align}
\inn{R} & =\frac{48M^{2}}{\Sigma^{6}}(r^{2}-a^{2}\cos^{2}\theta)(\Sigma^{2}-16r^{2}a^{2}\cos^{2}\theta)\ ;\nonumber \\
\innst{R} & =\frac{96M^{2}ra}{\Sigma^{6}}(3r^{2}-a^{2}\cos^{2}\theta)(r^{2}-3a^{2}\cos^{2}\theta)\cos\theta\ .\label{eq:*R.R_Kerr}
\end{align}
The structure of these invariants is graphed in Fig. \ref{fig:KerrInvariants}.
The zeros of $\inn{R}$ occur on the shells $r=\pm a\cos\theta$ and
$r=\pm(2\pm\sqrt{3})a\cos\theta$, signaling transitions between regions
of electric ($\inn{R}>0$) vs.\ magnetic ($\inn{R}<0$) dominance.
The zeros of $\innst{R}$ define purely electric/magnetic surfaces
and occur for $\theta=\pi/2$ and $r=\pm a\cos\theta/\sqrt{3}$ (purely
electric) and $r=\pm\sqrt{3}a\cos\theta$ (purely magnetic). Except
for the (purely electric) equatorial plane, all these surfaces lie
either inside the event horizon ($r\leq r_{+},\,r_{+}\equiv M+\sqrt{M^{2}-a^{2}}$)
or, in the case of the larger shells (given by $r=\pm(2+\sqrt{3})a\cos\theta$
when $\inn{R}=0$, and by $r=\pm\sqrt{3}a\cos\theta$ when $\innst{R}=0$),
they may, for large enough values of $a$, lie partly outside the
horizon,\footnote{In order to see this, one observes that, for the largest blue dashed
circle in Fig. \ref{fig:KerrInvariants} (the circle $r=\pm(2+\sqrt{3})a\cos\theta$),
the non-extreme condition $a/M<1$ implies $r_{{\rm max}}/r_{+}<2+\sqrt{3}$,
where $r_{{\rm max}}=r_{|_{\theta=0}}$ is the maximum value of the
coordinate $r$ along the circle. These regions shall be discussed
in detail elsewhere.} yet still very close to it. Hence, in the astrophysical applications
under discussion, which pertain to the ``post-Newtonian zone'' \cite{WillPoissonBook},
where $r\gg r_{+}$, we have $\inn{R}>0$ everywhere, and the only
surface where $\innst{R}=0$ is the (purely electric) equatorial plane:
\textcolor{black}{{} 
\begin{equation}
\left\{ \begin{array}{l}
{\displaystyle \mathbb{E}^{\alpha\gamma}\mathbb{E}_{\alpha\gamma}-\mathbb{H}^{\alpha\gamma}\mathbb{H}_{\alpha\gamma}=\frac{6M^{2}}{r^{6}}+\ensuremath{\O{6}{4}}>0}\ ;\\
\\
\mathbb{E}^{\alpha\gamma}\mathbb{H}_{\alpha\gamma}={\displaystyle \cos\theta\left[\frac{18JM}{r^{7}}+\ensuremath{\O{7}{4}}\right]\,(=0\mbox{ \ for\ }\theta=\frac{\pi}{2}),}
\end{array}\right.\label{eq:InvKerrPN}
\end{equation}
}which is a structure formally analogous to the electromagnetic counterpart
\eqref{eq:InvariantsSpinningCharge}. The equatorial plane being purely
electric means that there are therein observers for which $\mathbb{H}'_{\alpha\beta}=0$.
Their 4-velocities $u'^{\alpha}$ are obtained from Eqs. (\ref{eq:u'grav}),
(\ref{eq:wvector}), (\ref{eq:evec}), (\ref{eq:tgrav2}), and (\ref{eq:super-Poynting}).
For the auxiliary quantities involved, we have, in the equatorial
plane, $\inn{R}=48M^{2}/r^{6}$, $\xi=3M^{2}/(2r^{6})$, 
\[
\AD=\frac{(\Delta+a^{2})}{(\Delta-a^{2})}\ ,\qquad\mathcal{P}_{\alpha}=\frac{9aM^{2}\Delta(\Delta+a^{2})}{2r^{5}(\Delta-a^{2})^{5/2}}\delta_{\alpha}^{\phi}\ ,
\]
$\J=\mathbb{A}=-6M^{3}/r^{9}$, $\lambda=M/r^{3}$, and so 
\begin{align*}
t^{\alpha} & =u^{\alpha}+\frac{2\mathcal{P}^{\alpha}}{3\xi\AD(\AD+1)}=t^{0}\left[\delta_{0}^{\alpha}+\frac{a}{a^{2}+r^{2}}\delta_{\phi}^{\alpha}\right]\ ,\\
\eb^{\alpha} & =-\frac{\Delta}{r\sqrt{\Delta-a^{2}}}\delta_{r}^{\alpha}\ ,
\end{align*}
where $u^{\alpha}=(-g_{00})^{-1/2}\delta_{0}^{\alpha}$ and $t^{0}=(r-2M)(a^{2}+r^{2})\Delta^{-1}(\Delta-a^{2})^{-1/2}$.
Therefore, by (\ref{eq:u'grav}), 
\begin{equation}
u'^{\alpha}=(u'^{0},u'^{r},0,\frac{a}{a^{2}+r^{2}}u'^{0})\ ,\label{eq:uprimeKerr}
\end{equation}
corresponding to observers with angular velocity 
\begin{figure}
\begin{centering}
\includegraphics[width=0.95\columnwidth]{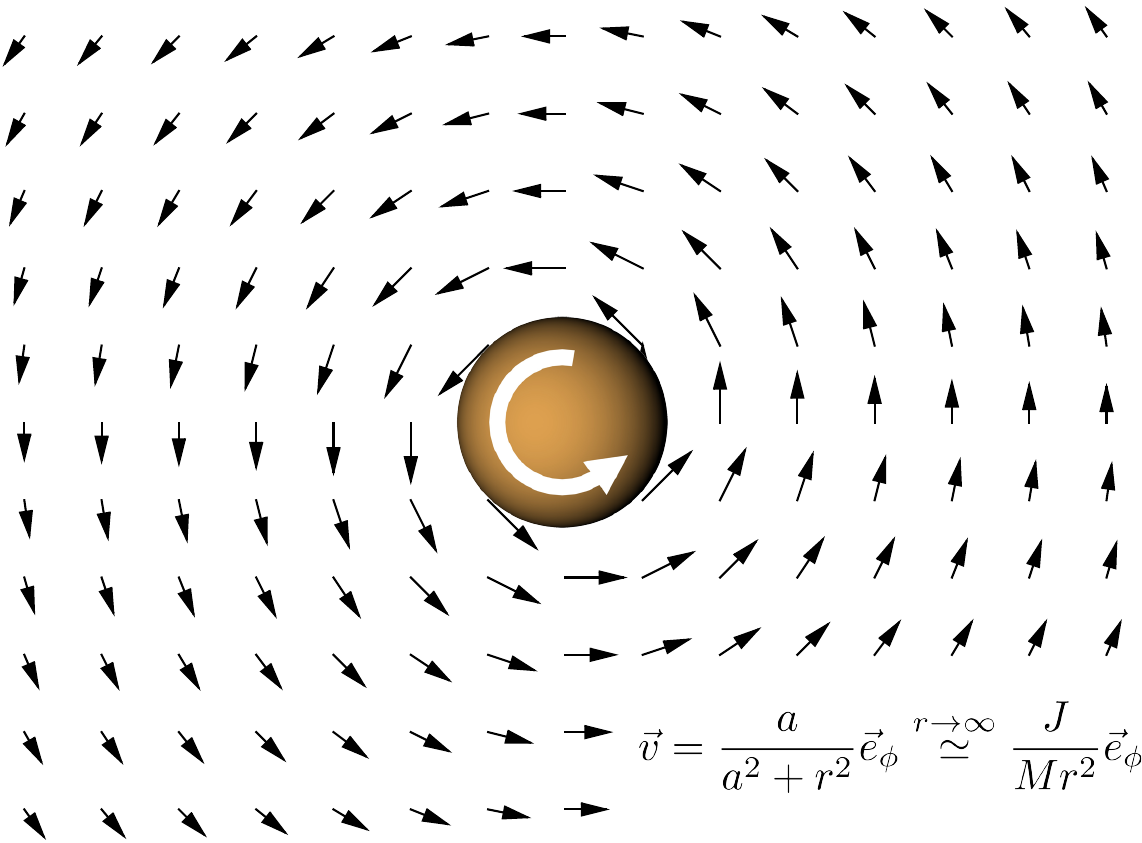} 
\par\end{centering}

\protect\protect\protect\protect\caption{\label{fig:KerrPlot}Observers for which the gravitomagnetic tidal
tensor $\mathbb{H}_{\alpha\beta}$ vanishes; their 3-velocity $\vec{v}$
is represented by black arrows (note that these are \emph{not} the
observers rigidly co-rotating with the body). $\vec{e}_{\phi}$ is
the coordinate basis vector $\vec{e}_{\phi}\equiv\vec{\partial}_{\phi}=\sqrt{g_{\phi\phi}}\vec{e}_{\hat{\phi}}$,
where $\vec{e}_{\hat{\phi}}$ has norm 1. Compare with the velocity
fields in Figs. \ref{fig:SchargeDots} and \ref{fig:Earth-Sun}.}
\end{figure}

\begin{equation}
\frac{d\phi}{dt}=\frac{u'^{\phi}}{u'^{0}}=\frac{a}{a^{2}+r^{2}}\label{velocityKerr}
\end{equation}
and an arbitrary radial velocity $dr/dt=u'^{r}/u'^{0}$ (subject only
to the normalization condition $u'^{\alpha}u'_{\alpha}=-1$). One
could check these results by computing explicitly $\mathbb{H}'_{\alpha\beta}$
(for an arbitrary $u'^{\alpha}$) as done in \cite{PaperGyros}. %
In the special case $u'^{r}=0$, one obtains the observers depicted
in Fig. \ref{fig:KerrPlot}, which coincide\footnote{We thank O. Semerák for pointing this out to us.}
with the so-called ``Carter canonical observers'' (e.g. \cite{SemerakStationaryFrames}).

Notice the similarity with the velocity field in Fig.~\ref{fig:SchargeDots},
which makes the magnetic field vanish in the analogous electromagnetic
problem: both velocities depend only on $r$ and on the ratio $a\equiv J/M$,
and asymptotically they match up to a factor of 2. Note also the similarities
with the velocity fields \eqref{eq:vNoH2BodyLimit} or \eqref{eq:v_field_2equalMasses}
for which $\mathbb{H}'_{\alpha\beta}=0$ in systems of two bodies
orbiting each other: in the post-Newtonian regime, $r\gg r_{+}\Rightarrow r\gg a$,
and, from Eq. \eqref{velocityKerr}, $v\simeq a/r\equiv J/(Mr)$;
hence, for large $r$, the velocities match up to a factor of three.
The vanishing of $\mathbb{H}'_{\alpha\beta}$ for such observers can
also be understood (in the PN regime) by the same reasoning we made
in Sec. \ref{sub:Two-masses,-planar}, by thinking about the rotating
body as a set of translating elements and adding up their individual
gravitomagnetic tidal tensors. This parallels what happens in electromagnetism,
where the vanishing of $\vec{B}'$ for some observers in the equatorial
plane of a spinning charge, Fig. \ref{fig:SchargeDots}, can be explained
by the same reasoning that explains its vanishing in the motion plane
of the system of two charges in Fig. \ref{fig:Two}. One thus concludes
that, although very different from a system of one single point source
of Sec \ref{sub:One-single-point-mass}, a spinning body is not, in
the PN regime, substantially different from the two-body systems of
Sec. \ref{sub:Two-masses,-planar}, in what pertains to the structure
of the curvature invariants (and the existence of observers for which
$\mathbb{H}'_{\alpha\beta}=0$).

Finally, we note that the velocity field in Fig. \ref{fig:KerrPlot}
provides another example of point \ref{enu:Gravpoint4} of Sec. \ref{sec:Interpretation-of-the-Gravitational}:
although there is no PN frame where all the mass currents are zero,
still $\innst{R}=0$ in a 2-D spatial surface (the equatorial plane),
where $\mathbb{H}'_{\alpha\beta}=0$ for certain observer congruences
that are \emph{not} PN frames.

\section{Non-vacuum examples\label{sec:Non-vacuum-examples}}

\subsection{Doubly aligned electrovacuum spacetimes --- the Kerr-Newman solution}

In electrovacuum gravitational fields $T_{\alpha\beta}$ reduces to
the stress-energy tensor of the electromagnetic field, $T_{\alpha\beta}=\frac{1}{4\pi}(F_{\alpha\gamma}F_{\beta}{}^{\gamma}-g_{\alpha\beta}\mathbf{F}\cdot\mathbf{F}/4)$,
and so the spatial mass/energy current (${\mathcal{J}'}^{\alpha}$)
as measured by an observer $\Oup$ consists of the relative Poynting
vector, 
\[
{\mathcal{J}'}^{\alpha}\equiv-{h'}_{\b}^{\a}T^{\beta\gamma}{u'}_{\gamma}=\frac{1}{4\pi}\epsilon_{\ \sigma\tau\beta}^{\alpha}{E'}^{\sigma}{B'}^{\tau}{u'}^{\beta}\equiv{p'}^{\alpha}\ .
\]
Condition (\ref{A2}) ($\Leftrightarrow{h'}_{\b}^{\a}T^{\beta\gamma}{u'}_{\gamma}=0$),
stating that ${u'}^{\alpha}$ is a Ricci eigenvector, thus holds if
and only if the observer measures no Poynting vector, ${p'}^{\alpha}=0$
(i.e., if it is an electromagnetic `principal observer' \cite{WylCosNat15}).
This is the case for all ${u'}^{\alpha}$ lying on the time-like principal
plane ($\Sigma_{F}$) of the (then non-null) Faraday tensor, which
is thus an eigenplane of $R_{\ \beta}^{\alpha}$. If the field is
``doubly aligned,'' the PND's and time-like principal planes of
the Faraday and (Petrov type D) Weyl tensors coincide ($\Sigma_{F}=\Sigma_{C}\equiv\Sigma$,
see e.g. the companion paper \cite{WylCosNat15}), so that condition
(2-a) of Sec. \ref{sub:Petrov-type-D} is satisfied. Such spacetimes
are generically described by the Pleba\'{n}ski-Demia\'{n}ski solutions
(e.g. \cite{GriffithsPodolsky2005,StephaniExact,Boos2015}), of which
the Kerr-Newman black hole is a special case. Hence observers measuring
$\mathbb{H}'_{\a\b}=0$ exist wherever the Weyl tensor is purely electric,
which (for type D) is ensured by the invariant conditions $\inn{C}>0$,
$\innst{C}=0$. Similarly to the Kerr case in Sec. \ref{sub:Kerr},
this occurs in the equatorial plane of the Kerr-Newman metric, and
the 4-velocities of the observers for which $\mathbb{H}'_{\a\b}=0$
analogously follow from the Weyl generalization of Eqs. (\ref{eq:u'grav}),
(\ref{eq:wvector}), (\ref{eq:evec}), (\ref{eq:tgrav2}), yielding
precisely the same result (\ref{eq:uprimeKerr})-(\ref{velocityKerr})
as in the Kerr spacetime (cf. \cite{WylCosNat15}, Sec. 6.2). The
special case $a=0$ yields the Reissner-Nordström solution, whose
Weyl (and Faraday) tensor is purely electric everywhere, and so everywhere
there are observers $\mathbb{H}'_{\a\b}=0$. Analogously, their 4-velocites
are of the same form \eqref{eq:4velSchw} as in the Schwarzschild
spacetime.

\subsection{Non-aligned electrovacuum spacetimes}

When the time-like principal planes of the Faraday and (Petrov type
D) Weyl tensors do not coincide ($\Sigma_{F}\ne\Sigma_{C}$), in general,
no observers will measure a vanishing $\mathbb{H}'_{\a\b}$, even
if both fields are purely electric. This is because in 4-D, two planes
generically intersect only at a point. The exceptional case where
they intersect along a timelike line, yields a \emph{unique} observer
$u'^{\alpha}\propto\Sigma_{F}\cap\Sigma_{C}$ for which $\mathbb{H}'_{\a\b}=0$.
An example are the solutions constructed in \cite{VdBCarm2020}. Therein
condition (2-c) of Sec. \eqref{sub:Petrov-type-D} reduces, in the
notation of \cite{VdBCarm2020}, to %
$\|C\|^{2}g^{2}\xi^{2}\varsigma^{2}/N>0$, with $g$, $\xi$, $\varsigma$
and $N>0$ real functions, being thus generically satisfied, and yielding
$q=1$. It follows from Eq. \eqref{up-para} that the 4-velocity of
the observer measuring $\mathbb{H}'_{\a\b}=0$ is $u'^{\alpha}=(k^{\alpha}+l^{\alpha})/\sqrt{2}$,
with $k^{\alpha}$ and $l^{\alpha}$ as given in Sec. 3 of \cite{VdBCarm2020}.

\subsection{Gödel universe and uniform Som-Raychaudhuri solution\label{sub:Godel-Som}}

Consider a line element in the form (``ultra-stationary spacetime'')
\begin{equation}
ds^{2}=-\left(dt-\mathcal{A}_{i}(x^{k})dx^{i}\right)^{2}+h_{ij}(x^{k})dx^{i}dx^{j}\ .\label{ultrastationary}
\end{equation}
 For the Gödel universe \cite{godel,HawkingEllis,NatarioGodel}, 
\begin{equation}
\mathcal{A}_{i}dx^{i}=e^{\sqrt{2}\omega x}dy\ ,\quad h_{ij}dx^{i}dx^{j}=dx^{2}+\frac{1}{2}e^{2\sqrt{2}\omega x}dy^{2}+dz^{2}\ ,\label{eq:Godel}
\end{equation}
where $\omega$ is a constant. It corresponds to a homogeneous universe
with negative cosmological constant, filled with dust rotating about
the $z$ axis in anti-clockwise direction (following our choice for
the sign of $\mathcal{A}_{i}$) \cite{NatarioGodel,JanztenIntrinsicDerivativeII}.
The Weyl tensor is purely electric and of Petrov type D, with \eqref{eq:Weyl-PE-D}
being satisfied for $\lambda=\omega^{2}/3$. The 4-velocities ${u'}^{\alpha}\equiv{u'}_{\mathcal{H=}0}^{\alpha}$
of the observers for which $\mathcal{H}'_{\alpha\beta}=0$ are obtained
from the Weyl generalization of Eqs. (\ref{eq:u'grav}), (\ref{eq:wvector}),
(\ref{eq:evec}), (\ref{eq:tgrav2}), and (\ref{eq:super-Poynting}).
We have, for the auxiliary quantities involved%
, $u^{\alpha}=\delta_{0}^{\alpha}$, $\mathcal{P}^{\alpha}=0\Rightarrow t^{\alpha}=u^{\alpha}$,
$\inn{C}=16\omega^{4}/3$, $\J=\mathbb{A}=-2\omega^{6}/9$, $\eb^{\alpha}=-\delta_{z}^{\alpha}$;
therefore, by (\ref{eq:u'grav})%
, 
\begin{equation}
{u'}_{\mathcal{H=}0}^{\alpha}=u'^{0}\delta_{0}^{\alpha}+u'^{z}\delta_{z}^{\alpha}\equiv(u'^{0},0,0,u'^{z})\ ,\label{eq:uGodelWeyl}
\end{equation}
with $u'^{z}$ arbitrary (under the normalization condition $u'^{\alpha}u'_{\alpha}=-1$).
We need now to check whether any of such observers also obeys (\ref{eq:extr-gravmag-curv}ii).
The null vectors generating the PNDs are obtained from (\ref{PNDs}),
$k^{\alpha}\equiv k_{+}^{\alpha}=(\delta_{0}^{\alpha}-\delta_{z}^{\alpha})/\sqrt{2}$,
$l^{\alpha}\equiv k_{-}^{\alpha}=(\delta_{0}^{\alpha}+\delta_{z}^{\alpha})/\sqrt{2}$,
yielding, for the remaining quantities involved in the criterion of
Sec. \ref{sub:Petrov-type-D}, $K^{\alpha}=L^{\alpha}=0$, $R_{kk}=R_{ll}=\omega^{2}$.
Hence, this spacetime corresponds to case (2-b) with $q=1$; therefore,
there is a unique observer for which $\mathbb{H}'_{\a\b}=0$, with
4-velocity ${u'}^{\alpha}\equiv{u'}_{\mathcal{\mathbb{H}=}0}^{\alpha}$
given by Eq. (\ref{up-para}), 
\begin{equation}
{u'}_{\mathcal{\mathbb{H}=}0}^{\alpha}=\delta_{0}^{\alpha}\equiv(1,0,0,0).\label{eq:uGodelRiemann}
\end{equation}

The Som-Raychaudhuri metrics \cite{Som-Raychaudhuri} are cylindrically
symmetric solutions corresponding to rigidly rotating charged dust
for which the Lorentz force vanishes everywhere. In the case where
the charge, mass, and electromagnetic field energy densities are uniform,
the metric reads (cf. Eq. (28) of \cite{Som-Raychaudhuri}) 
\begin{equation}
\mathcal{A}_{i}dx^{i}=\omega r^{2}d\phi\ ,\quad h_{ij}dx^{i}dx^{j}=dr^{2}+r^{2}d\phi^{2}+dz^{2}\ .\label{eq:Heisenberg}
\end{equation}
Again the Weyl tensor is purely electric and of Petrov type D, with
\eqref{eq:Weyl-PE-D} being satisfied for $\lambda=2\omega^{2}/3$.
The analysis is similar to that for the Godel universe, yielding the
same PNDs $k^{\alpha}$ and $l^{\alpha}$, and (\ref{eq:uGodelWeyl})
and (\ref{eq:uGodelRiemann}) holding for, respectively, the observers
measuring a vanishing $\mathcal{H}'_{\alpha\beta}$ and $\mathbb{H}'_{\a\b}$.

\subsection{Van Stockum cylinder\label{sub:Van-Stockum-cylinder}}

The van Stockum interior solution (e.g. \cite{Bonnor_Stockum,Cilindros}),
describes the gravitational field of an infinite rigidly rotating
cylinder of dust. Its line element is of the form (\ref{ultrastationary})
with
\begin{equation}
\mathcal{A}_{i}dx^{i}=\omega r^{2}d\phi\ ,\quad h_{ij}dx^{i}dx^{j}=r^{2}d\phi^{2}+e^{-\omega^{2}r^{2}}(dr^{2}+dz^{2})\ .\label{eq:Stockum}
\end{equation}
Therein $\innst{C}=0$, $\mathbb{M}_{C}=162\omega^{4}r^{4}f(r)/(2-9\omega^{2}r^{2})^{2}$,
where $f(r)\equiv1-4\omega^{2}r^{2}$; hence \eqref{eq:Weyl-PE-I}
is satisfied for $\omega r<1/2$, and so, within this regime, the
Weyl tensor is purely electric and of Petrov type I, as is well known
\cite{Bonnor_Stockum}. %
This means that, at each point, there is a unique observer for which
$\mathcal{H}_{\alpha\beta}=0$; the question is whether (\ref{eq:extr-gravmag-curv}ii)
is also satisfied for such observer. The Ricci tensor is of the perfect
fluid type, criterion (b) of Sec. \ref{sub:Rie-gen} and (b-2) of
Appendix \ref{app:extr-gravmag-curv} being obeyed with $\lambda=\omega^{2}e^{\omega^{2}r^{2}}$,
$w^{\alpha}\propto\delta_{0}^{\alpha}$. Hence, as shown in Appendix
\ref{app:extr-gravmag-curv} (see also \ref{sub:Alternative route}),
there is a unique observer for which (\ref{eq:extr-gravmag-curv}ii)
is satisfied, having 4-velocity parallel to $w^{\alpha}$: $u'^{\alpha}\equiv{u'}_{\mathcal{J=}0}^{\alpha}=\delta_{0}^{\alpha}$.
The physical interpretation is that this is the observer at rest with
respect to the dust; all other observers moving with respect to $u'^{\alpha}$
measure a non-vanishing spatial mass-energy current $\mathcal{J}^{\alpha}$.
The magnetic part of the Weyl tensor $\mathcal{H}'_{\alpha\beta}\equiv C_{\alpha\mu\beta\nu}u'^{\mu}u'^{\nu}$
as measured by this observer, however, is not zero, having non-vanishing
components $\mathcal{H}'_{rz}=\mathcal{H}'_{zr}=-w^{3}r$. Therefore
(\ref{eq:extr-gravmag-curv}i) is not obeyed, and so observers for
which $\mathbb{H}'_{\alpha\beta}=0$ do not exist.

\section{Dynamical implications of the invariants. Gravitomagnetism. \label{sec:Dynamical-implications-of}}

In the previous sections we made use of the insight that the analogy
$\{-\inn{F},\ -\innst{F}\}\leftrightarrow\{\inn{R},\ \innst{R}\}$
(see footnote~\ref{foot:signdiff}) between electromagnetic invariants
and gravitational invariants in vacuum gives us to interpret the structure
of the latter. It is crucial, however, to realize that this is a \emph{purely
formal} analogy. For in one case one is dealing with quantities built
on electromagnetic \emph{fields} $E^{\alpha},\ B^{\alpha}$; in the
other case with gravitational \emph{tidal tensors} $\mathbb{E}_{\alpha\beta},\ \mathbb{H}_{\alpha\beta}$;
and these objects do not play analogous dynamical roles. The fields
$E^{\alpha}$ and \textbf{$B^{\alpha}$ }govern effects like the Lorentz
force and the precession of a magnetic dipole, and have as closest
gravitational counterpart the so-called gravitoelectric ($G^{\alpha}$)
and gravitomagnetic ($H^{\alpha}$) \emph{inertial} fields, governing
effects like the (fictitious) inertial force that drives a particle
in geodesic motion, or the ``precession'' of a gyroscope. The tensors
$\mathbb{E}_{\alpha\beta},\ \mathbb{H}_{\alpha\beta}$, by contrast,
govern gravitational tidal effects, such as the geodesic deviation,
the spin-curvature force on a spinning particle, or the \emph{differential}
precession of spinning particles (and their electromagnetic analogues,
from a physical point of view, are the \emph{electromagnetic tidal
tensors} $E_{\alpha\beta},\ B_{\alpha\beta}$, as argued in \cite{CHPRD,PaperAnalogies}).
This means that the use, in some literature \cite{Gravitation and Inertia,Ciufolini LLR,LakeKerrInv},
of the formal analogy between the invariants to infer about effects
like the (inertial) gravitomagnetic force on a test particle or gyroscope
precession, is not a good \emph{physical} guiding principle. The effects
involved on both sides are different, and may actually be \emph{opposite},
as we shall exemplify next.

It is likewise crucial to distinguish and understand the relation
between the ``gravitoelectromagnetic'' inertial fields $G^{\alpha}$
and $H^{\alpha}$ (the ones involved in the frame-dragging effects
under debate in the literature) and the electric and magnetic parts
of the curvature $\mathbb{E}_{\alpha\beta},\ \mathbb{H}_{\alpha\beta}$,
as well as the invariants they form, which we shall also discuss next.

\subsection{A magnetic dipole in the field of a spinning charge vs. a gyroscope
in the Kerr spacetime\label{sub:Kerr vs Spinning charge}}

The equations of motion for a spinning particle with magnetic moment
$\mu^{\alpha}$ (and no charge nor electric dipole moment) in a electromagnetic
field in flat spacetime are, under the Mathisson-Pirani spin condition
(e.g. \cite{PaperGyros}), 
\begin{equation}
\frac{DP^{\alpha}}{d\tau}=B^{\beta\alpha}\mu_{\beta}\ ;\quad{\rm (a)}\qquad\frac{D_{F}S^{\alpha}}{d\tau}=\epsilon_{\ \beta\gamma\delta}^{\alpha}U^{\delta}\mu^{\beta}B^{\gamma}\ ,\quad{\rm (b)}\label{eq:EqsDipole}
\end{equation}
where $U^{\alpha}$, $P^{\alpha}$ and $S^{\alpha}$ are, respectively,
the particle's 4-velocity, 4-momentum, and spin angular momentum 4-vector;
$B^{\alpha}=\star F^{\alpha\beta}U_{\beta}$ and $B_{\alpha\beta}=\star F_{\alpha\gamma;\beta}U^{\gamma}$
are, respectively, the magnetic field and ``magnetic tidal tensor''
\cite{CHPRD} as measured by the particle; $D/d\tau=U^{\alpha}\nabla_{\alpha}$
is the usual (Levi-Civita) covariant derivative, and $D_{F}/d\tau$
is the Fermi-Walker covariant derivative, which reads, for a \emph{spatial}
vector $X^{\alpha}$ ($X^{\alpha}U_{\alpha}=0$), 
\[
\frac{D_{F}X^{\alpha}}{d\tau}=\frac{DX^{\alpha}}{d\tau}-X^{\beta}a_{\beta}U^{\alpha}\ .
\]

The equations of motion for a spinning pole-dipole particle in a gravitational
field are (under the same spin condition), e.g. \cite{PaperGyros},
\begin{equation}
\frac{DP^{\alpha}}{d\tau}=-\mathbb{H}^{\beta\alpha}S_{\beta};\quad{\rm (a)}\qquad\frac{dS^{\hat{\imath}}}{d\tau}=\left(\vec{S}\times\vec{\Omega}\right)^{\hat{\imath}}.\quad{\rm (b)}\label{eq:EqsGyro}
\end{equation}
Equation (\ref{eq:EqsGyro}a) is the spin-curvature force, which causes
the particle to deviate from geodesic motion; it consists of a coupling
between $S^{\alpha}$ and the gravitomagnetic tidal tensor as measured
by the particle, $\mathbb{H}_{\alpha\beta}=R_{\alpha\mu\beta\nu}U^{\mu}U^{\nu}$.
Equation (\ref{eq:EqsGyro}b) is the space part of equation $D_{F}S^{\alpha}/d\tau=0$
as measured in the particle's center of mass frame (stating that $S^{\alpha}$
is Fermi-Walker transported). The use of a simple derivative in (\ref{eq:EqsGyro}b)
manifests the fact that, by contrast with the Larmor precession in
(\ref{eq:EqsDipole}b), the so-called ``precession'' of a gyroscope
is not a covariant, \emph{locally} measurable effect. Indeed, $S^{\alpha}$
is fixed with respect to a comoving, locally non-rotating system of
axes (mathematically defined, precisely, as a Fermi-Walker transported
frame; for this reason one says that gyroscopes define the local ``compass
of inertia'', see e.g. \cite{Gravitation and Inertia,MassaZordan}).
The quantity $\vec{\Omega}$ in Eq.~(\ref{eq:EqsGyro}b) is thus
just the angular velocity of rotation of the spatial axes $\mathbf{e}_{\hat{\imath}}$
of the chosen frame relative to a locally non-rotating one. In the
context of the measurement of frame-dragging, the triad $\mathbf{e}_{\hat{\imath}}$
is chosen to be rotationally locked to the ``distant stars'' (how
such frame is constructed is discussed in Sec. \ref{sub:What-the-invariants say about GEM}
below); in such case $\vec{\Omega}$ yields \emph{minus} the precession
rate of the gyroscope with respect to the distant stars.

If the invariant conditions 
\begin{equation}
\innst{F}=0,\qquad-\inn{F}>0\label{eq:PEInvariantsEM}
\end{equation}
are satisfied in some region, then there are observers for which $B^{\alpha}=0$
everywhere, which by Eq.~(\ref{eq:EqsDipole}b) means that magnetic
dipoles carried by such observers do not undergo Larmor precession.
But it tells us nothing, a priori, about the force on the particle.
By contrast, what the conditions 
\begin{equation}
\innst{R}=0,\qquad\inn{R}>0\label{eq:PEInvariantsGrav}
\end{equation}
\emph{together with} \eqref{eq:grav-PEPMcond} tell us (in vacuum)
is that there are observers for which $\mathbb{H}_{\alpha\beta}=0$,
which by Eq. (\ref{eq:EqsGyro}a) means that gyroscopes comoving with
them feel no gravitational \emph{force}. It does not tell us (in general)
about gyroscope precession. Hence the effects at stake are different;
for seemingly analogous setups they may even be \emph{opposite}.

\begin{table}
\protect\protect\protect\protect\protect\protect\protect\caption{\label{tab:DipolevsGyroscope}Opposite effects: magnetic dipoles in
the equatorial plane of a spinning charge (where $\innst{F}=0$, $-\inn{F}>0$)
vs gyroscopes in the equatorial plane of a spinning celestial body
(where $\innst{R}=0$, $\inn{R}>0$).}

\centering{}%
\begin{tabular}{ll}
\hline 
Magnetic dipole moving  & Gyroscope moving\tabularnewline
\raisebox{0.7ex}{with angular velocity}  & \raisebox{0.7ex}{with angular velocity} \tabularnewline
\raisebox{3ex}{}\raisebox{2ex}{${\displaystyle \frac{d\phi}{dt}\equiv\frac{a}{2r^{2}}}$~~~(Fig.
\ref{fig:SchargeDots})}  & \raisebox{2ex}{${\displaystyle \frac{d\phi}{dt}=\frac{a}{a^{2}+r^{2}}}$~~~(Fig.
\ref{fig:KerrPlot})}\tabularnewline
\hline 
\hline 
No Larmor precession:  & Gyroscope precesses: \tabularnewline
$\vec{B}=0\,\Rightarrow{\displaystyle \frac{D\vec{S}}{dt}=0}$  & ${\displaystyle \frac{d\vec{S}}{dt}\ne0}$\tabularnewline
\raisebox{3.5ex}{}A force acts on it:  & No force: \tabularnewline
\raisebox{5ex}{}\raisebox{2ex}{${\displaystyle B_{\alpha\beta}\ne0\Rightarrow\frac{DP^{\alpha}}{d\tau}\ne0}$
~~~~~ }  & \raisebox{2ex}{${\displaystyle \mathbb{H}_{\alpha\beta}=0\Rightarrow\frac{DP^{\alpha}}{d\tau}=0}$}\tabularnewline
\hline 
\end{tabular}
\end{table}

A realization of this contrast is summarized in Table \ref{tab:DipolevsGyroscope}:
we have seen in Sec. \ref{sub:A-spinning-spherical} that, in the
equatorial plane of the spinning charge, conditions \eqref{eq:PEInvariantsEM}
are satisfied, implying that observers with angular velocity \eqref{eq:EMvplot}
measure no magnetic field. Hence magnetic dipoles comoving with them
do not undergo Larmor precession; they feel however a force, Eq. (\ref{eq:EqsDipole}a),
since $B_{\alpha\beta}\ne0$ \emph{always} for an observer moving
in a non-uniform field (due to the laws of electromagnetic induction)
as discussed in detail in \cite{PaperGyros}. We have also seen in
Sec. \ref{sub:Kerr} that, in the equatorial plane of the spacetime
around a spinning body, conditions \eqref{eq:PEInvariantsGrav} and
\eqref{eq:grav-PEPMcond}, are satisfied, implying that $\mathbb{H}_{\alpha\beta}=0$
for observers with angular velocity \eqref{velocityKerr}. This velocity
field has some similarities with \eqref{eq:EMvplot}; namely their
asymptotic limits match up to a factor of two. However, for gyroscopes
moving with these velocities, the situation is precisely the opposite:
by Eq. (\ref{eq:EqsGyro}a), no force is exerted on them, but they
\emph{precess} (with respect to the distant stars). This last point
deserves to be discussed in detail. To first post-Newtonian order,
in terms of the metric potentials in \eqref{eq:PNmetric}, the precession
frequency (let us denote it by $-\vec{\Omega}_{\star}$) of the spin
vector of a gyroscope with respect to a frame anchored to the distant
stars reads 
\begin{equation}
-\vec{\Omega}_{\star}=-\frac{1}{2}\vec{v}\times\vec{a}+\frac{3}{2}\vec{v}\times\nabla U-\frac{1}{2}\nabla\times\vec{\mathcal{A}}\label{eq:PrecessStarPN}
\end{equation}
(cf. e.g. Eqs. (40.33) of \cite{Gravitation}, Eqs. (3.4.38) of \cite{Gravitation and Inertia}),
where $\vec{v}$ is the gyroscope's velocity with respect to the PN
frame and $a^{i}=\nabla_{\mathbf{U}}U^{i}$ are the spatial components
of its \emph{covariant} acceleration. The first term, $\vec{a}\times\vec{v}/2=\vec{\Omega}_{{\rm Thomas}}$,
is the \emph{Thomas precession}; because of it, $\vec{\Omega}_{\star}$
depends on the gyroscope's acceleration. Hence, to determine $\vec{\Omega}_{\star}$
for gyroscopes moving with the velocities depicted in Fig. \ref{fig:KerrPlot},
we must say how they accelerate. It is natural to consider two cases:
i) gyroscopes in circular motion with angular velocity $d\phi/dt$
given by \eqref{velocityKerr}, and ii) gyroscopes at rest in boosted
PN frames \emph{momentarily} moving with $\vec{v}=(d\phi/dt)\vec{e}_{\phi}$.
To 1PN order, the Thomas precession is the same in both cases: in
case i), the exact acceleration of the gyroscope is $\vec{a}=[M/r^{2}-J^{2}/(M^{2}r^{3})]\vec{e}_{r}$;
hence, using $v\simeq J/(Mr)$, 
\begin{eqnarray*}
\vec{\Omega}_{{\rm Thomas}} & = & -\frac{1}{2}\vec{v}\times\vec{a}=\frac{J}{2r^{3}}\left(1-\frac{M}{r}\left[\frac{J}{M^{2}}\right]^{2}\right)\vec{e}_{z}\\
 & = & \frac{J}{2r^{3}}\left[1+O(\epsilon^{2})\right]\vec{e}_{z}\simeq\frac{\vec{J}}{2r^{3}}\ ,
\end{eqnarray*}
where $\vec{e}_{z}=-\vec{e}_{\hat{\theta}}$. In case ii), $\vec{v}\times\vec{a}=-\vec{v}\times\nabla U+O(\epsilon^{5}/L)$;
hence, to the accuracy at hand, $\vec{\Omega}_{{\rm Thomas}}$ is
the same. Observing that, in the equatorial plane, $\nabla\times\vec{\mathcal{A}}=2\vec{J}/r^{3}=2J\vec{e}_{z}/r^{3}$
and $\vec{v}\times\nabla U=\vec{e}_{z}vM/r^{2}$, the overall precession
with respect to the distant stars is (for both cases) 
\[
-\vec{\Omega}_{\star}=\frac{3\vec{J}}{2r^{3}}\ne0\ .
\]

A question that naturally arises is whether there are, in the equatorial
plane, velocity fields for which gyroscopes do not precess with respect
to the distant stars. The answer is affirmative, but, again, acceleration
dependent. If one considers gyroscopes comoving with boosted PN frames
(i.e., gyroscopes moving with constant coordinate velocity, $d\vec{v}/dt=0$),
then $-\vec{\Omega}_{\star}=2\vec{v}\times\nabla U-\frac{1}{2}\nabla\times\vec{\mathcal{A}}$,
and the condition $\vec{\Omega}_{\star}=0$ yields $v=J/(2Mr)$. This
is half the asymptotic limit of the velocity (\ref{velocityKerr})
for which $\mathbb{H}_{\alpha\beta}=0$, but is precisely the same
as the velocity (\ref{eq:EMvplot}) for which $\vec{B}=0$ in the
equatorial plane of a spinning charge (Fig. \ref{fig:SchargeDots}).
Indeed, this is \emph{physically} the analogue of the latter: $\vec{H}'=-4\vec{v}\times\nabla U+\nabla\times\vec{\mathcal{A}}=2\vec{\Omega}_{\star}$
is the gravitomagnetic field (see below) as measured in the PN rest
frame of the gyroscope; so solving for $\vec{\Omega}_{\star}=0$ amounts
to finding (at each point) a boosted PN frame where the gravitomagnetic
field $\vec{H}'$ vanishes. Its velocity is given by Eq. (\ref{eq:VelnoHPN})
below, analogous to Eq. (\ref{eq:explicitv}). Analogously to the
electromagnetic case, this can be cast as a cancellation between the
gravitomagnetic fields generated by the rotation and relative translation
of the source. One must note, however, that this has nothing to do
with the curvature invariants; it comes from the analogy (discussed
in Sec.~\ref{sub:PN-frames-where-H_0} below) between the transformation
laws for the GEM fields in the PN regime and the electromagnetic fields.

\subsection{``Gravitoelectromagnetic fields'' (GEM Fields)\label{sub:GEM-fields}}

The inertial GEM fields have been defined in different ways in the
literature, from the linearized theory approaches in e.g. \cite{Gravitation and Inertia,Gravitation and Spacetime,Harris1991,Ruggiero:2002hz,Wald et al 2010},
to the exact formulations in e.g. \cite{LandauLifshitz,Cattaneo1958,Black Holes,The many faces,GEM User Manual,JanztenIntrinsicDerivative,JanztenIntrinsicDerivativeII,SemerakInertial,ZonozBell,Natario,PaperAnalogies}.
Here we will follow the exact approach in \cite{PaperAnalogies},
which we believe to be physically motivated, and which leads, in the
corresponding limit, to the GEM fields usually defined in post-Newtonian
approximations, e.g.~\cite{DSX,SoffelKlioner,WillPoissonBook}.

Consider a congruence of observers of 4-velocity $u^{\alpha}$, and
a test particle of worldline $z^{\alpha}(\tau)$ and 4-velocity $dz^{\alpha}/d\tau=U^{\alpha}$.
Take it, for simplicity, to be a point-like monopole particle, and
assume that there are no external forces, so that $z^{\alpha}(\tau)$
is geodesic. Let $U^{\langle\alpha\rangle}\equiv h_{\beta}^{\alpha}U^{\beta}$
be the spatial projection of the particle's velocity with respect
to the observers, cf. Eq. \eqref{eq:SpaceProjector}; it can be interpreted
as the relative velocity of the particle with respect to the observers.
It is the variation of $U^{\langle\alpha\rangle}$ along $z^{\alpha}(\tau)$
that one casts as \emph{inertial forces}; the precise definition of
such variation involves some subtleties however. For that we need
a connection (i.e., a covariant derivative) for spatial vectors; however
the space projection of the spacetime (Levi-Civita) covariant derivative,
$h_{\beta}^{\alpha}\nabla_{\mathbf{U}}U^{\langle\beta\rangle}$, which
might seem the most obvious, is not the one we seek, as it yields
the Fermi-Walker derivative of $U^{\langle\alpha\rangle}$ (i.e.,
its variation with respect to a system of Fermi-Walker transported
axes). We seek a connection that yields the variation of $U^{\langle\alpha\rangle}$
with respect to a system of spatial axes undergoing a transport law
specific to the reference frame one chooses. Given a congruence of
observers, the most natural choice would be spatial triads \emph{co-rotating}
with the observers. That is, for an orthonormal basis $\mathbf{e}_{\hat{\alpha}}$,
whose general transport law along the observer congruence can be written
as (e.g. \cite{Gravitation}) 
\[
\nabla_{\mathbf{u}}\mathbf{e}_{\hat{\beta}}=\Omega_{\,\,\hat{\beta}}^{\hat{\alpha}}\mathbf{e}_{\hat{\alpha}};\quad\Omega^{\alpha\beta}=2u^{[\alpha}a^{\beta]}+\epsilon_{\ \ \nu\mu}^{\alpha\beta}\Omega^{\mu}u^{\nu}\ ,
\]
that amounts to choosing $\Omega^{\alpha}$ (the angular velocity
of rotation of the spatial axes relative to Fermi-Walker transport)
equal to the observer's vorticity: $\Omega^{\alpha}=\omega^{\alpha}$.
If the congruence is rigid, this ensures that the $\mathbf{e}_{\hat{\imath}}$
point to fixed neighboring observers (cf. Eq. (\ref{eq:connectVector})
below). One might argue \cite{MassaZordan,MassaII} that this is the
closest generalization of the Newtonian concept of reference frame;
we dub it \emph{the congruence adapted frame}. For more details we
refer to Sec. 3 of \cite{PaperAnalogies}. The connection that yields
the variation of a spatial vector $X^{\alpha}$ with respect to such
frame is $\tilde{\nabla}_{\alpha}X^{\beta}\equiv h_{\gamma}^{\beta}{\nabla}_{\alpha}X^{\gamma}+u_{\alpha}\epsilon_{\ \delta\gamma\lambda}^{\beta}u^{\gamma}X^{\delta}\omega^{\lambda}$,
cf. Eq. (51) of \cite{PaperAnalogies}; and the inertial or ``gravitoelectromagnetic''
force on a test particle is the variation of $U^{\langle\alpha\rangle}$
along $z^{\alpha}(\tau)$ with respect to $\tilde{\nabla}$, that
is $\tilde{\nabla}_{\mathbf{U}}U^{\langle\alpha\rangle}\equiv\tilde{D}U^{\langle\alpha\rangle}/d\tau$.
Since, for geodesic motion, $\nabla_{\mathbf{U}}U^{\alpha}=0$, it
follows, using \eqref{eq:SpaceProjector}, that $\tilde{\nabla}_{\mathbf{U}}U^{\langle\alpha\rangle}=-\gamma(\nabla_{\mathbf{U}}u^{\alpha}+\epsilon_{\ \delta\gamma\lambda}^{\beta}u^{\gamma}X^{\delta}\omega^{\lambda})$,
where $\gamma\equiv-U_{\alpha}u^{\alpha}$. Finally, from the decomposition
(cf. Eq. \eqref{eq:Kinematics}) 
\begin{equation}
\nabla_{\beta}u_{\alpha}\equiv u_{\alpha;\beta}=-u_{\beta}\nabla_{\mathbf{u}}u_{\alpha}-\epsilon_{\alpha\beta\gamma\delta}\omega^{\gamma}u^{\delta}+\sigma_{\alpha\beta}+\frac{\theta}{3}h_{\alpha\beta}\ ,\label{eq:Kinematics-Decomp}
\end{equation}
we have \cite{PaperAnalogies} (noting that $\nabla_{\mathbf{U}}u^{\alpha}=U^{\beta}\nabla_{\beta}u^{\alpha}$)
\begin{equation}
\frac{\tilde{D}U^{\langle\alpha\rangle}}{d\tau}=\gamma\left[\gamma G^{\alpha}+\epsilon_{\ \beta\gamma\delta}^{\alpha}u^{\delta}U^{\beta}H^{\gamma}-\sigma_{\ \beta}^{\alpha}U^{\beta}-\frac{\theta}{3}h_{\beta}^{\alpha}U^{\beta}\right],\label{eq:GEMforce}
\end{equation}
where 
\begin{equation}
G^{\alpha}=-\nabla_{\mathbf{u}}u^{\alpha}\ ;\qquad H^{\alpha}=2\omega^{\alpha}\label{eq:GEMfields}
\end{equation}
are, respectively, the ``gravitoelectric'' and ``gravitomagnetic''
fields. These are \emph{exact} GEM fields, herein defined in terms
of the kinematical quantities of the observers' congruence; they play
in \eqref{eq:GEMforce} a role analogous to the electric and magnetic
field in the Lorentz force. One should keep in mind that $G^{\alpha}$
\emph{is minus the observers' acceleration}, and $H^{\alpha}$ \emph{twice
their vorticity}. For the observers at rest ($u^{i}=0$) in a given
coordinate system, $G^{i}=\Gamma_{00}^{i}/g_{00}$, $H^{i}=-\epsilon_{\ k}^{ij}\Gamma_{0j}^{k}/g_{00}$;
hence, to first post-Newtonian (1PN) order,\footnote{Using the 1PN Christoffel symbols, e.g. Eqs. (8.15) of \cite{WillPoissonBook},
identifying $w\rightarrow U+\Psi$, $\mathcal{A}_{i}\rightarrow-4U_{i}$
in the notation therein.} 
\begin{equation}
\vec{G}=\nabla w-\frac{\partial\vec{\mathcal{A}}}{\partial t}+\O{6}\ ;\quad\vec{H}=\nabla\times\vec{\mathcal{A}}+\O{5}\ ,\label{eq:GEMfieldsPN}
\end{equation}
which match the GEM fields in Eqs. (3.21) of \cite{DSX}, or Eqs.
(2.5) of \cite{Kaplan}. The 1PN limit of \eqref{eq:GEMforce} takes
the form\footnote{Noting that, to 1PN, $\tilde{D}U^{\langle i\rangle}/d\tau=(U^{0})^{2}d^{2}x^{i}/dt^{2}+2v^{i}\partial_{t}U-v^{2}G^{i}+4v^{i}\vec{G}\cdot\vec{v}$,
$\sigma_{\alpha\beta}=0$, $\theta=3\partial_{t}U$ and $\gamma^{2}/(U^{0})^{2}=1-2U+\Os{4}$. } 
\begin{equation}
\frac{d^{2}\vec{x}}{dt^{2}}=(1+v^{2}-2U)\vec{G}+\vec{v}\times\vec{H}-3\frac{\partial U}{\partial t}\vec{v}-4(\vec{G}\cdot\vec{v})\vec{v}\ .\label{eq:GeoPN}
\end{equation}
which matches\footnote{Noting that by $\vec{G}\equiv G^{i}\partial_{i}$ we denote the spatial
components of $G^{\alpha}$ \emph{in the PN coordinate basis}, and
that, to 1PN order, $G_{i}=G^{j}(1+2U)\delta_{ij}=G^{i}+2UG^{i}$.} Eq. (7.17) of \cite{DSX}. Linearizing Eqs. \eqref{eq:GEMfieldsPN}-\eqref{eq:GeoPN}
one obtains (up to some factors depending on the conventions) the
GEM fields and geodesic equation of the linearized theory approaches
\cite{Gravitation and Inertia,Gravitation and Spacetime,Harris1991,Ruggiero:2002hz,Wald et al 2010,PaperCoriolis}.

\subsubsection{post-Newtonian frames where $\vec{H}=0$\label{sub:PN-frames-where-H_0}}

The transformation laws for the GEM fields in a change of reference
frame exhibit some similarities to their electromagnetic counterparts.
The exact forms are given in Eqs. (8.3) of \cite{The many faces}.
To 1PN order, the GEM fields of a boosted PN frame can be obtained
applying a post-Galilean coordinate transformation (e.g. Eqs. (13)
of \cite{WillNordvedt1972}) to the metric, and then computing expressions
\eqref{eq:GEMfieldsPN} for the boosted potentials. In the case of
$\vec{H}$ we have 
\begin{equation}
\vec{H}'=\vec{H}-4\vec{v}\times\vec{G}\label{eq:Hboost}
\end{equation}
(cf. Eqs. (5) of \cite{JantzenThomas}, Eq. (4.20b) of \cite{DSX}),
formally similar to the post-Coulombian limit of Eq. \eqref{Bvec'},
apart from the factor of 4 in the second term. It is clear from \eqref{eq:Hboost}
that when $\vec{G}\cdot\vec{H}=0$ and $\vec{G}^{2}>\vec{H}^{2}$
one can always find a boost velocity $\vec{v}$ such that 
\begin{equation}
\vec{H}'=0\ \Rightarrow\ 4\vec{v}\times\vec{G}=\vec{H}\ .\label{eq:Hboost_zero}
\end{equation}
This is in analogy with the situation in electromagnetism in Sec.
\ref{sub:Observers-with-no} for the vanishing of $\vec{B}$. Here
$\vec{v}$ is such that its component $\vec{v}_{\perp G}$ orthogonal
to $\vec{G}$ reads (taking the cross product of \eqref{eq:Hboost_zero}
with $\vec{G}$) 
\begin{equation}
\vec{v}_{\perp G}=\frac{\vec{G}\times\vec{H}}{4G^{2}}\ ,\label{eq:VelnoHPN}
\end{equation}
in analogy with Eq. \eqref{eq:explicitv}; and likewise no condition
is imposed on $\vec{v}_{\parallel G}$. An example is the case of
the equatorial plane of the field produced by a spinning body, where
$\vec{G}\perp\vec{H}$ (and $\vec{G}^{2}>\vec{H}^{2}$), and indeed,
as we have seen in Sec. \ref{sub:Kerr vs Spinning charge}, at each
point one can find PN frames where $\vec{H}'=0$ at that point.

However (contrary to what has been suggested in some literature \cite{Gravitation and Inertia,Ciufolini LLR,KopeikinInv,OConnel Inv,KopeikinFomlalont,PfisterKingBook,Pfister,Overduin}),
this has nothing to do with field invariants: firstly, $\vec{G}\cdot\vec{H}$
and $G^{2}-H^{2}$ are not frame invariant\footnote{Restricting ourselves to \emph{post-Newtonian frames}, we can still
say that $\vec{G}\cdot\vec{H}$ is invariant, \emph{to 1PN order},
under changes of PN frame, since, as follows from Eqs. (5) of \cite{JantzenThomas},
$\vec{G}\cdot\vec{H}=\vec{G}'\cdot\vec{H}'+\O{7}{2}$; however, $\vec{G}^{2}-\vec{H}^{2}$,
to that accuracy, is not.} ($\vec{G}$ and $\vec{H}$ are actually mere artifacts of the reference
frame, which vanish in a locally inertial one); secondly, they do
not have any obvious relation with the curvature invariants. Indeed,
as one may check computing $\star\mathbf{R}\cdot\mathbf{R}=16\mathbb{E}_{\alpha\beta}\mathbb{H}^{\alpha\beta}$
using Eqs. \eqref{eq:EijGEMPN}-\eqref{eq:HijGEMPN} below, one can
have e.g. $\vec{G}\cdot\vec{H}\ne0$ whilst $\star\mathbf{R}\cdot\mathbf{R}=0$,
or $\star\mathbf{R}\cdot\mathbf{R}\ne0$ whilst $\vec{G}\cdot\vec{H}=0$.

\subsubsection{Relation between GEM fields and tidal tensors}

It is of crucial importance to distinguish between the gravitational
tidal tensors $\mathbb{E}_{\alpha\beta}$, $\mathbb{H}_{\alpha\beta}$
and the inertial fields $\vec{G}$, $\vec{H}$. A first obvious difference
is that whereas $\mathbb{E}_{\alpha\beta}$ and $\mathbb{H}_{\alpha\beta}$
are physical fields, governing \emph{physical} forces such as the
spin-curvature force exerted on a gyroscope, Eq. (\ref{eq:EqsGyro}a)
(which is the covariant derivative of the 4-momentum), $\vec{G}$
and $\vec{H}$ are artifacts of the reference frame, governing fictitious
forces and torques, such as the inertial force in Eq. (\ref{eq:GEMforce}),
or the gyroscope ``precession'' in Eq. (\ref{eq:EqsGyro}a) (an
ordinary derivative of $\vec{S}$). The \emph{exact} relation between
the two types of objects is complicated in general; it is given by
Eqs. (109)-(110) of \cite{PaperAnalogies}. In this work we are interested
in two special cases where it becomes simpler: exact stationary fields,
and arbitrary fields to first post-Newtonian order.

In a rigid frame in a stationary spacetime we have (Eqs. (111)-(112)
of \cite{PaperAnalogies}), 
\begin{align}
\mathbb{E}_{ij} & =-\nabla_{j}^{\perp}G_{i}+G_{i}G_{j}+\frac{1}{4}\left(H^{2}h_{ij}-H_{j}H_{i}\right)\,;\label{Eij}\\
\mathbb{H}_{ij} & =-\frac{1}{2}\left[\nabla_{j}^{\perp}H_{i}+(\vec{G}\cdot\vec{H})h_{ij}-2G_{j}H_{i}\right]\ ,\label{Hij}
\end{align}
where $h_{\alpha\beta}$ is the spatial metric, cf. Eq. \eqref{eq:SpaceProjector},
and $\nabla^{\perp}$ the connection defined by Eq. \eqref{eq:nabla_perp},
whose restriction to the spatial directions (which equals that of
$\tilde{\nabla}$) yields the Levi-Civita connection of $h_{\alpha\beta}$.

In an arbitrary spacetime, to 1PN order, we have, from Eqs. (\ref{eq:EijPN})-(\ref{eq:HijPN})
and (\ref{eq:GEMfieldsPN}), 
\begin{align}
\mathbb{E}_{ij} & =-\nabla_{j}G_{i}+G_{i}G_{j}+\frac{1}{2}\epsilon_{ijk}\dot{H}^{k}-\ddot{U}\delta_{ij}+\ensuremath{\O{6}{2}};\label{eq:EijGEMPN}\\
\mathbb{H}_{ij} & =-\frac{1}{2}\nabla_{j}H_{i}-\epsilon_{ijk}\dot{G}^{k}+\ensuremath{\O{5}{2}}\ ,\label{eq:HijGEMPN}
\end{align}
in agreement with Eqs. (3.38) and (3.41) of \cite{DSX}.\footnote{To obtain Eq. (3.38) of \cite{DSX} from \eqref{eq:EijGEMPN}, one
notes that $\nabla_{j}G_{i}\simeq G_{i,j}-\Gamma_{ij}^{k}G_{k}=G_{i,j}-2G_{i}G_{j}+\delta_{ij}G^{2}$,
and $\nabla\times\vec{G}=-\partial\vec{H}/\partial t$, cf. Eq. (94)
of \cite{PaperAnalogies}.} In the\emph{ linear }regime, and when the fields are\emph{ stationary},
the gravitational tidal tensors reduce to \emph{derivatives} of the
GEM fields: $\mathbb{E}_{ij}\approx-G_{i;j}$, $\mathbb{H}_{ij}\approx-H_{i;j}/2$.

\subsubsection{Uniform gravitomagnetic fields\label{sub:UniformGMs}}

A pedagogical example that illustrates how crucial it is to distinguish
between the GEM inertial fields $\vec{G},\vec{H}$ and the GEM tidal
tensors $\mathbb{E}_{\alpha\beta},\mathbb{H}_{\alpha\beta}$ (showing
that there is no direct relation between $\star\mathbf{R}\cdot\mathbf{R}$
and $\vec{H}$) is to consider spacetimes with uniform gravitomagnetic
fields. Examples of such spacetimes are the Gödel universe and the
particular class of the Som-Raychaudhuri solutions considered in Sec.
\ref{sub:Godel-Som}. These metrics, described by the line elements
\eqref{ultrastationary}-\eqref{eq:Godel} and \eqref{eq:Heisenberg},
have special properties. Observers at rest in their coordinate systems,
$u^{\alpha}=\delta_{0}^{\alpha}$, form a rigid congruence ($\s_{\a\b}=0,\,\theta=0$)
with \emph{zero} acceleration {[}as in all metrics of the form \eqref{ultrastationary}{]},
and \emph{uniform} vorticity $\vec{\omega}=\omega\vec{e}_{z}$. That
is, in a frame adapted to such observers, the gravitoelectric field
vanishes, and there is a \emph{non-zero} uniform gravitomagnetic field
\cite{JanztenIntrinsicDerivativeII}, cf.~Eq.~(\ref{eq:GEMfields}):
\begin{equation}
\vec{G}=\vec{0}\ ,\qquad\vec{H}=2\omega\vec{e}_{z}\ .\label{eq:G_H_Uniform}
\end{equation}
In terms of the curvature, the situation is precisely the \emph{opposite}.
From Eqs. \eqref{Eij}-\eqref{Hij} one obtains 
\begin{equation}
\mathbb{E}_{ij}=\frac{1}{4}\left({\vec{H}}^{2}h_{ij}-H_{j}H_{i}\right)\ ;\qquad\mathbb{H}_{\alpha\beta}=0\ ,\label{Eij-Godel-SomRay}
\end{equation}
i.e., these observers measure a non-zero gravitoelectric tidal tensor
and a vanishing gravitomagnetic tidal tensor everywhere. Actually,
one may check that the only solutions to the system $R_{\a\g\b\d}x^{\g}x^{\d}=0$
are $x^{\a}\propto\delta_{z}^{\a}$ and thus spacelike, such that
for {\em all} observers the gravitoelectric tidal tensor is non-zero,
and so the Riemann tensor is purely electric according to the definition
in Sec.~\ref{sub:Rie-gen}. It then follows from Eq.~\eqref{eq:Pontry}
that $\star\mathbf{R}\cdot\mathbf{R}=0$ everywhere.

Thus, from the point of view of the curvature, the metrics \eqref{ultrastationary}-\eqref{eq:Heisenberg}
represent purely electric spacetimes, with $\mathbb{E}_{\alpha\beta}\ne0$
and $\mathbb{H}_{\alpha\beta}=0$ globally with respect to a rigid
congruence of observers, namely the rest observers in the coordinates
of \eqref{ultrastationary}; from the point of view of the GEM inertial
fields, by contrast, one would say that they are \emph{purely magnetic},
since $\vec{G}=\vec{0}$ and $\vec{H}\ne\vec{0}$ for the same observers.
It is actually impossible to make $\vec{H}$ vanish in any rigid frame,
which can be seen as follows. From Eq. (91) of \cite{PaperAnalogies},
we have, for a rigid frame, 
\[
\nabla^{\perp}\times\vec{H}=-16\pi\vec{\mathcal{J}}\ ,
\]
where $\mathcal{J}^{\alpha}$ is the spatial mass/energy current as
defined in Sec. \ref{sub:Rie-gen}. Hence, $\vec{H}=\vec{0}$ requires
$\vec{\mathcal{J}}=\vec{0}$ in that frame. The Ricci tensor of the
Gödel universe is of perfect fluid type, conditions (b) of Sec. \ref{sub:Rie-gen}
and (b-2) of Appendix \ref{app:extr-gravmag-curv} holding with $\lambda=\omega^{2}/2$,
$w^{\alpha}\propto\delta_{0}^{\alpha}$. Then, as shown in Appendix
\ref{app:extr-gravmag-curv} (see also \ref{sub:Alternative route}),
there is a unique congruence of observers for which $\mathcal{J}^{\alpha}=0$,
which are the observers at rest in the coordinates of (\ref{ultrastationary}),
$u_{(\mathcal{J}=0)}^{\alpha}=\delta_{0}^{\alpha}\propto w^{\alpha}$.
That is, the observers comoving with the fluid. Since $\omega^{\alpha}\ne0$
for those observers, no rigid congruences with vanishing vorticity
exist in this spacetime, that is, $\vec{H}\ne\vec{0}$ for frames
adapted to any rigid congruence of observers. A similar proof can
be made for the metric \eqref{ultrastationary} with \eqref{eq:Heisenberg},
only now the Ricci tensor is of Segre type {[}(11)1,1{]}{]}, conditions
(d) of Sec. \ref{sub:Rie-gen}, and (D) and (d2) of Appendix \ref{app:extr-gravmag-curv}
holding with $\lambda=3\omega^{2}/2$, $c=\omega^{2}$, thereby admitting
the single time-like eigenvector $\delta_{0}^{\alpha}$. More strongly,
it was actually shown in Sec.~II.C of \cite{WyllemanBeke} that,
for the Gödel universe, $\omega^{\a}\neq0$ and thus $\vec{H}\ne\vec{0}$
for frames adapted to any {\em shear-free} observer congruence
($\s_{\a\b}=0$). The same proof applies to the special Som-Raychaudhuri
metrics.\footnote{Both metrics \eqref{ultrastationary}-\eqref{eq:Heisenberg} are of
Petrov type D with the two Weyl PNDs spanned by the null vectors $\delta_{0}^{\a}\pm\delta_{z}^{\a}$;
in an adapted Newman-Penrose (complex null) frame, the relevant Newman-Penrose
Weyl scalars and spin coefficients are $\Psi_{2}=\overline{\Psi_{2}}\propto\omega^{2}$,
$\Psi_{k}=0$ for $k\neq2$, and $\kappa=\sigma=\tau=0$, $\nu=\lambda=\pi=0$,
$\rho=\mu={i}\omega/{\sqrt{2}}$; thus none of the criteria (1)-(5)
in proposition B.1 of \cite{WyllemanBeke} is fulfilled, and so the
spacetimes do not admit any shear- and vorticity-free observer congruence.} (These metrics thus possess a \emph{globally intrinsic }gravitomagnetic
field $\vec{H}$, according to the classification scheme proposed
in Sec. \ref{sub:criteria}.)

To see the consequences in terms of motion of test particles, consider
gyroscopes at rest ($U^{\alpha}=\delta_{0}^{\alpha}$) in the coordinate
system of \eqref{ultrastationary}. These feel no spin-curvature force,
since $\mathbb{H}_{\alpha\beta}=0$, cf. Eq. (\ref{eq:EqsGyro}a);
they will actually remain at rest in that coordinate system since,
moreover, they feel no gravitoelectric field (as the frame is freely
falling), cf. Eqs. \eqref{eq:G_H_Uniform}. However they precess relative
to the frame adapted to these observers (i.e., to the basis vectors
of the coordinate system of $h_{ij}$) with angular velocity $-\vec{\omega}=-\omega\vec{e}_{z}=-\vec{H}/2$,
like a magnetic dipole under a uniform magnetic field, cf. Eq. (\ref{eq:EqsGyro}b).
Moreover, there is no rigid (and even not a shear-free) frame relative
to which the gyroscopes do not precess. In fact, the only precession
effect that vanishes due to the vanishing of $\mathbb{H}_{\alpha\beta}$
is the so-called ``differential precession'', that is, the precession
of a gyroscope relative to a system of axes anchored to a set of neighboring,
infinitesimally close gyroscopes, as this is a tidal effect governed
precisely by $\mathbb{H}_{\alpha\beta}$, see Eq. (3.11) of \cite{DiffPrecession}
(cf. also Sec. 2.3 of \cite{PaperAnalogiesExtended}). \textcolor{black}{We
believe this to be enough to convince the reader about the importance
of distinguishing between GEM }\textcolor{black}{\emph{inertial}}\textcolor{black}{{}
and }\textcolor{black}{\emph{tidal }}\textcolor{black}{fields, and
that indeed the invariant $\star\mathbf{R}\cdot\mathbf{R}$ is not
a good test for ``intrinsic gravitomagnetic field''.}

\subsection{What the invariants say about the gravitomagnetic field\label{sub:What-the-invariants say about GEM}}

As is explicit from Eqs. \eqref{eq:inv-A}-\eqref{eq:inv-D}/\eqref{eq:Kretsch}-\eqref{eq:Pontry},
it is the electric and magnetic parts of the curvature (and their
possible vanishing for some observers), \emph{not} the GEM fields,
that are directly related with the curvature invariants. However,
still there are special cases where indeed from the curvature invariants
one can infer information about the gravitomagnetic field itself.

First let us discuss what can be understood as a physically meaningful
gravitomagnetic field, that can be identified with the effects that
have been under experimental and observational scrutiny. The gravitomagnetic
field $\vec{H}$ is an inertial field, i.e., a reference frame artifact,
that can always be gauged away by choosing a locally inertial frame.
Thus, \emph{locally}, it has no physical meaning; yet it may reflect
\emph{global} physical properties of a given spacetime. For instance,
the ``precession'' of a gyroscope (at a finite $r$) in the Kerr
spacetime with respect to a frame anchored to the distant stars, discussed
in Sec. \ref{sub:Kerr vs Spinning charge}, reflects an effect ---
frame-dragging --- which is physical, and intrinsic in the sense that
it distinguishes the Kerr metric from a static solution (e.g. the
Schwarzschild spacetime). Its \emph{non-local} nature is manifest
in the fact that in order to measure it one needs to lock the frame
to the distant stars by means of telescopes \cite{Polnarev}. Thus,
one can say that frame-dragging is manifest when at some point a system
of locally non-rotating axes (defined mathematically by the Fermi-Walker
transport law, or physically by guiding gyroscopes, see Sec. \ref{sub:Kerr vs Spinning charge})
rotates relative to an inertial frame at infinity. In other words,
when $\vec{H}$ is non-vanishing in a reference frame with axes rotationally
locked to an inertial frame at infinity (star-fixed axes). This is
however a concept that makes sense only in a special class of spacetimes.
In general, one has no way of determining the rotation of a system
of axes at one point relative to another system of axes at a different
point (since in a curved spacetime there is a priori no natural way
of comparing vectors in different tangent spaces). This is possible
only if (at least within some approximation) the spacetime admits
shear-free observer congruences. In order to see this, consider an
orthonormal tetrad frame $\mathbf{e}_{\hat{\alpha}}$, whose time
axis $\mathbf{e}_{\hat{0}}=\mathbf{u}$ is the 4-velocity of some
congruence of observers. Let $\xi^{\alpha}$ be a connecting vector
between the worldlines of the observers, $\mathcal{L}_{\mathbf{u}}\xi^{\alpha}=0$,
and $Y^{\alpha}=(h^{u})_{\ \beta}^{\alpha}\xi^{\beta}$ its space
projection; $Y^{\alpha}$ evolves in the tetrad as (Eq. (41) of \cite{PaperAnalogies})
\begin{equation}
\dot{Y}_{\hat{\imath}}=\left(\sigma_{\hat{\imath}\hat{\jmath}}+\frac{1}{3}\theta\delta_{\hat{\imath}\hat{\jmath}}+\omega_{\hat{\imath}\hat{\jmath}}-\Omega_{\hat{\imath}\hat{\jmath}}\right)Y^{\hat{\jmath}}\ .\label{eq:connectVector}
\end{equation}
If the congruence is rigid ($\sigma_{\hat{\imath}\hat{\jmath}}=\theta=0$),
and one chooses spatial triads $\mathbf{e}_{\hat{\imath}}$ co-rotating
with the observers, $\omega_{\hat{\imath}\hat{\jmath}}=\Omega_{\hat{\imath}\hat{\jmath}}$
(see Sec. \ref{sub:GEM-fields}), $Y^{\alpha}$ is constant in the
tetrad, $\dot{Y}_{\hat{\imath}}=0$. Hence the triads $\mathbf{e}_{\hat{\imath}}$
point to fixed neighboring observers. If the congruence is inertial
at infinity, this means (since it is rigid) that the $\mathbf{e}_{\hat{\imath}}$
are locked to an inertial frame at infinity. Hence, by measuring the
precession of a gyroscope with respect to the local axes $\mathbf{e}_{\hat{\imath}}$,
one is in fact measuring it with respect to the distant stars, and
it has thus a clear meaning in terms of frame-dragging. If the congruence
is not rigid but only expands (i.e., no traceless shear, $\sigma_{\hat{\imath}\hat{\jmath}}=0$),
then $\dot{Y}_{\hat{\imath}}=\theta Y_{\hat{\imath}}/3$; i.e., $Y^{\alpha}$,
albeit not constant, has a fixed direction on the tetrad, so similar
arguments still apply. When the congruence shears ($\sigma_{\hat{\imath}\hat{\jmath}}\ne0$),
however, one has no way of locking the frame to an inertial frame
at infinity, and therefore the gravitomagnetic field measured in a
frame adapted to such congruence generically has no relevant physical
meaning.

We thus conclude that if an asymptotically flat spacetime admits a
shear-free observer congruence which is inertial at infinity, the
frame adapted to it has axes fixed with respect to the distant stars,
and the gravitomagnetic field $\vec{H}$ measured therein has a meaning\footnote{This is not the only gravitomagnetic field that has a meaning in terms
of frame-dragging. For instance, the gravitomagnetic field $\vec{H}_{{\rm LNR}}$
measured in the so-called ``locally non-rotating frames'' considered
in \cite{PaperAnalogies,SemerakInertial,SemerakForcesGyro}, associated
to a shearing congruence (the zero angular momentum observers), and
where $\Omega_{\hat{\imath}\hat{\jmath}}\ne\omega_{\hat{\imath}\hat{\jmath}}$
(the triads $\mathbf{e}_{\hat{\imath}}$ are tied to the background
symmetries), signals frame-dragging and vanishes in a static spacetime;
however, this frame is not tied to the distant stars, thus it does
not correspond to the gravitomagnetic field under experimental scrutiny.} in terms of precession of gyroscopes and deflection of test particles
with respect to the distant stars {[}this is the case of any post-Newtonian
frame to 1PN order, as $\sigma_{ij}=0$ for the rest observers in
the 1PN metric \eqref{eq:PNmetric}{]}. Now, the connection with the
curvature invariants and with $\mathbb{H}_{\alpha\beta}$ is the following.
Since $\vec{H}$ is twice the vorticity of the observers, cf. Eq.
(\ref{eq:GEMfields}), the vanishing of $\vec{H}$ requires the congruence
to be vorticity-free (also known as a \emph{``normal''} congruence).
If the spacetime is conformally flat (i.e., $C_{\a\b\g\d}=0$ everywhere)
then there are as many shear-free normal congruences as in flat spacetime,
as follows from Eq.~(6.15) in \cite{StephaniExact}. Assume now the
generic case that $C_{\a\b\g\d}\ne0$. From Eq. (110) of \cite{PaperAnalogies},
we have, in the %
tetrad frame above,
\begin{equation}
\mathcal{\mathcal{H}}_{\hat{\imath}\hat{\jmath}}=-\nabla_{(\hat{\jmath}}^{\perp}\omega_{\hat{\imath})}+\delta_{\hat{\imath}\hat{\jmath}}\nabla^{\perp}\cdot\vec{\omega}+2G_{(\hat{\jmath}}\omega_{\hat{\imath})}+\nabla_{\hatl}^{\perp}\s_{\hatm(\hati}\epsilon_{\ \ \hat{\jmath})}^{\hat{l}\hat{m}}\;.\label{eq:Weyltetrad}
\end{equation}
Hence, relative to shear- and vorticity-free observer congruences
($\s_{\a\b}=\omega^{\a}=0$), the magnetic part of the Weyl tensor
vanishes, $\mathcal{H}_{\alpha\beta}=0$ (cf. also \cite{Trumper},
Theorem 3). Thus the Weyl tensor is necessarily purely electric, which
comes down to the conditions 
\begin{equation}
\star\mathbf{C}\cdot\mathbf{C}=0\ ,\qquad\ \mathbf{C}\cdot\mathbf{C}>0,\label{eq:InvPEWeyl}
\end{equation}
plus the Weyl generalization of \eqref{eq:grav-PEPMcond} (see Sec.
\ref{sub:Rie-gen}), or equivalently to one of the conditions \eqref{eq:Weyl-PE-D}
or \eqref{eq:Weyl-PE-I}. These are not however, in general, \emph{sufficient}
conditions for the existence of shear- and vorticity-free congruences
(they only ensure that $\mathcal{H}_{\alpha\beta}=0$ for some $u^{\alpha}$).
For instance, as we have seen in Sec.~\ref{sub:UniformGMs} the metrics
\eqref{ultrastationary}-\eqref{eq:Heisenberg} are Riemann and thus
Weyl purely electric {[}see \eqref{Eij-Godel-SomRay} and \eqref{eq:extr-gravmag-curv}{]}
but do not admit a shear- and vorticity-free observer congruence.
Only in the special case of vacuum (or Einstein, $R_{\alpha\beta}=\Lambda g_{\alpha\beta}$)
Petrov type D solutions, it is known (see \cite{WyllemanBeke}, Theorem
2.1, Appendix B and proposition B.1 therein) that the invariant conditions
(\ref{eq:InvPEWeyl}), when they hold in some \emph{open 4-D spacetime
region}, are indeed sufficient to ensure the existence of shear- and
vorticity-free congruences. And since, in vacuum, $\mathbf{C}=\mathbf{R}$%
, one can say that when $\innst{R}=0,\,\inn{R}>0$ in some open 4-D
region, a shear-free normal congruence exists therein. If such congruence
is inertial at infinity, then this means that there is a frame rotationally
locked to the distant stars (and where frame dragging is a well-defined
notion) where $\vec{H}$ globally vanishes.

This is all one can say about $\vec{H}$ based on the curvature invariants.
It is of limited applicability for the astrophysical systems under
discussion. Among the systems studied in the present paper, only the
fields of a single non-spinning/spinning body can be seen as Petrov
type D vacua, as they are approximately described by the Schwarzschild/Kerr
solutions (as for the two-body metric in Sec. \ref{sub:Two-masses,-planar},
although the exact solution is not known, its post-Newtonian limit
is already incompatible with the type D at any point off the Earth-Sun
axis, as we have seen therein). Schwarzschild's solution is purely
electric everywhere, so there are indeed shear-free normal congruences
everywhere, one of them the static observers $\mathbf{u}\propto\partial/\partial t$.
That is, $\vec{H}=0$ everywhere relative to the static observers.
In the case of Kerr spacetime, the only purely electric region outside
the horizon is the equatorial plane; this is a 3-D hypersurface, not
an \emph{open} 4-D spacetime region. Hence, in spite of $\star\mathbf{R}\cdot\mathbf{R}=0$
at the equatorial plane, there is no congruence which is shear- and
vorticity-free therein; and the fact that $\star\mathbf{R}\cdot\mathbf{R}\ne0$
elsewhere implies that such congruences do not exist at all in this
spacetime. This means that $\vec{H}\ne0$ in a frame adapted to any
non-shearing congruence of observers in the Kerr spacetime.

\subsection{New criteria for intrinsic/extrinsic gravitomagnetism}

\label{sub:criteria}

Given the interest on these notions in the literature, and the unsatisfactory
character of the existing ones, in this section we propose new criteria
for extrinsic/intrinsic gravitomagnetism. Similarly to previous approaches
in the literature \cite{Gravitation and Inertia,Ciufolini LLR}, we
start from the observation of the situation for electromagnetic fields
in flat spacetime to get insight, but devise criteria that are more
\emph{physically} motivated and that make sense in view of knowledge
gathered in the previous sections.

For electromagnetism in \emph{flat} spacetime, the following classification
seems reasonable: 
\begin{enumerate}
\item[{\bf a)}] globally extrinsic (intrinsic) magnetic field: there is (there is
not) a globally inertial frame where $\vec{B}=0$ everywhere in the
region of interest. Example of globally extrinsic $\vec{B}$: Coulomb
field of a point charge. 
\item[{\bf b)}] Locally extrinsic (intrinsic) magnetic field: there are (there are
not), at the given point, observers measuring $\vec{B}=0$. Amounts
to the notion of ``purely electric'' field, given by the invariant
conditions ii) of Sec. \ref{sec:Electromagnetic-Scalar-Invariant}.
Examples of globally intrinsic but locally extrinsic magnetic field:
equatorial plane of a spinning charge; motion plane of two charged
bodies in co-planar motion. Example of (globally, and at every point
locally) intrinsic magnetic field: field of spinning charge outside
the equatorial plane. 
\end{enumerate}
Note that a) implies b), but not the other way around. The distinction
between globally/locally extrinsic, and casting globally inertial
frames as preferred in this context, seems to make sense from the
analysis in Sec. \ref{sec:Interpretation_Invariants_EM}, as indeed
there is a substantial difference between e.g. the Coulomb field of
a point charge and the field in the equatorial plane of a spinning
charge. In the former, $\vec{B}=0$ everywhere in the inertial rest
frame of the charge; this may be cast as the vanishing of $\vec{B}$
everywhere for a family of observers all with the ``same'' 4-velocity
(the observers ``at rest with respect to the charge''), since in
flat spacetime we have a well-defined notion of parallelism,\footnote{Namely parallelism with respect to the Levi-Civita connection. It
amounts to saying that two observers have the same 4-velocity if $u^{\alpha}=u'^{\alpha}$
in a rectangular coordinate system.} and can thus talk about the relative velocity of distant observers.
In the case of a spinning charge, as we have seen in Sec. \ref{sub:A-spinning-spherical},
$\vec{B}$ can be made to vanish everywhere in the equatorial plane,
but \emph{not in an inertial frame}; only with respect to shearing
observer congruences (of angular velocity~\eqref{eq:EMvplot}). With
respect to an inertial frame, $\vec{B}$ vanishes only at a point
(different in general for different inertial frames). That is, observers
exist for which $\vec{B}=0$, but their 4-velocity differs from point
to point.

To generalize this to curved spacetime, the obvious difficulty is
that there are no globally inertial frames, and the parallelism of
vectors (thus the relative velocity of observers) at different points
is not a well-defined notion. There is a \emph{local} notion of difference
(with respect to the Levi-Civita connection) between the 4-velocities
of (infinitesimally close) neighboring observers in a congruence,
which is given by, cf. Eqs. (\ref{eq:Kinematics}),\,(\ref{eq:Kinematics-Decomp}),
\begin{equation}
\nabla_{\mathbf{X}}u^{\alpha}=-\epsilon_{\alpha\beta\gamma\delta}X^{\beta}\omega^{\gamma}u^{\delta}+\sigma_{\alpha\beta}X^{\beta}+\frac{\theta}{3}X^{\alpha}\ ,\label{eq:DiffVelCongruence}
\end{equation}
for any spatial vector $X^{\alpha}$ orthogonal to $u^{\alpha}$ ($X^{\alpha}u_{\alpha}=0$).
That tells us that the observer's 4-velocity differs from that of
its neighbors when the congruence has shear, expansion or vorticity.
However, congruences where they all vanish do not exist in general,
as is well known %
(%
in vacuum, in particular, this would require the spacetime to be locally
static, see Theorem 4 in \cite{Trumper} and Theorem 2 in \cite{Barnes}
for Petrov type I, and Theorem 2.2 in \cite{WyllemanBeke} for Petrov
type D). We propose generalizing criteria a)-b) to general relativity
by replacing ``globally inertial frames'' by ``shear-free frames''
(i.e., allowing the preferred frame to have vorticity and expansion,
but no traceless shear). The justification is that such replacement,
in flat spacetime, leaves the above classification unchanged for all
the examples studied (which would not be the case for vorticity-free
or expansion-free frames\footnote{For instance the velocity field (\ref{eq:EMvplot}) for which $\vec{B}=0$
in the equatorial plane of a spinning charge is vorticity and expansion
free, see Footnote \ref{fn:Kinematics_obs_noB}; however $\vec{B}\ne0$
(except at a point) in this plane with respect to any inertial frame.}). Moreover, shearfree (not vorticity-free or expansion-free) frames
are the case of post-Newtonian frames (to 1PN order), which may be
regarded as the closest entity in a curved spacetime to the globally
inertial frame of flat spacetime.

In this generalized form, the criteria can be closely mirrored for
the gravitational field. 
\begin{table*}
\begin{tabular}{|c|c|l|}
\hline 
\multicolumn{2}{|c|}{\raisebox{3.5ex}{}\raisebox{0.5ex}{\textbf{Electromagnetism} (flat
spacetime)}} & \raisebox{0.5ex}{Examples}\tabularnewline
\hline 
\hline 
\multicolumn{2}{|c|}{\raisebox{8ex}{}\raisebox{3ex}{%
\begin{tabular}{c}
Globally extrinsic $\vec{B}$\tabularnewline
{\footnotesize{}{}{}(inertial frames exist where $\vec{B}=0$ globally)}\tabularnewline
\end{tabular}}} & \raisebox{3ex}{%
\begin{tabular}{c}
$\bullet$ EM field of any static charge distribution\tabularnewline
(e.g Coulomb field)\tabularnewline
\end{tabular}} \tabularnewline
\hline 
\raisebox{-1.8ex}{Globally intrinsic $\vec{B}$}  & \raisebox{0.7ex}{%
\begin{tabular}{c}
Locally extrinsic\tabularnewline
{\footnotesize{}{}{}(observers measuring $\vec{B}=0$)}\tabularnewline
\end{tabular}}  & \raisebox{0.5ex}{%
\begin{tabular}{l}
$\bullet$ equatorial plane of spinning charge\tabularnewline
$\bullet$ motion plane of 2-body systems\tabularnewline
\end{tabular}}\tabularnewline
\cline{2-3} 
\raisebox{8ex}{}\raisebox{5ex}{%
\begin{tabular}{c}
{\footnotesize{}{}{}(no inertial frames}\tabularnewline
{\footnotesize{}{}{}where $\vec{B}=0$ globally)}\tabularnewline
\end{tabular}}  & \raisebox{3.5ex}{%
\begin{tabular}{c}
Locally intrinsic\tabularnewline
{\footnotesize{}{}{}($\vec{B}\ne0$ for all observers)}\tabularnewline
\end{tabular}}  & \raisebox{3.5ex}{%
\begin{tabular}{l}
$\bullet$ spinning charge outside equatorial plane\tabularnewline
$\bullet$ 2-body systems outside motion plane \tabularnewline
\end{tabular}}\tabularnewline
\hline 
\hline 
\multicolumn{2}{|c|}{\raisebox{4ex}{}\raisebox{0.7ex}{\textbf{Gravity}}} & \raisebox{0.7ex}{Examples}\tabularnewline
\hline 
\multicolumn{2}{|c|}{%
\begin{tabular}{c}
Globally extrinsic $\mathbb{H}_{\alpha\beta}$\tabularnewline
{\footnotesize{}{}{}(shearfree frames exist where $\mathbb{H}_{\alpha\beta}=0$
globally)}\tabularnewline
\end{tabular}} & %
\begin{tabular}{l}
$\bullet$ any locally static spacetime (e.g Schwarzschild)\tabularnewline
$\bullet$ vacuum spacetimes with globally extrinsic $\vec{H}$\tabularnewline
$\bullet$ spacetimes with uniform $\vec{H}$ (e.g. Gödel)\tabularnewline
$\bullet$ FLRW metrics\tabularnewline
\end{tabular}\tabularnewline
\hline 
\raisebox{-2ex}{Globally intrinsic $\mathbb{H}_{\alpha\beta}$}  & \raisebox{0.7ex}{%
\begin{tabular}{c}
Locally extrinsic\tabularnewline
{\footnotesize{}{}{}(observers measuring $\mathbb{H}_{\alpha\beta}=0$)}\tabularnewline
\end{tabular}}  & \raisebox{0.5ex}{%
\begin{tabular}{l}
$\bullet$ equatorial plane of spinning body\tabularnewline
$\bullet$ orbital plane of 2-body systems\tabularnewline
\end{tabular}}\tabularnewline
\cline{2-3} 
\raisebox{8ex}{}\raisebox{5ex}{%
\begin{tabular}{c}
{\footnotesize{}{}{}(no shearfree frames}\tabularnewline
{\footnotesize{}{}{}where $\mathbb{H}_{\alpha\beta}=0$ globally)}\tabularnewline
\end{tabular}}  & \raisebox{3.5ex}{%
\begin{tabular}{c}
Locally intrinsic\tabularnewline
{\footnotesize{}{}{}($\mathbb{H}_{\alpha\beta}\ne0$ for all observers)}\tabularnewline
\end{tabular}}  & \raisebox{3.5ex}{%
\begin{tabular}{l}
$\bullet$ spinning body outside equatorial plane\tabularnewline
$\bullet$ 2-body systems outside orbital plane \tabularnewline
\end{tabular}}\tabularnewline
\hline 
\multicolumn{2}{|c|}{\raisebox{8ex}{}\raisebox{3ex}{%
\begin{tabular}{c}
Globally extrinsic $\vec{H}$\tabularnewline
{\footnotesize{}{}{}(shearfree frames exist where $\vec{H}=0$ globally)}\tabularnewline
\end{tabular}}} & \raisebox{3.2ex}{%
\begin{tabular}{l}
$\bullet$ any locally static or spherically symmetric\tabularnewline
 spacetime (e.g Schwarzschild)\tabularnewline
$\bullet$ \emph{vacuum} spacetimes with globally extrinsic $\HH_{\a\b}$\tabularnewline
$\bullet$ all conformally flat spacetimes (e.g.~FLRW)\tabularnewline
\end{tabular}} \tabularnewline
\hline 
\raisebox{-4ex}{Globally intrinsic $\vec{H}$}  & Exact theory  & %
\begin{tabular}{l}
$\bullet$ Kerr spacetime everywhere\tabularnewline
$\bullet$ any vacuum open spacetime region with\tabularnewline
locally intrinsic $\mathbb{H}_{\alpha\beta}$\tabularnewline
$\bullet$ Gödel universe\tabularnewline
\end{tabular}\tabularnewline
\cline{2-3} 
\raisebox{0ex}{%
\begin{tabular}{c}
{\footnotesize{}{}{}(no shearfree frames}\tabularnewline
{\footnotesize{}{}{}where $\vec{H}=0$ globally)}\tabularnewline
\end{tabular}}  & %
\begin{tabular}{c}
PN theory: locally extrinsic\tabularnewline
{\footnotesize{}{}{}(PN frames where $\vec{H}=0$ at a point)}\tabularnewline
\end{tabular} & %
\begin{tabular}{l}
$\bullet$ equatorial plane of spinning body\tabularnewline
$\bullet$ orbital plane of 2-body systems\tabularnewline
\end{tabular}\tabularnewline
\cline{2-3} 
 & %
\begin{tabular}{c}
PN theory: locally ``intrinsic''\tabularnewline
{\footnotesize{}{}{}($\vec{H}\ne0$ in all PN frames)}\tabularnewline
\end{tabular} & %
\begin{tabular}{l}
$\bullet$ spinning body outside equatorial plane\tabularnewline
$\bullet$ 2-body systems outside orbital plane \tabularnewline
\end{tabular}\tabularnewline
\hline 
\end{tabular}\protect\protect\protect\protect\protect\protect\caption{\label{tab:NClassification}Proposed classification scheme for magnetic
field $\vec{B}$, gravitomagnetic tidal tensor $\mathbb{H}_{\alpha\beta}$,
and gravitomagnetic field $\vec{H}$. Note that ``globally extrinsic''
implies ``locally extrinsic'' everywhere in the region of interest,
but not the other way around. Locally intrinsic implies globally intrinsic,
but not the other way around. (The examples given pertain to the systems
studied in this paper, thus are not exhaustive.) }
\end{table*}

Starting with the curvature tensor, 
\begin{enumerate}
\item[{\bf c)}] globally extrinsic (intrinsic) gravitomagnetic \emph{curvature}:
there is (there is not) a non-shearing congruence of observers measuring
$\mathbb{H}_{\alpha\beta}=0$ everywhere within the region of interest.
Examples of globally extrinsic: all (locally or globally \cite{Cilindros})
static spacetimes, e.g. Schwarzschild; FLRW metrics; uniform gravitomagnetic
fields (e.g. Gödel universe). 
\item[{\bf d)}] Locally extrinsic (intrinsic) gravitomagnetic \emph{curvature} --
there are (there are not), at the given point, observers measuring
$\mathbb{H}_{\alpha\beta}=0$. In (non-flat) vacuum amounts to the
notion of ``purely electric curvature'', given by condition ii)
of Sec.~\ref{sub:Rie-vac}; in general, it coincides with one of
the possibilities \ref{sub:Petrov-type-O}-\ref{sub:Petrov-type-I}
in the criterion of Sec.~\ref{sub:Rie-gen}. Examples of globally
intrinsic but locally extrinsic magnetic curvature: equatorial plane
of spinning body; orbital plane of 2-body systems. Example of (globally,
and at every point locally) intrinsic magnetic curvature: Kerr spacetime
outside the equatorial plane. 
\end{enumerate}
Note that c) implies d), but not the other way around. Both globally
and locally extrinsic gravitomagnetic curvature require the Weyl tensor
to be zero or purely electric and are thus of Weyl-Petrov type O,
D or I, see the criterion of Sec.~\ref{sub:Rie-gen} for the local
case. In Appendix \ref{app:extr-gravmag-curv} we have determined
the observers measuring $\HH_{\a\b}=0$ in case criterion d) is satisfied;
if it holds over a 4-D region then, in those subcases where a {\em
unique} observer congruence measuring $\HH_{\a\b}=0$ exists {[}namely
(b2), (d2), and (e) of Appendix \ref{app:extr-gravmag-curv}, (2-b),
(2-c) of Sec. \ref{sub:Petrov-type-D}, and Sec. \ref{sub:Petrov-type-I}{]},
it is easy to test criterion c), i.e., whether this congruence is
shear-free; if this holds true then the spacetime exhibits globally
extrinsic (else globally intrinsic) gravitomagnetic curvature; in
the other subcases criterion c) may be more difficult to test.

As for the gravitomagnetic \emph{field} $\vec{H}$: 
\begin{enumerate}
\item[{\bf e)}] globally extrinsic (intrinsic) gravitomagnetic \emph{field}: there
is (there is not) a non-shearing congruence of observers measuring
$\vec{H}=0$ everywhere inside the region of interest. It amounts
to the existence of shear- and vorticity-free observer congruences
(Sec. \ref{sub:What-the-invariants say about GEM}). Examples of globally
extrinsic: all static and all spherically symmetric spacetimes (see
Sec.~II.C of \cite{WyllemanBeke}), e.g. Schwarzschild; all conformally
flat spacetimes (e.g.~FLRW), cf.~Sec.~\ref{sub:What-the-invariants say about GEM}.
Example of globally intrinsic: Gödel universe. 
\item[{\bf f)}] Locally extrinsic (intrinsic) gravitomagnetic \emph{field} (only
for PN approximation): there are (there are not) PN frames where $\vec{H}=0$
at the given point. It amounts to the conditions $\vec{G}\cdot\vec{H}=0$,
$\vec{G}^{2}>\vec{H}^{2}$ (see Sec. \ref{sub:PN-frames-where-H_0}).
Examples of globally intrinsic but locally extrinsic $\vec{H}$: equatorial
plane of spinning body; orbital plane of 2-body systems. Example of
(globally, and at every point locally) intrinsic $\vec{H}$: metric
of spinning body outside the equatorial plane.
\end{enumerate}
Criterion e) does not translate into a condition on the invariants,
although it has a relation with the invariants of the Weyl tensor,
in the sense that shear- and vorticity-free observer congruences exist
only when the Weyl tensor is purely electric (and thus of Petrov type
D or I); but not the other way around, cf. Sec. \ref{sub:What-the-invariants say about GEM}.
An easy way to test criterion e) was given in Proposition B.1 of \cite{WyllemanBeke}
for the Petrov type D case, while for Petrov type I one simply needs
to check whether the Weyl generalization of \eqref{eq:PEPM-I} is
satisfied, and if so whether the unique observer measuring vanishing
${\cal H}_{\a\b}$ (with 4-velocity proportional to $\tI^{\a}$ given
by the Weyl generalization of \eqref{eq:u'-grav-I}) is shear- and
vorticity-free. 

\emph{Extrinsic gravitomagnetic curvature} $\mathbb{H}_{\alpha\beta}$
\emph{vs. extrinsic gravitomagnetic field} $\vec{H}$.--- In the presence
of sources, a spacetime can have globally extrinsic $\mathbb{H}_{\alpha\beta}$
whilst not globally extrinsic $\vec{H}$%
; examples are the Gödel universe or the Som-Raychaudhuri metrics
studied in Sec. \ref{sub:UniformGMs}. And the other way around; examples
are conformally flat spacetimes having a Ricci tensor not obeying
criteria (a)-(e) of Sec \ref{sub:Rie-gen} {[}so that condition (\ref{eq:extr-gravmag-curv}ii)
is not obeyed for any observer, whilst there being as many shear-
and vorticity-free observer congruences as in flat spacetime{]}, e.g.
the pure radiation metrics in \cite{BradleyEdgarRamos}. However,
in a vacuum or Einstein spacetime ($R_{\a\b}=\Lambda g_{\a\b}$),
these notions are {\em equivalent.} Indeed, by \eqref{eq:HLambda},
a globally extrinsic $\mathbb{H}_{\alpha\beta}$ ($\vec{H}$) then
comes down to the existence of a shear-free observer congruence for
which ${\cal H}_{\a\b}=0$ ($\vec{H}=2\vec{\omega}=0$). Taking an
orthonormal frame ``adapted'' to the observers and substituting
${\cal H}_{\hat{\imath}\hat{\jmath}}=\sigma_{\hat{\imath}\hat{\jmath}}=0$
together with the Einstein space conditions $\rho+p={\cal J}_{\hat{\imath}}=0=\pi_{\hat{\imath}\hat{\jmath}}$
into the differential Bianchi identity \eqref{eq:DivHij} one obtains
$\mathcal{E}_{\hat{\imath}\hat{\jmath}}\omega^{\hat{\jmath}}=0$;
if $\omega^{\hat{\jmath}}\neq0$ then ${\cal Q}_{~\b}^{\a}={\cal E}_{~\b}^{\a}$,
seen as an operator in the rest space of the observer, would have
an eigenvalue 0, in contradiction with a result of Brans~\cite{Brans}\footnote{Brans stated his result for pure vacuum, but the proof is unaltered
when adding a cosmological constant.}; hence $\omega^{\hat{\imath}}=H^{\hat{\imath}}/2=0$. Conversely,
by (\ref{eq:Weyltetrad}), $\omega^{\hat{\imath}}=0=\sigma_{\hat{\imath}\hat{\jmath}}$
implies ${\cal H}_{\hat{\imath}\hat{\jmath}}=0$, and the equivalence
is established. This generalizes the proof made in \cite{HerreraFrameDragging1997}
considering the special case of rigid congruences in vacuum axistationary
spacetimes. In general, when $\vec{H}$ is globally extrinsic, it
means that there is a non-shearing frame (the frame adapted to the
shear- and vorticity-free congruence) relative to which all gyroscopes
whose center of mass is at rest do \emph{not} precess. If such frame
is inertial at infinity, then it means that no gyroscope at rest in
such frame precesses with respect to the distant stars.%

Criterion f) has no relation with any field invariants (see Sec. \ref{sub:PN-frames-where-H_0})
and is a notion that makes sense only in the framework of the post-Newtonian
approximation. One might argue that no inertial fields should ever
be dubbed ``locally intrinsic'', as they can always be made to vanish
by switching to a locally inertial frame. Still this notion (as long
as limited to the PN framework), seems useful to distinguish the situation
in static spacetimes from e.g. the equatorial plane of the field of
a spinning body, or the orbital plane of a 2-body system. The gravitomagnetic
field in this regime is formally very similar to the magnetic field,
and what is said in point b) above applies to $\vec{H}$ and to the
analogous gravitational systems, replacing inertial frames by PN frames.
Hence a formally analogous criterion seems to make sense. Moreover,
$\vec{H}$ in this framework always has a meaning in terms of precession
of gyroscopes with respect to the distant stars, since the basis vectors
of PN coordinate systems are locked to inertial frames at infinity.

Regarding the astrophysical setups of interest, these criteria clearly
distinguish between the gravitational field of a single translating
non-spinning body, which has globally extrinsic gravitomagnetic curvature
and field, and the fields of a spinning body or of a system of two
bodies orbiting each other; but not between these last two fields,
as they both have gravitomagnetic curvature and field which is globally
intrinsic, locally extrinsic in the equatorial/orbital planes, and
locally intrinsic, generically, elsewhere. The proposed scheme is
summarized in Table \ref{tab:NClassification}.

\section{Conclusion}

Motivated by the recent interest in the curvature invariants and their
formal analogies with the invariants of the Maxwell tensor, in the
context of the debate on the notions of ``intrinsic''/``extrinsic''
gravitomagnetism and their detection in solar system based experiments
and astronomical observations, we thoroughly discussed in this work
the invariants, their mathematical meaning and physical interpretation,
and what they actually tell us about the motion of test particles.

We started with a rigorous discussion of the algebraic meaning of
the invariants. The quadratic invariants of the Maxwell tensor give
conditions for the existence of observers for which one of the fields
(magnetic or electric) vanishes; an explicit expression {[}Eq. (\ref{eq:explicitv}){]}
for their velocities%
{} was derived. The invariants of the Riemann tensor in vacuum, and
of the Weyl tensor in general, are analogously related with conditions
for the existence of observers for which the corresponding magnetic
or electric parts vanish. The explicit expressions for the velocities
of such observers were also obtained, which, for a special class of
spacetimes {[}Petrov type D vacua, Eq. (\ref{eq:vGrav}){]}, exhibit
a strong formal analogy with the electromagnetic counterpart. In the
gravitational case, however, the quadratic invariants are not sufficient.
For the vacuum Riemann tensor (or the Weyl tensor in general) such
conditions involve also the cubic invariants, and the invariants are
even insufficient if a certain relation between them holds {[}see
\eqref{eq:distinctII-D} and \eqref{eq:grav-PEPM1}-\eqref{eq:grav-PEPMcond}
in Sec.\ \ref{sub:Rie-vac}{]}. For the Riemann tensor in the presence
of sources, we have derived a set of conditions for the vanishing
of $\HH_{\a\b}$, which involve moreover the scalar invariants of
the Ricci tensor (whose algebraic classification is reformulated),
and fully determined the corresponding observers. These are seen to
consist of the intersection between the set of observers measuring
a vanishing magnetic part of the Weyl tensor and the time-like eigenspace
of the Ricci tensor. But here the curvature invariants do not suffice
either, a fortiori. A consequence of this, concerning the proposal
in the literature \cite{Gravitation and Inertia,Ciufolini LLR,KopeikinInv}
of using the Chern-Pontryagin invariant $\star\mathbf{R}\cdot\mathbf{R}$
as a probe for intrinsic gravitomagnetism, is that even though its
non-vanishing has a clear meaning, implying that $\mathbb{H}_{\alpha\beta}\ne0$
for all observers, the converse is not true, i.e., the condition $\star\mathbf{R}\cdot\mathbf{R}=0$
\emph{alone} does not have a special significance (it does not ensure
that $\mathbb{H}_{\alpha\beta}=0$ for some observer, even in vacuum).
Thus, even prior to physical considerations, one notes that such criteria
are based on incomplete conditions.

Then we investigated the physical principles behind the behavior of
the invariants in different systems, with emphasis on $\star\mathbf{F}\cdot\mathbf{F}$
and $\star\mathbf{R}\cdot\mathbf{R}$, and the question of why $B^{\alpha}$
and $\mathbb{H}_{\alpha\beta}$ vanish for certain observers in some
systems and not in others. An explanation based on a loose notion
of relative motion has been suggested in \cite{Gravitation and Inertia}
(p. 358): \emph{``spacetime geometry and the corresponding curvature
invariants are affected and determined, not only by mass-energy, but
also by mass-energy currents relative to other mass, that is, mass-energy
currents not generable nor eliminable by any Lorentz transformation''}
(with a similar explanation for the electromagnetic case). This is
not, however, satisfactory, since in a curved spacetime the relative
motion of distant objects is not possible to define unambiguously
(as there is no global notion of parallelism). Different definitions
of relative velocity have been proposed in the literature (see \cite{BolosIntrinsic,BolosLightlike});
however, a direct relation of any of these with the curvature invariants
seems to be ruled out by simple arguments.\footnote{From the intrinsic relative velocities proposed in \cite{BolosIntrinsic},
only one yields a symmetric notion of rest (i.e., $A$ being comoving
with $B$ implies $B$ to be comoving with $A$), and is not transitive
(i.e. $A$ being comoving with $B$, and $B$ being comoving with
$C$, does \emph{not} mean that $A$ comoves with $C$). } We looked instead into the field equations --- the Maxwell equations
for $B^{\alpha}$, in its general form for arbitrary frames in arbitrary
spacetimes, and, on the gravitational side, the so-called ``higher
order field equations'' for $\mathcal{H}_{\alpha\beta}$ and $\mathbb{H}_{\alpha\beta}$
--- since they are always valid, and checked what one can say about
the invariants based on them (points \ref{enu:Empoint1}-\ref{enu:EMpoint4}
of Secs. \ref{sec:Interpretation_Invariants_EM} and \ref{sec:Interpretation-of-the-Gravitational}).
Concerning the explanation above, our results show that it is \emph{partially}
correct, but as a feature of the weak field slow motion approximation:
if the system is such that a post-Newtonian frame exists where mass
currents $\vec{\mathcal{J}}$ vanish everywhere, then (to 1PN accuracy)
$\star\mathbf{R}\cdot\mathbf{R}=0$; the converse, however, is \emph{not
true} (i.e., when there is no PN frame where $\vec{\mathcal{J}}=0$
everywhere, that does not ensure $\star\mathbf{R}\cdot\mathbf{R}\ne0$).
This is in close analogy with the electromagnetic invariants in \emph{flat}
spacetime: if the setup is such that an inertial frame exists where
all currents (charge and displacement) are zero, then $\star\mathbf{F}\cdot\mathbf{F}=0$;
but the converse is not true. In the more general cases of electromagnetic
fields in a curved spacetime, or gravity outside the PN regime, things
are more complicated because one has no inertial or PN frames, and
in generic frames the observer's vorticity, shear and expansion act
as sources of the fields (in addition to the currents).

We studied and physically interpreted the structure of the invariants
in the astrophysical setups of interest --- the field of a single
non-spinning body (the one effectively involved in the measured gravitomagnetic
effects in binary pulsars, and in the geodetic precession of the Earth-Moon
system in the Sun's field), described by the Schwarzschild solution,
the field of a system of two bodies to first post-Newtonian order
(involved in the LLR measurements of the Moon's orbit), and the exact
Kerr field (which approximately describes the field of a spinning
body, namely the Earth) --- and their electromagnetic analogues: the
exact fields of single non-spinning and spinning charged bodies, and
two-body systems to first ``post-Coulombian'' approximation. The
electromagnetic analogy proved illuminating to explain the structure
of the gravitational invariants; to post-Newtonian order, in particular,
a similar reasoning can be employed. We found that the invariant structure
of the field of a single non-spinning body (where $\star\mathbf{R}\cdot\mathbf{R}=0$
everywhere) is clearly different from that of a spinning body, in
agreement with the analysis in \cite{Gravitation and Inertia,Ciufolini LLR};
but that the latter (contrary to the claim in \cite{Ciufolini LLR})
is not substantially different from that of a system of two bodies
orbiting each other: $\star\mathbf{R}\cdot\mathbf{R}=0$ in the equatorial/orbital
plane, $\star\mathbf{R}\cdot\mathbf{R}\ne0$ generically elsewhere.
In the post-Newtonian framework, this structure can actually be physically
explained in both cases using the same reasoning. This closely mirrors
the situation in electromagnetism and can be traced back to the fact
that $\mathbb{H}_{\alpha\beta}$ is linear to 1PN order, such that
a superposition principle applies (like in electromagnetism) and one
can treat for these matters a spinning body as an assembly (in the
spirit of \cite{MurphyNordtvedtPRL1}) of translating mass elements.
We hope this may shed some light on this issue.

However, in spite of the insight it gives into the invariant structures,
it is crucial to realize that the analogy between the invariants of
$F_{\alpha\beta}$ and $R_{\alpha\beta\gamma\delta}$ is purely \emph{formal},
as it relates objects that do not play analogous \emph{physical} roles
in the two theories. The effects involved are different, and may actually
be opposite (Sec. \ref{sub:Kerr vs Spinning charge}). Focusing on
magnetism/gravitomagnetism (and taking, as probes, magnetic dipoles/gyroscopes),
the invariants of $F_{\alpha\beta}$ give conditions for the existence
of velocity fields for which the magnetic \emph{field} $B^{\alpha}$
vanishes, i.e., for which magnetic dipoles do not undergo Larmor precession
(but they feel a force in general, as the magnetic \emph{tidal tensor}
$B_{\alpha\beta}$ is non-vanishing for a particle moving in an inhomogeneous
field); the invariants of $R_{\alpha\beta\gamma\delta}$ give (in
vacuum) conditions for the existence of velocity fields for which
the gravitomagnetic tidal tensor $\mathbb{H}_{\alpha\beta}$ vanishes,
i.e., a gyroscope feels no force (not that it does not precess relative
to the ``distant stars''). Hence the use of the invariants and the
electromagnetic analogy in the discussion \cite{Gravitation and Inertia,Ciufolini LLR,KopeikinInv}
of gyroscope precession and the gravitomagnetic deflection of test
particles (that have been under experimental and observational scrutiny)
is essentially misguided, as these are effects governed by the gravitomagnetic
field $H^{\alpha}$ (the dynamical analogue of $B^{\alpha}$), not
the tidal tensor $\mathbb{H}_{\alpha\beta}$. One should not confuse
GEM \emph{inertial fields} with \emph{tidal tensors}; a pedagogical
example are spacetimes with uniform $H^{\alpha}$, e.g. the Gödel
universe (Sec. \ref{sub:UniformGMs}), where one has, in a rigid frame,
$\mathbb{E}_{\alpha\beta}=0$ and $\mathbb{H}_{\alpha\beta}=0$ everywhere,
whilst $G^{\alpha}=0$ and $H^{\alpha}\ne0$ (being purely electric
from the point of view of the curvature, and exactly the opposite
in terms of the GEM fields).

The curvature invariants are locally measurable quantities that are
built on GEM tidal tensors, \emph{not} on inertial fields, \textcolor{black}{which
are reference frame artifacts (that vanish in locally inertial frames),
and as such cannot be directly reflected in invariants}. Yet still
there are special cases where one can infer about the gravitomagnetic
field $\vec{H}$ from the invariants. $\vec{H}$ has a clear meaning
in terms of gyroscope precession and test particle deflection relative
to the distant stars if it is measured in a shear-free frame which
is inertial at infinity. As discussed in Sec. \ref{sub:What-the-invariants say about GEM},
in a vacuum Petrov type D spacetime, shear-free frames where $\vec{H}$
is globally zero (i.e., shear- and vorticity-free observer congruences)
exist if and only if $\star\mathbf{R}\cdot\mathbf{R}=0$ and $\mathbf{R}\cdot\mathbf{R}>0$
hold in an \emph{open 4-D region}. Concerning the astrophysical setups
of interest, this tells us that, in the Kerr spacetime (describing
approximately the field of a spinning body), $\vec{H}\ne0$ with respect
to any non-shearing frame, in any open 4-D region; and that in the
Schwarzschild spacetime there are shear-free frames (e.g. the one
adapted to the static observers, $\mathbf{u}\propto\partial/\partial t$,
which is star fixed), where $\vec{H}=0$ everywhere. Hence, in these
two special cases, one can indeed imply, based on the curvature invariants,
a distinction between them in terms of gravitomagnetic field $\vec{H}$/frame
dragging, which, to some extent, supports the claim in \cite{Gravitation and Inertia}.
But since, amongst the systems under discussion, these are the only
ones of Petrov type D (it does not apply to the 2-body system, or
others in general), this is all one can tell, from the invariants
(with the present knowledge), about $\vec{H}$.

We note (Sec. \ref{sub:PN-frames-where-H_0}), on the other hand,
that in the post-Newtonian regime (which pertains to all the gravitomagnetic
effects detected to date, and those that one hopes to detect in the
near future), criteria for the vanishing or not of $\vec{H}$ \emph{in
PN frames}, based on the ``scalars'' $\vec{G}^{2}-\vec{H}^{2}$
and $\vec{G}\cdot\vec{H}$, formally analogous to the electromagnetic
invariants, can be devised. These quantities, however, are not field
invariants, nor do they have a straightforward relation with the curvature
invariants. Such analogy originates instead from the similarity between
the transformation laws for the post-Newtonian GEM fields and those
for the electromagnetic fields.

Concluding, curvature invariants tell us about the gravitomagnetic
tidal field (magnetic curvature); they do not tell us directly (or
at all, in general), about the gravitomagnetic field $H^{\alpha}$
and the frame-dragging effects that have been under experimental scrutiny,
which are based on spin precession and orbital effects (including
precession) caused by the gravitomagnetic ``force'' $\vec{v}\times\vec{H}$
on test particles in (approximately) \emph{geodesic} motion. \textcolor{black}{Appropriate
probes to measure magnetic curvature would be the }\textcolor{black}{\emph{force}}\textcolor{black}{{}
on a gyroscope, or gravity gradiometers, as proposed in e.g. \cite{Polnarev,Gradiometers}.}

\subsection*{Acknowledgments}

We thank I.~Ciufolini, M.~Soffel, S.~Klioner and G.~Anglada for
the discussions that motivated this work, E. Minguzzi for defining
discussions in its early stages, and N. Van den Bergh for useful remarks
on this manuscript. We are especially indebted to O. Semerák for his
generous and careful reading of the manuscript, and the many discussions,
suggestions and remarks that decisively contributed to it. L.F.C.
and J.N. were supported by FCT/Portugal through projects UID/MAT/04459/2019,
UIDB/MAT/04459/2020, and UIDP/MAT/04459/2020. L. F. C. was funded
by FCT through grant SFRH/BDP/85664/2012.

\appendix

\section{Algebraic classification of the Ricci tensor\label{app:extr-gravmag-curv}}

We revise here the algebraic classification of the Ricci tensor. We
take the eigenvalue degeneracy of the traceless Ricci operator $S_{~\b}^{\a}\equiv R_{~\b}^{\a}-\tfrac{1}{4}R\delta_{~\b}^{\a}$
on (complexified) tangent space at $p$, using the invariants (\ref{Ric-invars}).
The characteristic polynomial of $S_{~\b}^{\a}$ is 
\[
k(x)\equiv x^{4}-\tfrac{1}{2}I_{6}x^{2}-\tfrac{1}{3}I_{7}x+\tfrac{1}{8}I_{6}{}^{2}-\tfrac{1}{4}I_{8}.
\]
Its discriminant equals $(\IS{}^{3}-\JS{}^{2})/432$, so there is
a degenerate eigenvalue if and only if\footnote{The factors 64 and 27 in the corresponding equation (2.9) of \cite{JolyMacC90}
should be omitted.} 
\begin{equation}
\IS{}^{3}=\JS{}^{2}.\label{discr=00003D0}
\end{equation}
Suppose $\l$ is such eigenvalue. Because of $S_{~\a}^{\a}=0$ the
four eigenvalues can then be written as $\l,\l,-\l+\AA$ and $-\l-\AA$,
and one has 
\begin{align*}
 & I_{6}=4\l^{2}+2\AA^{2},\quad I_{7}=-6\l\AA^{2},\quad I_{8}=4\l^{4}+12\AA^{2}\l^{2}+2\AA^{4},\\
 & \IS=4(\AA^{2}-4\l^{2})\,,\quad\JS=8(\AA^{2}-4\l^{2})^{3}\,,
\end{align*}
which explicitly verifies \eqref{discr=00003D0} and implies $\JS+2I_{6}\IS=24\AA^{2}(\AA^{2}-4\l^{2})^{2}$.
This leads to four main cases, fully characterized by the invariant
conditions indicated in square brackets, and with several subcases
based on the real or non-real character of the eigenvalues and the
minimal polynomial $m(x)$ of $S_{~\b}^{\a}$:\footnote{\label{foot:energycond} Including also case (E), some subcases cannot
occur due to the Lorentzian signature of the metric, namely $m(x)=k(x)$
in (A), (C) and (D) with $\AA^{2}<0$, and one pair resp.~two pairs
of non-real complex conjugate eigenvalues in (C) and (E). If the weak
energy condition is assumed then those cases with two non-real eigenvalues,
i.e., (D) with $\AA^{2}<0$ and (E) with $\IS^{3}<\JS^{2}$, cannot
occur either (see Ch.~5 of \cite{StephaniExact}).} 
\begin{enumerate}
\item[(A)] $\l=\AA=0$ {[}$\IS=\JS=0=I_{6}${]}. There is one quadruple eigenvalue
$\l=0$. The allowed minimal polynomials are $m(x)=x^{p}$ with $p\in\{1,2,3\}$. 
\item[(B)] $\AA^{2}=4\l^{2}\neq0$ {[}$\IS=\JS=0\neq I_{6}$, implying $I_{6}>0${]}.
There is one triple, real eigenvalue $\l=-I_{7}/(2I_{6})\neq0$ and
one simple, real eigenvalue $-3\l$. The allowed minimal polynomials
are $m(x)=(x-\l)^{p}(x+3\l)$ with $p\in\{1,2,3\}$. 
\item[(C)] $\AA=0\neq\l$ {[}$\JS+2I_{6}\JS=I_{7}=0\neq I_{6}$, implying $I_{6}>0${]}.
There are two double, real eigenvalues $\l=\pm\sqrt{I_{6}}/2\neq0$
and $-\l$. The allowed minimal polynomials are $m(x)=(x-\l)^{p}(x+\l)$
with $p\in\{1,2\}$. 
\item[(D)] $(\AA^{2}-4\l^{2})\AA\neq0$ {[}$\IS{}^{3}=\JS{}^{2},\,\JS+2I_{6}\IS\neq0${]}.
There is one double, real eigenvalue $\l=-I_{7}\IS/(\JS+2I_{6}\IS)$
and two simple eigenvalues $-\l\pm c$, with $\AA^{2}=I_{6}/2-2\l^{2}$.
The simple eigenvalues are real (non-real and complex conjugate) when
$\AA^{2}>0$ ($\AA^{2}<0$), and the allowed polynomials $m(x)$ are
$(x-\l)^{p}((x+\l)^{2}-\AA^{2})$ with $p\in\{1,2\}$ ($p=1$). 
\end{enumerate}
Finally one has the non-degenerate case 
\begin{enumerate}
\item[(E)] $\IS{}^{3}\neq\JS{}^{2}$. All eigenvalues are simple, such that
$m(x)=k(x)$. If $\IS{}^{3}>\JS{}^{2}$ all four eigenvalues are real,
while if $\IS{}^{3}<\JS{}^{2}$ two are real and two non-real and
complex conjugate. 
\end{enumerate}
\noindent Considering the further split of the subcases b) with $p=1$
and d) with $p=1,\,\AA^{2}>0$ into two branches each (see below)
we retrieve the fifteen algebraic Ricci types listed in Table 5.1
of \cite{StephaniExact}.

The condition that $S_{~\beta}^{\alpha}$ is diagonalizable corresponds
to $p=1$ in (A)-(D) and is automatic in (E) {[}see footnote \ref{foot:minimal-poly});
adding the condition that all eigenvalues are real {[}which only gives
a further restriction in cases (D) and (E){]} one arrives at the conditions
(a)-(e) in Sec. \ref{sub:Rie-gen}.\footnote{Note that in the respective cases (b)-(d) of Sec. \ref{sub:Rie-gen}
the polynomials $m(x)=(x-\l)(x+3\l),\,x^{2}-I_{6}/4$ and $(x-\l)((x+\l)^{2}-\AA^{2})$
are annihilating and of degree less than 4, such that there is at
least one degenerate eigenvalue by footnote \ref{foot:minimal-poly}.
The conditions $I_{6}\neq0$ in (b), (c) and $\JS+2I_{6}\IS\neq0$
in (d) of Sec. \ref{sub:Rie-gen} now ensure that there are at least
two, resp.~three different eigenvalues, such that $m(x)$ must be
the minimal polynomial, having the distinct eigenvalues as roots (see
again footnote \ref{foot:minimal-poly}). The complete list of eigenvalues
then follows from $S_{~\a}^{\a}=0$; e.g., when $m(x)=(x-\l)(x+3\l)$
the complete list must be $[\l,\,\l,\,\l,-3\l]$ and we are in case
(b) of the above Ricci classification.} In the extended Segre notation of \cite{StephaniExact} these cases
precisely cover Ricci type $[111,1]$ and its degenerations, as indicated
below. Here symbols 1 enclosed by round brackets refer to coinciding
eigenvalues, and the 1 after the comma refers to the eigenvalue corresponding
to the unique timelike eigenspace $\TT$ of $S_{\ \b}^{\a}$, which
is also indicated and contains the 4-velocities of precisely those
observers $\Oup$ for which (\ref{eq:extr-gravmag-curv}ii)-\eqref{A2}
holds.
\begin{enumerate}
\item[(a)] coincides with type $[(111,1)]$ (Einstein space type, $R_{\alpha\beta}=\Lambda g_{\alpha\beta}$),
where $\TT$ is the full 4-D tangent space.
\item[(b)] splits into types $[1(11,1)]$ and $[(111),1]$. $\l$ is the triple
eigenvalue. Take any $y^{\a}$ such that $w^{\a}\equiv S_{~\b}^{\a}y^{\b}-\l y^{\a}\neq0$;
then $w^{\a}$ spans the eigendirection of the simple eigenvalue $-3\l$,
and the type is 

\begin{enumerate}
\item[(b1)] $[1(11,1)]$ (tachyonic fluid type) if $w^{\a}w_{\a}>0$, where $\TT$
is the $\l$-eigenspace, i.e., the 3-D orthogonal complement of $w^{\a}$; 
\item[(b2)] $[(111),1]$ (perfect fluid type) if $w^{\a}w_{\a}<0$, where $\TT$
is 1-D and spanned by $w^{\a}$. 
\end{enumerate}
\item[(c)] is type $[(11)(1,1)]$ (non-null Einstein-Maxwell type), where $\l_{\pm}=\pm\sqrt{I_{6}}/2$
are the double eigenvalues and $\TT$ is 2-D. Compute $w^{\a}=S_{~\b}^{\a}u^{\b}-\l_{+}u^{\a}$
for any timelike $u^{\a}$; if $w^{\a}w_{\a}<0$ then $\l_{-}$ corresponds
to $\TT$, else $\l_{+}$. 
\item[(d)] splits into types $[11(1,1)]$ and $[(11)1,1]$. $\lambda$ is the
double eigenvalue and $-\lambda\pm\AA$ are the simple ones. Take
any $y_{\pm}^{\a}$ such that $w_{\pm}^{\a}\equiv(S_{~\b}^{\a}+(\l\pm\AA)\delta_{~\b}^{\a})(S_{~\g}^{\b}y_{\pm}^{\g}-\l y_{\pm}^{\b})\neq0$;
then $w_{+}^{\a}$ and $w_{-}^{\a}$ span the eigendirections of the
simple eigenvalues, and the type is 

\begin{enumerate}
\item[(d1)] $[11(1,1)]$ if $w_{+}^{\a}(w_{+})_{\a}w_{-}^{\b}(w_{-})_{\b}>0$,
where both $w_{+}^{\a}$ and $w_{-}^{\a}$ are spacelike 
and $\TT$ is the 2-D orthogonal complement of $w_{+}^{\a}$ and $w_{-}^{\a}$; 
\item[(d2)] $[(11)1,1]$ if $w_{+}^{\a}(w_{+})_{\a}w_{-}^{\b}(w_{-})_{\b}<0$,
where one of $w_{+}^{\a},\,w_{-}^{a}$ is timelike (the other one
being spacelike) and spans the 1-D space $\TT$. 
\end{enumerate}
\item[(e)] is type $[111,1]$, where all eigenvalues $\l_{i},\,i=1..4$ are
simple. For each $i$ take $y_{i}^{\a}$ such that $w_{i}^{\a}\equiv\prod_{j\neq i}S_{~\b}^{\a}y_{i}^{\b}-\l_{j}y_{i}^{\a}\neq0$;
then one of the vectors $w_{i}^{\a}$ is timelike (the other ones
being spacelike) and spans the 1-D space $\TT$. %

\end{enumerate}

\subsection{Locally extrinsic gravitomagnetic curvature --- alternative route\label{sub:Alternative route}}

The existence of observers $u^{\alpha}$ for $\mathbb{H}_{\alpha\beta}=0$
at a point $p$ (locally extrinsic gravitomagnetic curvature) requires
one of the conditions (a)-(e) of Sec. \ref{sub:Rie-gen} to hold in
any case, i.e., the Ricci type must be $[111,1]$ or one of its degenerations
{\em regardless of the Weyl-Petrov type}. This can be easily tested
and it is useful to do this a priori, along with verifying the necessary
Weyl condition $C_{\a\b\g\d}=0$ or \eqref{eq:Weyl-PE-D} or \eqref{eq:Weyl-PE-I}. 

Alternatively, for each allowed Ricci type, we can look when there
are observers with 4-velocity satisfying (\ref{eq:extr-gravmag-curv}ii)
(i.e., those in the timelike Ricci eigenspace ${\cal T}$) that also
satisfy (\ref{eq:extr-gravmag-curv}i); this is especially useful
when the Weyl tensor is of Petrov type I, where the use of $\tI^{\a}$
may be cumbersome due to the possibly intricate expressions for the
Weyl eigenvalues \eqref{eq:L-alpha}. If the Ricci type is $[(111,1)]$
(vacuum type) the criterion reduces to the validity of either \eqref{eq:PEPM-I}
or \eqref{eq:PEPM-D} or $R_{\a\b\g\d}=0$. If the Ricci type is $[(111),1]$
(perfect fluid type), $[(11)1,1]$ or $[111,1]$ then there is only
one 4-velocity $u'^{\a}$ that verifies \eqref{A2}, namely the one
proportional to $w^{\a}$ constructed in (b2), the timelike $w_{+}^{\a}$
or $w_{-}^{\a}$ in (d2), or the timelike $w_{i}^{\a}$ in (e), respectively;
hence {\em if the spacetime is of one of these Ricci types at $p$
then it has locally extrinsic gravitomagnetic curvature if and only
if (\ref{eq:extr-gravmag-curv}i) or equivalently $\star{C}_{\a\b\g\d}w^{\g}w^{\d}=0$
holds}, where $w^{\a}$ is joint notation for the timelike vector
mentioned in each case; this alternative criterion is especially useful
for Ricci types $[(111),1]$ or $[(11)1,1]$, where the construction
of $w^{\a}$ is straightforward, whereas type $[111,1]$ may be more
difficult to treat if Descartes' method for solving quartics is to
be used to find the distinct Ricci eigenvalues $\l_{i}$. For Ricci
types $[(11)(1,1)]$ (Einstein-Maxwell type) or $[11(1,1)]$ the space
${\cal T}$ is 2-D and the vectors $u'^{\a}$ satisfying \eqref{A2}
are given by 
\begin{equation}
u'^{\a}=(q\kt^{\a}+\lt^{\a})/\sqrt{2q}\,,\qquad q>0,\label{u'-Ric}
\end{equation}
where the null vectors $\kt^{\a}$ and $\lt^{\a}$ are chosen along
the null directions of ${\cal T}$ and normalized by $\kt^{\a}\lt_{\a}=-1$.
In terms of the symmetric tensors 
\begin{align*}
 & A_{\a\b}\equiv\star{C}_{\a\g\b\d}\kt^{\g}\kt^{\d},\quad B_{\a\b}\equiv\star{C}_{\a\g\b\d}(\kt^{\g}\lt^{\d}+\lt^{\g}\kt^{\d}),\\
 & C_{\a\b}\equiv\star{C}_{\a\g\b\d}\lt^{\g}\lt^{\d},\quad A^{\a}\equiv A^{\a\b}\lt_{\b},\quad C^{\a}=C^{\a\b}\kt_{\b}
\end{align*}
the condition (\ref{eq:extr-gravmag-curv}i) for such $u'^{\a}$ becomes
\begin{equation}
A_{\a\b}q^{2}+B_{\a\b}q+C_{\a\b}=0.\label{cond-Ric-D}
\end{equation}
By contraction with $\kt^{\a}$ (or $\lt^{\a}$) and use of $A_{\a}=-B_{\a\b}\kt^{\b}$
(or $C_{\a}=-B_{\a\b}\lt^{\b}$) this implies 
\[
qA^{\a}=C^{\a}.
\]
Note that the vectors $A^{\a}$ and $C^{\a}$ are orthogonal to ${\cal T}$,
and are thus spacelike or zero. Moreover, if $A^{\a}=C^{\a}=0$ then
$B_{\a\b}=0$ (because of $\star{C}_{~\a\g\b}^{\g}=0$) and $A_{\a\b}$
and $C_{\a\b}$ are orthogonal to ${\cal T}$ (i.e., $A_{\a\b}\kt^{\b}=A_{\a\b}\lt^{\b}=0$
and thus $A_{\a\b}=0$ or $A^{\a\b}A_{\a\b}>0$, and analogously for
$C_{\a\b}$). It follows that {\em a spacetime that is of Ricci
type $[(11)(1,1)]$ or $[11(1,1)]$ at $p$ has locally extrinsic
gravitomagnetic curvature if and only if one of the following holds}:\\
 \vspace{-0.2cm}

(i) $A_{\a\b}=B_{\a\b}=C_{\a\b}=0$;

(ii) $A^{\a}=C^{\a}=0\;(\Rightarrow B_{\a\b}=0)$, $A^{\a\b}C_{\a\b}<0$,
\eqref{cond-Ric-D}

with $q=-\frac{A^{\a\b}C_{\a\b}}{A^{\g\d}A_{\g\d}}$ (i.e., $A^{\g\d}C_{\g\d}A_{\a\b}=A^{\g\d}A_{\g\d}C_{\a\b}$);

(iii) $A^{[\a}C^{\b]}=0$, $A^{\a}C_{\a}>0$, \eqref{cond-Ric-D}
with $q=\frac{A^{\a}C_{\a}}{A^{\b}A_{\b}}$.\\
 \vspace{-0.2cm}

In case (i) the Weyl-Petrov type is necessarily O or D, where the
Weyl principal plane $\Sigma$ equals ${\cal T}$ for type D; \eqref{eq:extr-gravmag-curv}
is satisfied by {\em all} observers $\Oup$ with 4-velocity \eqref{u'-Ric}
in this case, and by the {\em unique} observer given by the respective
indicated values of $q$ in cases (ii) and (iii). Finally, for Ricci
type $[1(11,1)]$ (tachyonic fluid type) one needs to suitably parameterize
the 2-D variety of the unit timelike vectors $u'^{\a}$ within the
3-D space ${\cal T}$ (i.e., the orthogonal complement of the unique,
spacelike simple eigenvector $w^{\alpha}$) and check when at least
one of these vectors solves \eqref{eq:extr-gravmag-curv}; for purely
electric Petrov type D {[}i.e., when \eqref{eq:Weyl-PE-D} is verified
to hold{]} this is the case for all (respectively exactly one) $u'^{\alpha}\in\Sigma$
when $w^{\a}$ is orthogonal to $\Sigma$ (respectively the projection
of $v^{\a}$ onto $\Sigma$ is spacelike, the unique observer lying
in $\Sigma$ and orthogonal to it). In the Petrov type I case this
is more involved and left to be treated on a case-by-case basis.

\end{document}